\documentclass[12pt,english]{article}
\usepackage[toc,page]{appendix}
\usepackage[small]{titlesec}
\usepackage[english]{babel}
\usepackage{amsmath,amssymb,amscd,amsfonts}
\usepackage{mathtools}
\usepackage[small,nohug,heads=vee]{diagrams}
\usepackage{graphicx}
\usepackage{appendix}
\usepackage{url}
\usepackage{xspace}
\usepackage{enumitem}
\usepackage{cite}
\diagramstyle[labelstyle=\scriptstyle]
\numberwithin{equation}{section}
\newcommand{\addtotoc}[2]{
    \phantomsection
    \addcontentsline{toc}{section}{#1}
    #2 \clearpage
}
\date{}

\newcommand{\cB}{{\mathcal B }}

\newcommand{\cD}{{\mathcal D }} 
\newcommand{\cE}{{\mathcal E }} 
\newcommand{\cF}{{\mathcal F }}            
\newcommand{\cG}{{\mathcal G }}
\newcommand{\cH}{{\mathcal H }}

\newcommand{\cN}{{\mathcal N }}            
            
\newcommand{\cP}{{\mathcal P }} 
\newcommand{\cQ}{{\mathcal Q }}

\newcommand{\cT}{{\mathcal T }} 
 
\newcommand{\cV}{{\mathcal V }}    
\newcommand{\cW}{{\mathcal W }}

\newcommand{\cZ}{{\mathcal Z }} 
\newcommand{\cd}{\mathord{\cdot}}
\def\eq{\begin{equation}}
\def\en{\end{equation}}
\def\eqa{\begin{eqnarray}}
\def\ena{\end{eqnarray}}
\def\aeq#1{\begin{align}#1\end{align}}  % for amsmath
\def\ateq#1#2{\begin{alignat}{#1}#2\end{alignat}}  % for amsmath

\def\expval#1{\langle \, #1 \,\rangle}

\def\Complexes{\mathbb{C}}		%Complex numbers
\def\Reals{\mathbb{R}}			%Reals
\def\Integers{\mathbb{Z}}		%Integers
\DeclareMathOperator{\Ker}{Ker}	%Kernel

\def\integral{\mathrm{int}}

\def\flat{\mathit{flat}}
\def\nat{\mathit{nat}}
\def\tot{\mathit{tot}}
\newcommand{\overbar}[1]{\mkern2mu\overbracket[0.25pt][-1pt]{\mkern-2mu#1\mkern-5mu}\mkern 5mu}

\DeclarePairedDelimiter\norm{\lVert}{\rVert}%
\mathchardef\mhyphen="2D

\def\Ex{\cE_{\partial\xi}}
\def\ExC{\cE_{\partial\xi}^{\Complexes}}
\def\Ezero{\cE_{0}}

\def\QSigma{\cQ(\Sigma)}
\def\QSigmaplus{\cQ(\Sigma_{+})}
\def\QxZ{\cQ_{\Integers\partial\xi}}
\def\PB{\cP\cB}

\def\QM{\cQ(M)}

\def\PBM{\PB(M)}
\def\ZQ{\cZ(\cQ^{\integral}_{1})}

\def\hc{C}

\DeclareMathOperator{\Iso}{\mathbf{Iso}}
\DeclareMathOperator{\Aut}{\mathbf{Aut}}

\DeclareMathOperator{\Ima}{Im}

\def\PoneC{\mathbb{C}\cup\{\infty\}}
\def\IM#1#2{I_{M}\expval{#1,#2}}
\def\ISigma#1#2{I_{\Sigma}\expval{#1,#2}}
\def\IQ#1#2{I_{\cQ}\expval{#1,#2}}
\def\Conf{\mathbf{Conf}}

\def\xiint{\xi^{{\mathit{int}}}}
\newcounter{MQ}

\title
{\vspace{-1cm}
\vspace{2cm}
\LARGE Quantum field theories of extended objects
}

\author
{
Daniel Friedan
\\
\\
New High Energy Theory Center\\
and Department of Physics and Astronomy\\
Rutgers, The State University of New Jersey\\
Piscataway, New Jersey 08854-8019, USA.\\
\\
The Science Institute\\
The University of Iceland\\
Reykjavik, Iceland
}

\begin{document}

\maketitle
\thispagestyle{empty}

\begin{center}
{\bf Abstract}
\end{center}

First steps are taken in a project to construct a general class of
conformal and perhaps, eventually, non-conformal quantum field theories of
$(n{-}1)$-dimensional extended objects in a $d{=}2n$ dimensional
conformal space-time manifold $M$.  The fields live on the spaces
$\Ex$
% $\Ex=\cD^{\integral}_{n-1}(M)_{\partial\xi}\subset \cD^{\integral}_{n-1}(M)$ 
of relative integral $(n{-}1)$-cycles in $M$.
These are the integral $(n{-}1)$-currents of given boundary $\partial\xi$.
Each $\Ex$ is a
complete metric space geometrically analogous to a Riemann
surface $\Sigma$.
For example, if  $M=S^{d}$,
$\Sigma = S^{2}$.
The quantum fields on $\Ex$ are to be mapped to observables in 
a 2d CFT on $\Sigma$.
The correlation functions on $\Ex$  are to be given 
by the 2d correlation functions on $\Sigma$.
The goal is to construct a CFT of extended objects in
$d{=}2n$ dimensions for every 2d CFT, 
and eventually a non-conformal QFT of extended objects for every 
non-conformal 2d QFT, 
so that
all the technology of 2d QFT can be applied to the construction and
analysis of quantum field theories of extended objects.
The project depends crucially on settling some
mathematical questions about analysis in the spaces
$\Ex$.
The project also depends on extending the observables of 2d CFT
from the finite sets of points in a Riemann surface to the integral 0-currents.

% This is the first step in a program to construct a general class of
% conformal field theories of $(n{-}1)$-dimensional extended objects in
% $d=2n$ space-time dimensions, comprising
% at a minimum one such conformal field 
% theory in dimension $d=2n$
% for every
% two-dimensional conformal field theory.

\newpage
\tableofcontents
\newpage

\section{Introduction}

This paper reports the first steps of a project to construct 
and analyze a general class of quantum field theories of 
$(n{-}1)$-dimensional extended objects in a space-time manifold of 
dimension $d=2n$.
Much  still remains to be done.

My referencing is surely inadequate.
I hope to do better in future revisions.
Section \ref{sect:hist-refs-questions} asks for advice on
history and references.

% This section of the paper sketches the program.
% The next section asks questions about history and references.
% The rest of the paper elaborates on the first steps of the 
% construction.
\subsection{The free $n$-form on a conformal space-time manifold $M$ of 
dimension $d=2n$}
\label{sect:intro1}

The first step is to reformulate the conformally invariant quantum
field theory of a free $n$-form $F(x)$ on a space-time manifold $M$ of
dimension $d=2n$ as the quantum field theory of a free $1$-form on a
space of $(n{-}1)$-dimensional extended objects.
The general project extrapolates from this example.

Space-time is taken to be a compact real manifold $M$ of dimension $d=2n$ with an
orientation and with a conformal class of riemannian metrics.  The main
example is euclidean $\Reals^{d}$ or, rather, its conformal compactification, 
the $d$-sphere $S^{d} = \Reals^{d}\cup\{\infty\}$.
Actually, all that is used of the conformal structure on $M$ is the Hodge 
$*$-operator acting in the middle dimension, on
$n$-forms.  This might be less structure than a conformal
class of riemannian metrics.

The free $n$-form theory is the generalization to $d=2n$ dimensions of free quantum
electrodynamics in 4\mbox{-}d  
\cite{Maxwell:1865zz,Born:1926:QIG,Dirac:1931kp}.
The field equations of the free $n$-form $F(x)$ are
\eq
dF =0
\,,\qquad
d F^{*}=0
\,,\qquad
F^{*} = i^{-1}{*}F
\,,
\en
where $*$ is the conformally invariant Hodge $*$-operator acting on 
$n$-forms,
\eq
*\omega_{\mu_{1}\ldots \mu_{n}}(x) = 
\epsilon_{\mu_{1}\ldots \mu_{n}}{}^{\nu_{1}\ldots \nu_{n}}(x)
\,\omega_{\nu_{1}\ldots \nu_{n}}(x)
\,,\qquad
*^{2} = (-1)^{n}
\,.
\en
The integral of the $n$-form $F$ over an $n$-surface is the magnetic charge.
The integral of $F^{*}$ is the electric charge.
The gauge potential $A(x)$ 
and the dual gauge potential $A^{*}(x)$ are $(n{-}1)$-forms constructed by integrating
\eq
dA = F
\,,\qquad
d A^{*} = F^{*}
\,.
\en
The gauge potentials are determined up to gauge transformations
\eq
A\rightarrow A + df
\,,\qquad
A^{*}\rightarrow A^{*} + df^{*}
\,,
\en
given by $(n{-}2)$-forms $f$, $f^{*}$.

The extended objects of the free $n$-form theory 
are described by fields
\eq
V_{p,p^{*}}(\xi)
=
e^{ip\int_{\xi}A + i p^{*}\int_{\xi}A^{*}}
\,.
\en
which live on $(n{-}1)$-currents $\xi$ in space-time.
They carry electric charge $p$ and magnetic charge $p^{*}$.
Taking the gauge group $G$ to be compact, $G=U(1){\times} U(1)$,
the charges lie in integer lattices,
\eq
p= \frac{m}{R}\,,\qquad p^{*} = \frac{m^{*}}{R^{*}}\,,
\qquad m,m^{*}\in\Integers
\,.
\en
The Dirac quantization condition
\eq
R R^{*} =1
\en
follows from the requirement that the correlation functions of the
fields $V_{p,p^{*}}(\xi)$ should be single-valued.

\subsection{Currents in $M$ and the boundary operator $\partial$ on 
currents}
\label{sect:intro2}

Currents will be the basic mathematical objects of this enterprise.
A $k$-current $\xi$ in the space-time manifold $M$ is a linear 
function -- a distribution -- on $k$-forms 
$\omega$,
\eq
\int_{\xi}\omega
= \int_{M} d^{d}x \; \xi^{\mu_{1}\ldots \mu_{k}}(x) \,\omega_{\mu_{1}\ldots \mu_{k}}(x)
\,.
\en
$\cD_{k}(M)$ is the linear space of $k$-currents in $M$.
A $k$-current $\xi$ is called {\it smooth} when
$\xi(x)^{\mu_{1}\ldots\mu_{k}}$ is smooth.
Equivalently, $\xi$ is smooth when it is represented by a smooth 
$(d-k)$-form $\omega_{\xi}$,
\eq
\int_{\xi}\omega = \int_{M} \omega_{\xi} \wedge \omega\,.
\en
Since we will be considering fields that live on spaces of currents,
it will be more congenial to write differential forms as linear functions 
of currents,
\eq
\omega(\xi) = \int_{\xi}\omega
\,.
\en
When the $n$-form $F(x)$ is considered as a quantum field,
it is a distributional $n$-form
acting as a linear function
$\xi \mapsto F(\xi)$ on smooth $n$-currents in $M$.

The boundary operator on currents is
dual to the exterior derivative on forms,
\eq
\partial\colon \cD_{k}(M) \rightarrow \cD_{k-1}(M)
\,,
\en
\eq
\int_{\partial\xi}  \omega = \int_{\xi} d\omega
\,,\qquad
(\partial\xi)^{\mu_{2}\ldots \mu_{k}}(x) 
= -\partial_{\mu_{1}} \xi^{\mu_{1}\ldots \mu_{k}}(x)
\,.
\en

Examples of $k$-currents are given by $k$-dimensional 
submanifolds in $M$.  The corresponding linear function on a $k$-form 
$\omega$ is the integral of $\omega$ over the submanifold.  The 
boundary of the $k$-current corresponds to the boundary of the submanifold.

\subsection{Reformulating the free $n$-form field theory}

The reformulation of the free $n$-form theory stems from the realization that
the space of $(n{-}1)$-currents in $M$ on which the $n$-forms live
can be decomposed into a union of spaces $\Ex$
% (describing the extended objects)
each of which has
the property that its tangent space at each point
is a linear space of $n$-currents in space-time
which is closed under the action of the Hodge $*$-operator.
% and the cotangent space is the linear space of 
% $n$-forms on the space-time manifold $M$.
The spaces $\Ex$ will be described in more detail 
shortly.

% Given the identification of the cotangent spaces of $\Ex$ with the 
% space of $n$-forms on $M$,
Since tangent vectors in $\Ex$ are $n$-currents in space-time,
the $n$-form fields $F$ and $F^{*}=i^{-1}{*}F$ on the space-time manifold $M$
become 1-forms $j$ and $j^{*}=i^{-1}{*}j$ on $\Ex$,
where the Hodge $*$-operator on $n$-forms has become
a linear operator acting on the 1-forms on $\Ex$.
The field equations become
\eq
dj =0
\,,\qquad
d j^{*} =0
\,.
\en
Scalar fields, i.e., 0-forms, $\phi$ and $\phi^{*}$ on $\Ex$ are constructed by 
integrating
\eq
d \phi = j
\,,\qquad
d \phi^{*} = j^{*}
\,.
\en
The scalar fields on $\Ex$ express the gauge potentials on space-time,
\eq
\phi(\xi) = \int_{\xi}A
\,,\qquad
\phi^{*}(\xi) = \int_{\xi}A^{*}
\,.
\en
The extended objects are described by
the ``vertex operator'' fields on $\Ex$,
\eq
V_{p,p^{*}}(\xi)
=
e^{ip\phi(\xi)+ip^{*}\phi^{*}(\xi)}
\,.
\en
The free $n$-form theory on $M$ is thus reformulated as a free 
$1$-form theory on $\Ex$.
The free $n$-form theory begins to look
formally analogous to the conformal field theory of a free 
$1$-form on a two-dimensional manifold.
% The conformal field theory of a free $1$-form in two dimensions
% is traditionally called
% the {\it 2\mbox{-}d gaussian model.}

%
%Now 

\subsection{The path of generalization}

The path of generalization will retrace
the historical development of the general class of two-dimensional conformal and non-conformal quantum
field theories starting from the theory of the free 1-form.
The 2d conformal field theories that were 
constructed in that development include the theories of several
1-form fields, the orbifolds of the 1-form theories, 
and the theories constructed from all these by conformal perturbation
theory.
Notable  special cases are the 2d conformal field theories containing non-abelian current algebras.
Along a sideline are the holomorphic conformal field theories made 
from 1-form theories, including the Monster 2d CFT.
Finally, there are the non-conformal 2d quantum field theories constructed by
perturbation theory governed by the 2d renormalization group.  
If these constructions on the free 1-form in two dimensions can 
indeed be carried out
on the free 1-form on the spaces $\Ex$ of $(n{-}1)$-currents in 
space-time,
then such a
menagerie of examples will strongly suggest that it should be possible to formulate
axiomatically a general class of quantum field theories of extended
objects in one-to-one correspondence with the 2d quantum field
theories.
I will try here to describe the setting in which 
such quantum field theories of extended objects
might be formulated,
and develop mathematical conjectures
that would provide
a construction of such theories.

\subsection{Geometric Measure Theory -- flat currents and integral currents}
\label{sect:GMT}

The quantum field theories of extended objects
envisioned here are all to be theories of fields on certain spaces
$\Ex$ of currents in space-time.
The tools for defining $\Ex$ and for doing analysis in $\Ex$
come from a branch of mathematics called 
Geometric Measure Theory (GMT).
Given a  manifold $M$, Geometric Measure Theory
defines the complete metric space 
of {\it integral} $k$-currents in $M$ \cite{MR0123260}.
One of several equivalent definitions starts from maps
$\sigma$  from the oriented $k$-simplex $\Delta^{k}$ 
to $M$.
The 
$k$-current $[\sigma]$ defined by 
\eq
\int_{[\sigma]} \omega 
= \int_{\Delta^{k}} \sigma^{*} \omega
\en
is a delta-function $k$-current 
concentrated on the image $\sigma(\Delta^{k})$ in $M$.
A linear combination $\sum_{\alpha} n_{\alpha} [\sigma_{\alpha}]$
of such $k$-currents with integer coefficients $n_{\alpha}$
is the $k$-current representing the singular $k$-chain
$\sum_{\alpha} n_{\alpha} \sigma_{\alpha}$ in $M$.
The  singular $k$-currents
realize a naive idea of the $k$-dimensional objects in $M$.

Next, a certain norm, called the {\it flat} norm, is put on 
$k$-currents.
At each point $x\in M$, the riemannian metric at $x$ is used to define a euclidean length 
function $|\omega(x)|$ on the vector space of $k$-forms $\omega(x)$ at $x$.
The {\it co-mass} $M_{k}^{*}(\omega)$ of a $k$-form is defined as
\eq
M_{k}^{*}(\omega) = \sup_{x\in M} |\omega(x)|
\,.
\en
The {\it mass} $M_{k}(\xi)$ of a $k$-current is then defined as
\eq
M_{k}(\xi) =  \sup_{M_{k}^{*}(\omega) =1} |\omega(\xi)|
\,.
\en
When $\xi$ is the characteristic current of a submanifold of $M$,
the mass $M_{k}(\xi)$ is the $k$-volume of the submanifold.
The flat norm of a $k$-current is defined as
\eq
\lVert{\xi}\rVert_{\mathit{flat}}
= \inf_{\xi'}\left [ M_{k}(\xi-\partial\xi') + M_{k+1}(\xi') 
\right]
\,,
\en
where $\xi'$ ranges over all $(k{+}1)$-currents.
The flat norm gives the {\it flat metric}
on the space of $k$-currents,
\eq
d_{\mathit{flat}}(\xi_{1},\xi_{2}) = \lVert{\xi_{1}-\xi_{2}}\rVert_{\mathit{flat}}
\,,
\en
The completion in the flat metric of the vector space of finite norm $k$-currents
is the vector space of {\it flat $k$-currents}.
Roughly, a flat current is a distribution on $k$-forms that takes no 
derivatives.

The flat norm is a physically reasonable measure of the
size of a $k$-dimensional object.
A $k$-current $\xi$ of small flat norm 
is physically small in the sense that it can be shrunk away to nothing with 
little effort.
If a singular $k$-current $\xi$ is small in the flat norm, then 
there is a $(k{+}1)$-current $\xi'$ such that both
$M_{k+1}(\xi')$ and $ M_{k}(\xi-\partial\xi')$ are small.
So part of $\xi$ can easily be shrunk away through
$\xi'$, which has small $(k{+}1)$-volume,
and what remains of $\xi$ has small $k$-volume
and can easily be shrunk away through itself.
So the flat metric is a physically reasonable measure of the difference 
between two $k$-dimensional objects.
The definition of the flat metric requires a notion of distance in 
$M$, i.e.,  a choice of riemannian metric on $M$, but the flat metric 
topology on currents is the same for any choice of 
riemannian metric on $M$.

Completing the space of singular $k$-currents in the
flat metric gives the space of {\it integer rectifiable $k$-currents}.
Finally, an {\it integral $k$-current} is defined to be an integer rectifiable 
$k$-current $\xi$ whose boundary $\partial\xi$ has finite mass.
The integral currents that are not singular currents are fractal objects.

We will write $\cD_{k}^{\integral}(M)$ for the space of integral 
$k$-currents in $M$.
A basic theorem of  \cite{MR0123260} states that 
$\cD_{k}^{\integral}(M)$
is a complete metric space
and that the boundary operator  takes
integral $k$-currents to
integral $(k{-}1)$-currents,
\eq
\partial \colon \cD_{k}^{\integral}(M) \rightarrow 
\cD_{k-1}^{\integral}(M)
\,,
\en
acting continuously in the metric topology.
Moreover,
$\cD_{k}^{\integral}(M)$ is a normed abelian group under addition
of currents, i.e.,
the addition law is continuous in the flat metric topology.

We will be using only the $\cD^{\integral}_{k}(M)$ with 
$k=n-2,n-1,n,n+1$.
In order to handle the case $d=2$, $n=1$, we need to define
the $(-1)$-currents,
\eq
\cD^{\integral}_{-1}(M) = \Integers
\,,\qquad
\cD_{-1}(M) = \Complexes
\,,
\en
\eq
\label{eq:boundaryzero}
\partial\colon \cD^{\integral}_{0}(M) \rightarrow \cD^{\integral}_{-1}(M)
\,,\qquad
\partial \eta = \int_{\eta} 1 = 1(\eta)
\,.
\en
Also, there is a distinguished $d$-current which is the oriented manifold $M$ 
itself, which acts on $d$-forms $\omega$ by
\eq
\omega(M) = \int_{M}\omega
\,.
\en
We will use this $d$-current for the case $d=2$,
when it lies in $\cD^{\integral}_{n+1}(M)$.
So we define
\eq
\cD^{\integral}_{d+1}(M) = \Integers
\,,\qquad
\cD_{d+1}(M) = \Complexes
\,,
\en
\eq
\partial\colon \cD^{\integral}_{d+1}(M) \rightarrow \cD^{\integral}_{d}(M)
\,,\qquad
\partial 1 = M
\,.
\en
Now we have the augmented de Rham complex of currents
\eq
\begin{diagram}
0 	&\lTo	& \cD^{\integral}_{-1}(M) &\lTo^{\partial}	& \cD^{\integral}_{0}(M) 
&\lTo^{\partial}	& \cdots
&\lTo^{\partial}	& \cD^{\integral}_{d}(M) 
&\lTo^{\partial}	& \cD^{\integral}_{d+1}(M) 
&\lTo	& 0
\\
&& \dInto && \dInto&& &&  \dInto&&  \dInto
\\
0 	&\lTo	& \cD_{-1}(M) &\lTo^{\partial}	& \cD_{0}(M) 
&\lTo^{\partial}	& \cdots
&\lTo^{\partial} & \cD_{d}(M) 
&\lTo^{\partial}	& \cD_{d+1}(M) 
&\lTo	& 0
\end{diagram}
\en
The $\cD_{k}(M)$ will be the complex currents, but they can be taken to be real for 
$n$ odd.  
The precise characterization of the linear space of complex currents --- smooth, flat, 
distributional, etc. ---
will mostly be left unspecified, to be determined by the context.

\subsection{The bundle $\cE\xrightarrow{\partial}\cB$ of extended 
objects:  $\cE = \cD^{\integral}_{n-1}(M)$, 
$\cB=\partial\cD^{\integral}_{n-1}(M)$}
%\subsection{The space $\cE = \cD^{\integral}_{n-1}(M)$ of extended objects}

I am proposing to take as the space of extended
objects
the space
\eq
\cE=\cD_{n-1}^{\integral}(M)
\en
of integral $(n{-}1)$-currrents in $M$.
The space $\cE$ forms a bundle
\eq
%\label{eq:nfiberbundle}
\begin{CD}
\cE\\
@VV\partial V \\
\cB
\end{CD}
\en
over the space of $(n{-}2)$-boundaries,
\eq
\cB = \partial\cD_{n-1}^{\integral}(M) \subset \cD_{n-2}^{\integral}(M)
\,.
\en
The spaces $\Ex$ are the fibers of the bundle, the spaces of relative 
$(n{-}1)$-cycles,
\eq
\Ex = \partial^{-1}(\partial\xi)
=
\left \{
\xi' \in \cD_{n-1}^{\integral}(M)\colon
\partial \xi' = \partial \xi
\right \}
\,.
\en
The special fiber $\Ezero =\cD^{\integral}_{n-1}(M)_{0} $ is the space of integral 
$(n{-}1)$-cycles.

We will see that the geometry of currents in each $\Ex$
is analogous to the geometry of currents in a Riemann surface.
I will call such spaces ``quasi Riemann surfaces''.

\subsection{Disclaiming rigor}

I want to do quantum field theory on the space of extended objects,
so I need calculus and tensor analysis on $\cD_{n-1}^{\integral}(M)$.
But the spaces $\cD_{k}^{\integral}(M)$
are  not, as far as I can tell, differentiable manifolds.
On the other hand, there is a construction of
currents --- and flat currents and integral currents --- in any complete metric space \cite{MR1794185}.
So a calculus of currents in $\cD_{n-1}^{\integral}(M)$ is available,
and thus a calculus of differential forms as the duals of currents.
The spaces $\cD_{k}^{\integral}(M)$
have nice properties that lend themselves to geometric analysis --- each is a normed abelian group that is
generated as an abelian group by an arbitrarily small $\epsilon$-ball 
around $0$,
and each is embedded in a normed vector space (of flat currents).
The special case  $\cD_{n-1}^{\integral}(M)$
has even nicer properties which will be described below.

It may be that the mathematical basis for calculus and tensor analysis
on the spaces $\cD_{k}^{\integral}(M)$ already exists,
but I cannot tell
--- mathematical analysis is not exactly my cup of tea.
I will proceed naively,
without trying for mathematical rigor, blithely 
optimistic that rigor can be achieved eventually.
I will explore the possibilities of achieving the project of 
constructing quantum field theories of extended objects from 2d 
quantum field theories to the point where I can formulate a 
more or less well-posed mathematical conjecture on which the program can be based.
The mathematical conjecture is that there is a classification of equivalence 
classes of
``quasi Riemann surfaces'' that
is analogous to and  extends the classification of 
ordinary Riemann surfaces.
I have almost no evidence for the conjecture.
Its appeal is that it makes feasible the construction of quantum 
field theories of extended objects from 2d quantum field theories in 
the simplest and most direct fashion that I can imagine.
So I call it a ``wishful'' conjecture.

\subsection{Constructing a QFT of extended objects from a 2d QFT}

If the conjecture holds, then there will be an essentially unique map
from the space $\Ex$ of integral relative $(n{-}1)$-cycles in the 
space-time $M$ to the space of integral relative $0$-cycles
in a two-dimensional space $\Sigma$.  
The space $\Sigma$ will be a Riemann surface or 
something akin to one.  For example, when the space-time is
$M=S^{d}=\Reals^{d}\cup\{\infty\}$, then $\Sigma$ 
will be the Riemann sphere $S^{2}=\PoneC$.
A quantum field $\Phi(\xi)$ on the space of extended objects will correspond to a 
2d observable in the 2d quantum field theory on 
$\Sigma$ located on the integral 0-current in $\Sigma$ that 
corresponds to $\xi$.
The correlation functions of the extended object fields
will be given by the correlation functions in the 2d quantum field 
theory on $\Sigma$.
In this way,
the quantum field theory of extended objects will be constructed from 
the 2d quantum field theory.

\section{Geometry of the space
$\cD^{\integral}_{k}(M)$
of integral $k$-currents in $M$}
\label{sect:GeometryofDkM}

\subsection{The principal fiber bundle
$\cD^{\integral}_{k}(M)\xrightarrow{\partial} \partial \cD^{\integral}_{k}(M)
\subset \cD^{\integral}_{k-1}(M)$}

The space $\cD_{k}^{\integral}(M)$ is a fiber bundle
\eq
\label{eq:fiberbundle}
\begin{CD}
\cD^{\integral}_{k}(M) & \\
@VV\partial V && \\
\partial\cD^{\integral}_{k}(M) & \;\;\subset\;
\cD^{\integral}_{k-1}(M)_{0} & \;\;\subset\;
\cD^{\integral}_{k-1}(M) 
\end{CD}
\en
over the space of integral $(k{-}1)$-boundaries.
The special fiber $\cD_{k}^{\integral}(M)_{0}$
over $0\in \cD^{\integral}_{k-1}(M) $
is the space of integral $k$-cycles in $M$,
the space of integral $k$-currents without boundary,
\eq
\cD_{k}^{\integral}(M)_{0}
= 
\left \{
\xi \in \cD_{k}^{\integral}(M),\; \partial\xi =0
\right \}
\,.
\en
$\cD_{k}^{\integral}(M)_{0}$ is closed under addition,
thus an abelian group.
The other fibers $\cD^{\integral}_{k}(M)_{\xi_{0}} = 
\partial^{-1}(\xi_{0})$,
$\xi_{0}\ne 0$,
are the spaces of 
integral {\it relative} $k$-cycles in $M$.
That is, if $\xi_{1}$ is a $k$-current in the fiber over $\xi_{0}$,
%i.e., $\partial \xi_{1} = \xi_{0}$, 
then every $\xi$ in the same fiber
differs from $\xi_{1}$ by a $k$-cycle,
\eq
\partial(\xi-\xi_{1}) = 0
\,.
\en
Each fiber is isomorphic to the space of integral $k$-cycles
$\cD_{k}^{\integral}(M)_{0}$,
but not in a canonical way.
The isomorphism depends on the choice of  $\xi_{1}$ in the fiber.
The abelian group $\cD_{k}^{\integral}(M)_{0}$  acts by addition on 
$\cD^{\integral}_{k}(M)$,
preserving the fibers,
acting transitively and faithfully on each fiber.
So the fiber bundle (\ref{eq:fiberbundle}) is a principle fiber bundle 
with structure group the abelian group $\cD_{k}^{\integral}(M)_{0}$ 
of integral $k$-cycles.

\subsection{The local metric geometry of $\cD^{\integral}_{k}(M)$ is 
the same at every point}
Translation in $\cD_{k}^{\integral}(M)$ as abelian group  takes any point
to any other.
In particular, translation by $\xi$ takes $0$ to $\xi$,
for any $\xi\in \cD^{\integral}_{k}(M)$.
Translation by $\xi$ preserves the fibers of the bundle,
taking the fiber over $\partial\xi'$ to the fiber over 
$\partial(\xi'+\xi)$.
Translation by $\xi$ takes the $\epsilon$-ball $B_{\epsilon}$ around $0$ to
the $\epsilon$-ball $\xi + B_{\epsilon}$ around $\xi$.
So the local metric geometry of the principal fiber bundle
is the same at every point.

\subsection{The tangent spaces of the fibers $\cD^{\integral}_{k}(M)_{\partial\xi}$
all equal $\cV_{k+1}\subset\cD^{\flat}_{k+1}(M)$}

% \subsection{Vertical tangent vectors in 
% a fiber $\cD^{\integral}_{k}(M)_{\partial\xi}$
% are flat $(k{+}1)$-currents}
%the space of integral relative $k$-cycles}

The space $\cD_{k}^{\integral}(M)_{\partial\xi}$ of integral relative $k$-cycles
is the fiber of the principal fiber bundle $\cD^{\integral}_{k}(M)$ over
the $(k{-}1)$-boundary $\partial\xi$.
A tangent vector $\dot \xi$ in $\cD_{k}^{\integral}(M)_{\partial\xi}$ is 
a vertical tangent vector  in the principal fiber bundle.
By translation in the abelian group, the vertical tangent space is the same  at every point in the bundle,
\eq
T_{\xi}(\cD^{\integral}_{k}(M)_{\partial\xi}) = 
T_{0}(\cD^{\integral}_{k}(M)_{0})
\,,
\en
equal to the tangent space to the fiber at the distinguished point 
$0\in \cD_{k}^{\integral}(M)$.

Suppose $\dot \xi$ is the tangent vector at $\epsilon=0$ to an infinitesimal curve 
$\epsilon\mapsto \xi(\epsilon)$ in the fiber 
$\cD_{k}^{\integral}(M)_{\partial\xi}$.
Then $\xi(\epsilon)-\xi(0)$ is an
infinitesimally small integral $k$-cycle.
There is a unique integral $(k{+}1)$-current $\xi_{S}(\epsilon)$ 
of minimal mass solving
\eq
\partial \xi_{S}(\epsilon) = \xi(\epsilon)-\xi(0)
\,.
\en
It satisfies
\eq
\label{eq:secant-norm}
\lVert\xi(\epsilon)-\xi(0) \rVert_{\mathit{flat}}
= M_{k+1}(\xi_{S}(\epsilon))
= \lVert\xi_{S}(\epsilon)\rVert_{\mathit{flat}}
\,.
\en
This minimal integral $(k{+}1)$-current $\xi_{S}(\epsilon)$ can be thought of as the 
secant between $\xi(0)$ and $\xi(\epsilon)$.
The vertical tangent vector is represented by the flat $(k{+}1)$-current
\eq
\dot \xi = \lim_{\epsilon\rightarrow 0} 
\frac{\xi_{S}(\epsilon)}{\epsilon}
\,.
\en
Equation (\ref{eq:secant-norm}) implies that
the $(k{+}1)$-current  $\dot \xi$ faithfully represents the tangent 
vector to the curve $\xi(\epsilon)$,
i.e., the map from vertical tangent vectors to flat $(k{+}1)$-currents is 
injective.
So the tangent spaces to the fibers are all
equal to a certain subspace of the flat $(k{+}1)$-currents,
\eq
T_{\xi}(\cD^{\integral}_{k}(M)_{\partial\xi}) = 
T_{0}(\cD^{\integral}_{k}(M)_{0})
= \cV_{k+1} \subset \cD^{\flat}_{k+1}(M)
\,.
\en
The question then is:  exactly what subspace $\cV_{k+1}$ of flat $(k{+}1)$-currents
is formed by the tangent vectors in the space of
integral relative $k$-cycles?

The easiest  examples of vertical tangent vectors are the 
delta-function $(k{+}1)$-currents.
Work in coordinates 
$x^{\mu}$ on $M$ and let $\hat e_{\mu_{1}},\ldots, \hat 
e_{\mu_{k+1}}$ be unit tangent vectors
at $x_{0}\in M$
along $k+1$ different axes.
Let $\xi_{S}(\epsilon)$ be the $(k{+}1)$-current representing the $(k{+}1)$-parallelotope with vertex at 
$x_{0}$ and edges $\epsilon^{\frac1{k+1}}\hat e_{\mu_{1}}, \ldots, 
\epsilon^{\frac1{k+1}} \hat e_{\mu_{k+1}}$.
The tangent vector at $\xi(0)$ to the curve $\xi(\epsilon) = \xi(0) + \partial 
\xi_{S}(\epsilon)$ is the flat $(k{+}1)$-current
\eq
\label{eq:deltafntangents}
\dot \xi = 
\lim_{\epsilon\rightarrow 0} 
\frac{\xi_{S}(\epsilon)}{\epsilon}
= \delta^{d}(x-x_{0}) \hat e_{\mu_{1}}\wedge \cdots \wedge \hat 
e_{\mu_{k+1}}
\en
which is supported at the point $x_{0}$.
The value of a $(k{+}1)$-form $\omega$ on this tangent vector is
\eq
\omega(\dot \xi) = \omega(x_{0})_{\mu_{1}\cdots\mu_{k+1}}
\,.
\en
So, naively, there are at least enough tangent vectors to detect all 
$(k{+}1)$-forms.

The vector space $\cV_{k+1}$ is
the Gromov-Hausdorff tangent space.
Let $B_{\epsilon}(\cD^{\integral}_{k}(M)_{0},0)$ be the 
$\epsilon$-ball around $0$ in the space $\cD^{\integral}_{k}(M)_{0}$ of integral $k$-cycles.
The secant map described above is a map
\eq
S\colon B_{\epsilon}(\cD^{\integral}_{k}(M)_{0},0) \rightarrow 
B_{\epsilon}(\cD^{\integral}_{k+1}(M),0)
\subset \cD^{\flat}_{k+1}(M)
\,.
\en
The image of $S$ lies in the vector space 
$\cD^{\flat}_{k+1}(M)$, so it makes sense to construct
the tangent space $\cV_{k+1}$ by dilating the image of $S$ within that vector space.
The unit ball in $\cV_{k+1}$ is
\eq
B_{1}(\cV_{k+1},0) = \lim_{\epsilon\rightarrow 0}
\frac1{\epsilon} 
S(B_{\epsilon}(\cD^{\integral}_{k}(M)_{0},0))
\,,
\en
where the limit is taken in the Gromov-Hausdorff metric on metric 
spaces.
A tangent vector, an element of $\cV_{k+1}$ is a flat $(k{+}1)$-current
in $M$ whose support set is the same as the support set
of an infinitesimally small integral $(k{+}1)$-current.
%I do not know how to describe these sets of support precisely.
Naively, these should be the support sets of the integral $k'$-currents with 
$k'<k+1$.

\section{The Hodge $*$-operator on 
the tangent spaces $T_{\xi}(\cE_{\partial\xi})=\cV_{n}\subset 
\cD^{\flat}_{n}(M)$}
% \section{Does the Hodge $*$-operator act on 
% $T_{\xi}\cD^{\integral}_{n-1}(M)_{\partial\xi}=\cV_{n}\subset 
% \cD^{\flat}_{n}(M)$?}
\label{sect:Hodgestaracts}

As above, the tangent spaces to the fibers
$\Ex = \cD^{\integral}_{n-1}(M)_{\partial\xi}$
in the bundle of extended objects $\cE\rightarrow \cB$
are all equal to the vector space $\cV_{n}\subset\cD^{\flat}_{n}(M)$.
The crucial question is: does the Hodge $*$-operator
act on $\cV_{n}$?
The Hodge $*$-operator obviously acts on the tangent vectors that are 
linear combinations of the delta-function $n$-currents,
\eq
\dot \xi = 
\delta^{d}(x-x_{0}) \hat e_{\mu_{1}}\wedge \cdots \wedge \hat e_{\mu_{n}}
\;\mapsto\;
{*}\dot \xi = 
\delta^{d}(x-x_{0}) 
\epsilon_{\mu_{1}\ldots \mu_{n}}{}^{\nu_{1}\ldots \nu_{n}}(x_{0})
\hat e_{\nu_{1}}\wedge \cdots \wedge \hat  e_{\nu_{n}}
\,,
\en
since any $n$-vector can multiply the delta-function.
But consider a tangent vector $\dot \xi$ whose support set is at the 
opposite extreme, the support of an integral current of dimension 
$n-1$.
For simplicity, 
suppose $d=4$, $n=2$, and suppose space-time is euclidean 
$\Reals^{4}$. Let $\xi_{S}(\epsilon)$ be the 2-current 
\eq
\xi_{S}(\epsilon) = \delta(x^{1})\delta(x^{2}) \theta_{[0,1]}(x^{3}) \theta_{[0,\epsilon]}(x^{4})
\hat e_{3}\wedge \hat e_{4}
\,,
\en
representing 
a $1\times \epsilon $ rectangle in the 3-4 plane.
Here $\theta_{[a,b]}$ is the characteristic function of the 
interval $[a,b]\subset \Reals$.
Then $\xi(\epsilon) = \partial \xi_{S}(\epsilon) $ is the 1-current 
representing the boundary 
of the rectangle.
The tangent vector to the curve $\xi(\epsilon)$ is the flat $2$-current
\eq
\dot \xi = 
\lim_{\epsilon\rightarrow 0} 
\frac{\xi_{S}(\epsilon)}{\epsilon}
=\delta(x^{1})\delta(x^{2}) \theta_{[0,1]}(x^{3}) \delta(x^{4})
\hat e_{3}\wedge \hat e_{4}
\,,
\en
whose support is the interval $[0,1]$ in the 3-axis.
The $*$-operator acts on this tangent vector to give
\eq
\label{eq:badtangent}
* \dot \xi = 
\delta(x^{1})\delta(x^{2}) \theta_{[0,1]}(x^{3}) \delta(x^{4})
\hat e_{1}\wedge \hat e_{2}
\,.
\en
It is clear that the $2$-current $*\dot\xi$ cannot be the tangent vector to a 
curve of singular $1$-currents, to a curve of naive one dimensional objects.

Appendix \ref{app:GOT} contains the construction of a curve of {\it integral} $1$-currents that 
has the flat $2$-current $*\dot \xi$ of (\ref{eq:badtangent}) as tangent vector.
So this  $*\dot \xi$ does lie in the tangent space $\cV_{n}$.
The construction depends essentially on
the metric completion of the space of integral currents.
The possibility of this construction 
was the main motivation for taking the extended objects
to be the integral $(n{-}1)$-currents in the general case.
Appendix \ref{app:GOT} goes on to explain how
this example might serve as the germ of a rigorous proof that the Hodge $*$-operator 
takes all of $\cV_{n}$ to itself, for any manifold $M$ of any 
dimension $d=2n$.
Essentially, the construction of Appendix \ref{app:GOT} gives a basis for
showing that the vertical tangent space $\cV_{n}$ consists of 
{\it all} flat $n$-currents supported on integral $(n{-}1)$-currents.
Then, since $*$ acts continuously on flat $n$-currents and does not 
change their supports, $*$ would act on $\cV_{n}$.

I will assume that the Hodge $*$-operator does act on  $\cV_{n}$ and 
thus on all the tangent spaces of the $\Ex$.  The whole enterprise rests on this
assumption, so it is especially urgent that the mathematical question be settled
one way or the other.

\section{Currents in $\cD^{\integral}_{k}(M)$}
\label{sect:CurrentsinDkM}

We will suppose that \cite{MR1794185},
which constructs the space of currents
in a complete metric space,
gives a calculus of currents in $\cD^{\integral}_{k}(M)$
that includes
\begin{itemize}
\item spaces $\cD^{\integral}_{j}(\cD_{k}^{\integral}(M))$ of 
integral $j$-currents in $\cD^{\integral}_{k}(M)$
contained in the spaces
$\cD_{j}(\cD_{k}^{\integral}(M))$
of complex currents,
with properties analogous to those of the
currents in $M$,
and
\item spaces of $j$-forms on $\cD^{\integral}_{k}(M)$
dual to the spaces 
of $j$-currents in $\cD^{\integral}_{k}(M)$,
\item natural linear maps
\ateq{2}{
\partial_{*}: \qquad &&\partial_{*}^{j,k} \colon \cD^{\integral}_{j}(\cD_{k}^{\integral}(M))  &\rightarrow
\cD^{\integral}_{j}(\cD_{k-1}^{\integral}(M))
\\[1ex]
\Pi_{*}: \qquad &&\Pi_{*}^{j,k}\colon \cD^{\integral}_{j}(\cD_{k}^{\integral}(M)) &\rightarrow
\cD^{\integral}_{j+k}(M)
}
which satisfy
\eq
\partial (\Pi_{*}^{j,k} \eta) = 
\Pi_{*}^{j-1,k} (\partial \eta)
+ \Pi_{*}^{j,k-1} (\partial_{*}^{j,k} \eta)
\,,\quad
j\ge 1
\,.
\en
\end{itemize}

\subsection{$\partial_{*} \colon \cD^{\integral}_{j}(\cD_{k}^{\integral}(M))  \rightarrow
\cD^{\integral}_{j}(\cD_{k-1}^{\integral}(M))$
}

Composing with the boundary operator in $M$, $\partial\colon \cD^{\integral}_{k}(M)
\rightarrow \cD^{\integral}_{k-1}(M)$, takes an integral $j$-current in 
$\cD^{\integral}_{k}(M)$ to an integral $j$-current in 
$\cD^{\integral}_{k-1}(M)$,
\eq
\partial_{*}^{j,k} \colon \cD^{\integral}_{j}(\cD_{k}^{\integral}(M))  \rightarrow
\cD^{\integral}_{j}(\cD_{k-1}^{\integral}(M))
\,.
\en

\subsection{$\Pi_{*}\colon \cD^{\integral}_{j}(\cD_{k}^{\integral}(M)) \rightarrow
\cD^{\integral}_{j+k}(M)$}

There is natural map ``pushing down'' an integral $j$-current in 
$\cD_{k}^{\integral}(M)$ to give an integral $(j{+}k)$-current in $M$,
\eq
\Pi_{*}^{j,k}\colon \cD_{j}^{\integral}(\cD_{k}^{\integral}(M)) \rightarrow
\cD_{j+k}^{\integral}(M)
\,,
\en
based on the fact that a map from the simplex $\Delta^{j}$ to the 
space of maps from $\Delta^{k}$ to $M$ is a map from $\Delta^{j}\times \Delta^{k}$ 
to $M$, which is represented by an integral $(j{+}k)$-current in $M$ 
\cite{MR0146835}.
The pushdown operation extends to the vector space of (flat) currents in 
$\cD^{\integral}_{k}(M)$,
\eq
\Pi_{*}^{j,k}\colon \cD_{j}(\cD_{k}^{\integral}(M)) \rightarrow
\cD_{j+k}(M)
\,.
\en
The interaction with the boundary operator is
\eq
\label{eq:boundarypushdown}
\partial (\Pi_{*}^{j,k} \eta) = 
\Pi_{*}^{j-1,k} (\partial \eta)
+ \Pi_{*}^{j,k-1} (\partial_{*}^{j,k} \eta)
\,,\quad
j\ge 1
\,,
\en
which follows from $\partial (\Delta^{j}\times \Delta^{k}) = 
(\partial \Delta^{j}) \times \Delta^{k} + \Delta^{j} \times (\partial  
\Delta^{k})$.

The pushdown operation $\Pi_{*}^{j,k}$ is translation invariant 
for $j\ge 1$,
\eq
\Pi_{*}^{j,k} (T_{\xi} \eta) = \Pi_{*}^{j,k} (\eta)
\,,\quad
j\ge 1
\en
where $T_{\xi}$ is translation by the integral $k$-current $\xi$.
Roughly, the pushdown operation takes the $j$-current $\eta$ to 
the $(j{+}k)$-dimensional region swept out by the $j$-parameter 
family of $k$-currents.
In the translated current $T_{\xi} \eta$, the $j$-parameter family of 
$k$-currents keeps $\xi$ constant, so nothing additional is swept out,
for $j\ge 1$.

\subsection{$\Pi_{*}\colon
\cD^{\integral}_{j}(\cD_{k}^{\integral}(M)_{\partial\xi}) \rightarrow
\cD^{\integral}_{j+k}(M)$ commutes with $\partial$}

Now restrict to the space 
$\cD^{\integral}_{k}(M)_{\partial\xi}$ of integral relative $k$-cycles.
For an integral $j$-current $\eta$ in the space 
$\cD^{\integral}_{k}(M)_{\partial\xi}$ of integral relative $k$-cycles,
\eq
\eta\in \cD^{\integral}_{j}(\cD_{k}^{\integral}(M)_{\partial\xi})
\,,\quad
j\ge 1
\,,
\en
the composition with the boundary operator vanishes,
\eq
\partial_{*}^{j,k} \eta = 0
\,,\quad
j\ge 1\,,
\en
because  composing with the boundary operator takes the $j$-current 
$\eta$ to the single point $\partial\xi$ in $\cD_{k-1}^{\integral}(M)$.
Therefore the pushdown operation commutes with the boundary 
operator,
\eq
\partial (\Pi_{*}^{j,k} \eta) = \Pi_{*}^{j-1,k} (\partial \eta)
\,,\quad
j\ge 1
\,.
\en

\subsection{$\Pi_{*}$ is an isomorphism on the homology groups}

The map
\eq
\Pi_{*}\colon
\cD^{\integral}_{j}(\cD_{k}^{\integral}(M)_{\partial\xi}) \rightarrow
\cD^{\integral}_{j+k}(M)
\en
induces a map of homology groups
\eq
\Pi_{*}\colon
H_{j}(\cD_{k}^{\integral}(M)_{\partial\xi}) 
\rightarrow
H_{j+k}(M)
\en
which is an isomorphism \cite{MR0146835}.

\subsection{Tangent vectors as infinitesimal 1-currents}

Suppose $\epsilon\mapsto \xi(\epsilon)$ is an infinitesimal curve in 
$\cD_{k}^{\integral}(M)$, at $\xi(0) = \xi$.
It could be directed vertically, along a fiber, but need not be.
The curve is a map from the interval $[0,\epsilon]$ to 
$\cD_{k}^{\integral}(M)$,
so it is represented by an infinitesimal integral 1-current 
$\eta(\epsilon)$ in $\cD_{k}^{\integral}(M)$
or, equivalently,
as the flat 1-current supported at $\xi$,
\eq
\dot \xi = \lim_{\epsilon\rightarrow 0} 
\frac{\eta(\epsilon)}{\epsilon}
\,.
\en
The tangent space at $\xi$ is the space of flat 1-currents supported 
at $\xi$,
or, equivalently,
the space of infinitesimal integral 1-currents at $\xi$.

This is just the usual idea of tangent vectors
expressed in terms of currents.
The boundary of $\eta(\epsilon)$ is the 0-current
\eq
\partial \eta(\epsilon) = \delta_{\xi(\epsilon)} - 
\delta_{\xi(0)}
\,.
\en
If $f$ is a function on $\cD_{k}^{\integral}(M)$, i.e., a 0-form, 
then the derivative of $f$ along the tangent vector at $\xi$ is
\eq
%{\frac{d}{dt}}_{/t=0} f(\xi(t)) =
%{\frac{d}{dt} f(\xi(t))}_{/t=0} =
\lim_{\epsilon\rightarrow 0} 
\frac{f(\delta_{\xi(\epsilon)})-f(\delta_{\xi(0)})}{\epsilon}
= \lim_{\epsilon\rightarrow 0} 
\frac{f(\partial \eta(\epsilon))}{\epsilon}
= f(\partial \dot \xi)
= df (\dot \xi)
\,.
\en

The pushdown map $\Pi^{1,k}_{*}$ takes the infinitesimal 
integral 1-current $\eta$ to an integral $(k{+}1)$-current in $M$,
and the flat 1-current $\dot \xi$ to a flat $(k{+}1)$-current in $M$.
If we restrict to vertical tangent vectors,
the map $\Pi^{1,k}_{*}$ is injective on the vertical tangent vectors 
at $\xi$, giving the space $\cV_{k+1}$ described previously.

From the same point of view, the
$j$-vectors at $\xi$,
i.e., the linear combinations of antisymmetric  $j$-fold products of 
tangent vectors,
are the infinitesimal integral $j$-currents at 
$\xi$, which are the flat $j$-currents supported at $\xi$.
The pushdown map $\Pi^{j,k}_{*}$ takes an infinitesimal integral $j$-current
to infinitesimal integral $(j{+}k)$-current in $M$.

\subsection{The Hodge $*$-operator on 1-currents in 
$\Ex$}

Given our assumption 
that the Hodge $*$-operator acts on the vertical tangent spaces $\cV_{n}$, 
it will act on the infinitesimal 1-currents in the fiber $\Ex= 
\cD^{\integral}_{n-1}(M)_{\partial\xi}$
and therefore on all 1-currents in $\Ex$.
Since $\Pi_{*}^{1,n-1}\colon\cD_{1}(\Ex) \rightarrow \cD_{n}(M)$ identifies the infinitesimal 1-currents in 
the fibers with $\cV_{n}\subset \cD^{\flat}_{n}(M)$,
\eq
*\, \Pi_{*}^{1,n-1} = \Pi_{*}^{1,n-1} *
\,.
\en

\section{An analog of a 2d conformal field theory on each $\Ex$}

\subsection{$n$-forms on space-time as $1$-forms on $\cE_{\partial\xi}$}

The $n$-forms $F$ and $F^{*}$ on $M$ pull up to
1-forms $j$ and $j^{*}$ on each 
$\cE_{\partial\xi}$,
\eq
j(\eta) = F(\Pi_{*}\eta)
\,,\qquad
j^{*}(\eta) = F^{*}(\Pi_{*}\eta)
\,,\qquad
\eta\in \cD_{1}(\cE_{\partial\xi})
\,.
\en
Here $\eta$ is a 1-current in the fiber $\cE_{\partial\xi}$.
Its pushdown $\Pi_{*}\eta$ is an $n$-current in $M$.
Since Hodge $*$ commutes with the pushdown,
\eq
j^{*} = i^{-1} {*} j
\,.
\en
The 1-forms $j$ and $j^{*}$ on the fibers are closed,
\eq
dj = 0 
\,,\qquad
d j^{*} = 0\,,
\en
by the equations of motion on $F$ and $F^{*}$,
\eq
dj(\eta) = j(\partial\eta)
= F(\Pi_{*}\partial\eta)
= F(\partial \Pi_{*}\eta)
= dF(\Pi_{*}\eta)
=0
\,,
\en
and similarly for $j^{*}$.

The 1-forms $j$ and $j^{*}$ on the fibers 
of $\cE\rightarrow \cB$
are invariant under translations
in the whole abelian group $\cE$.

\subsection{Scalar fields and vertex operators on $\Ex$}

On each $\Ex$, integrate
\eq
\label{eq:phiintegrals}
j = d \phi
\,,\qquad
j^{*} = d \phi^{*}
\,.
\en
to get 0-forms $\phi$ and $\phi^{*}$.
Consider the 0-forms as functions on the fibers, as scalar fields $\phi(\xi)$ 
and
$\phi^{*}(\xi)$.
Depend on context to distinguish scalar fields written as 
functions on the fiber from scalar fields written as linear functions 
on 0-currents,
\eq
\phi(\xi) = \phi(\delta_{\xi})
\,.
\en
Define ``vertex operators'' on each fiber,
\eq
\label{eq:vertexops}
V_{p,p^{*}}(\xi) = e^{ip\phi(\xi)+ip^{*}\phi^{*}(\xi)}
\,.
\en
The general observable on a fiber $\cE_{\partial\xi}$ is a product of 
vertex operators
\eq
\label{eq:observables}
V_{p_{1},p_{1}^{*}}(\xi_{1})
\cdots
V_{p_{N},p_{N}^{*}}(\xi_{N})
\,,\qquad
\partial \xi_{1} = \partial \xi_{2}\cdots = \partial\xi_{N} = \partial\xi
\,.
\en
On each $\cE_{\partial\xi}$  there is a formal analog of the 2d conformal field 
theory of a free $1$-form.

\subsection{Global symmetry on $\Ex$}

Each of the scalar fields $\phi$ and $\phi^{*}$ is determined
on each $\Ex$ up to a constant of 
integration,
except for the special fiber $\Ezero$ where there is a natural 
normalization condition $\phi(0)=\phi^{*}(0)=0$.
So, on each non-special fiber 
$\cE_{\partial\xi}$, $\partial\xi\ne 0$,
the gauge symmetry of the space-time symmetry
becomes a global symmetry, shifting $\phi$ and $\phi^{*}$
by constants $f$ and $f^{*}$,
\eq
\label{eq:globalsymmetryEx}
\phi(\xi) \rightarrow \phi(\xi)+ f(\partial\xi)
\,,\qquad
\phi^{*}(\xi) \rightarrow \phi^{*}(\xi)+ f^{*}(\partial\xi)
\,,
\en
The constants $f$ and $f^{*}$ depend on the fiber $\Ex$, so they are 
functions on the base space $\cB=\partial\cD^{\integral}_{n-1}(M)$.

The vertex operators transform under the symmetry group of the fiber 
as operators of charges $p$, $p^{*}$,
\eq
\label{eq:vertexopgaugetransf}
V_{p,p^{*}}(\xi) \rightarrow V_{p,p^{*}}(\xi) e^{ipf(\partial\xi)}
e^{ip^{*}f^{*}(\partial\xi)}
\,.
\en

\subsection{Space-time gauge symmetries are special collections of 
global symmetries}

The product of the global symmetry groups of the fibers $\Ex$
is considerably larger than the local gauge group in space-time.
The space-time gauge potentials $A$ and $A^{*}$ pull up to scalar 
fields $\tilde \phi = \Pi^{*} A$, $\tilde \phi^{*} = \Pi^{*} A^{*}$
on $\cE$ that, restricted to each fiber $\Ex$, are
solutions of (\ref{eq:phiintegrals}).
They are special solutions,
characterized by the additional condition of additivity in $\cE$,
\eq
\tilde \phi (\xi_{1}+\xi_{2}) =
\tilde \phi (\xi_{1}) + \tilde \phi (\xi_{2})
\,,\qquad
\tilde \phi^{*} (\xi_{1}+\xi_{2}) =
\tilde \phi^{*} (\xi_{1}) + \tilde \phi^{*} (\xi_{2})
\,.
\en
The collection of scalars $\phi$, $\phi^{*}$ on the fibers $\Ex$ are not so
constrained.

The space-time gauge symmetries
$A\rightarrow A+ d\tilde f$,
$A^{*}\rightarrow A^{*}+ d\tilde f^{*}$
are given by $(n{-}2)$-forms $\tilde f$ and $\tilde f^{*}$ on $M$
which pull up to 0-forms on $\cB=\partial\cD^{\integral}_{n-1}(M)$ which give global symmetries
in each $\Ex$, as in (\ref{eq:globalsymmetryEx}),
satisfying the additional additivity condition
\eq
\tilde f (\xi_{1}+\xi_{2}) =
\tilde f (\xi_{1}) + \tilde f (\xi_{2})
\,,\qquad
\tilde f^{*} (\xi_{1}+\xi_{2}) =
\tilde f^{*} (\xi_{1}) + \tilde f^{*} (\xi_{2})
\,.
\en

\subsection{An analog of the $U(1){\times}U(1)$ 2d gaussian model on 
each $\Ex$}
%with symmetry $G=U(1){\times}U(1)$}
Identify
\eq
\phi(\xi) \sim \phi(\xi) + 2\pi R
\,,\qquad
\phi^{*}(\xi) \sim \phi^{*}(\xi)  + 2\pi R^{*}
\en
for positive real numbers $R$, $R^{*}$.
The  symmetry group of the theory on $\Ex$ becomes the compact group
$U(1){\times} U(1)$,
\eq
f(\partial\xi) \sim f(\partial\xi) + 2\pi R
\,,\qquad
f^{*}(\partial\xi) \sim f^{*}(\partial\xi) + 2\pi R^{*}
\,,
\en
\eq
U(1){\times}U(1) = (\Reals/2\pi R \,\Integers )
{\times}(\Reals/2\pi R^{*}\, \Integers )
\,.
\en
The ``momenta'' of the vertex operators are quantized,
\eq
p= \frac{m}{R}\,,\qquad p^{*} = \frac{m^{*}}{R^{*}}\,,
\qquad m,m^{*}\in\Integers
\,.
\en
Now there is on each $\Ex$ a formal analog of the 2d conformal field 
theory of a free $1$-form with compact symmetry group $G=U(1){\times} 
U(1)$.  This 2d conformal field theory is the 2d gaussian model.

The compactification of the symmetry group
on each fiber expresses
the compactification of the global gauge group
of the space-time $n$-form theory to $U(1){\times} U(1)$.

\section{Synopsis}

The goal now is to flesh out the analogy between the 2d field theory and 
the field theory of extended objects on each of the fibers 
$\cE_{\partial\xi}$,
the end being to develop machinery to
translate the construction of any 2d quantum field theory 
into the actual construction of a quantum field theory of extended 
objects.

I can imagine
replacing the analog of the
2d gaussian model on each fiber 
with analogs other 2d quantum field theories
by performing on each 
fiber the known constructions on the 2d gaussian model, 
including, for example,
(1)  constructing the twist fields of the $\Integers_{2}$ orbifold 
of the 2d gaussian model, or
(2) specializing to $R=R^{*}=1$ and 
constructing  the $SU(2){\times}SU(2)$ current algebra of the
self-dual gaussian model, or
(3) constructing the analog of 2d conformal perturbation theory
and then the analog of 2d non-conformal perturbation theory.
But what is really wanted is a 
general machine that implements the analogy for an arbitrary
2d quantum field theory,
by actually
constructing a quantum field theory of extended objects from the 
data of the 2d quantum field theory.

Sections \ref{sect:moreoncurrentsinM}
%, \ref{sect:qftofthefreenform},
% \ref{sect:ExCalmostcomplex},
% and 
-- \ref{sect:theqftonEx}
take up the quantum field theory of the free $n$-form.
The quantization is expressed by the
Schwinger-Dyson equations on the two-point correlation functions of the
space-time $n$-forms and the gauge potentials.  
Section \ref{sect:moreoncurrentsinM} develops the geometric 
structures on $M$ used in section \ref{sect:qftofthefreenform} to write the 
Schwinger-Dyson equations in terms of currents.
The geometric structures consist of a linear operator $J=\epsilon_{n} *$ on 
$n$-currents in $M$ satisfying $J^{2}=-1$,
which is a small modification of the Hodge $*$-operator,
and a skew-hermitian form $I_{M}\expval{\bar\xi_{1},\xi_{2}}$ on currents
in $M$, which is a similarly small modification of the usual bilinear intersection form on 
currents.
The small modifications are needed to make the properties of these 
geometric structures independent of the parity of $n$.
The operator $J$ is imaginary for $n$ even,
so $\Ex$ must be complexified
in order for $J$ to act on its tangent vectors.
This is done in section~\ref{sect:ExCalmostcomplex}.
The complexification $\ExC$ is then an almost-complex space.
In section \ref{sect:theqftonEx}, the $J$-operator and 
the skew-hermitian form
pulled up to $\ExC$ are used to write the Schwinger-Dyson equations
for the fields of the free 1-form quantum field theory on $\ExC$.
The S-D equations on $\Ex$ are formally identical to 
the S-D equations for the free 1-form on a Riemann surface.

Section \ref{sect:quasi2d} points out the geometric resemblance of
$\ExC$ to a Riemann surface.
The linear operator $J$ acting on 1-currents in $\ExC$
resembles
resembles the linear operator $J$ acting on 1-currents in a Riemann 
surface that expresses the almost-complex structure of the Riemann 
surface.
The push-down map $\Pi_{*}$ takes $k$-currents in $\ExC$ to 
$(n{-}1{+}k)$-currents in $M$,
so the pulled-up skew-hermitian form
on currents in $\ExC$,
$\Pi^{*}I_{M}\expval{\bar\eta_{1},\eta_{2}}$,
pairs a $k_{1}$-current $\eta_{1}$
and a $k_{2}$-current $\eta_{2}$
when $k_{1}+k_{2}=2$,
exactly as does the skew-hermitian intersection form on the currents in a Riemann 
surface.

Section \ref{sect:quasiRiemannsurfaces}
tries to capture
the geometry of currents
shared by the spaces $\ExC$ and by ordinary Riemann surfaces
in the definition of a {\it quasi Riemann surface}.
The quasi Riemann surfaces are to be the geometric settings for
the general class of quantum field theories of extended objects.
A {\it quasi holomorphic curve} is defined to be a morphism
of quasi Riemann surfaces
from an ordinary Riemann surface $\Sigma$
to one of the $\ExC$.
A local q-h curve is a q-h curve where the Riemann surface $\Sigma$ is the open unit 
disk in the complex plane.

Section~\ref{sect:2dCFTonaqhc} describes how the free 1-form quantum 
field theory on $\ExC$ pulls back along a quasi-holomorphic curve to 
give the 2d CFT of the free 
1-form on the Riemann surface $\Sigma$.
The q-h curves serve as 2d probes of the extended objects.
The local q-h curves probe the local structure of the extended 
objects.

In section \ref{sect:awishfulconjecture}, I propose a strong conjecture
on the classification of quasi Riemann surfaces:
that quasi Riemann surfaces are isomorphic iff they have the same homology 
data
---
the skew-hermitian intersection form
and the complex structure on the integral homology in the middle dimension.
The conjecture identifies the spaces $\ExC$ of integral 
$(n{-}1)$-currents in $M$
with the space $\cD^{\integral}(\Sigma)_{0}$ of integral $0$-currents 
in a two-dimensional space $\Sigma$ that has the same homology data as $M$.
When $M$ has the homology data of a Riemann surface,
the 2d space $\Sigma$ will be a Riemann surface.
For example, when $M$ is the conformal manifold $S^{d}=\Reals^{d}\cup\{\infty\}$, 
the 2d space $\Sigma$ will be the Riemann sphere $S^{2}=\PoneC$.
The bundle $\cE\rightarrow\cB$ is replaced by a bundle $\QM 
\rightarrow \PBM$ of quasi Riemann surfaces over
the integral projective space of $\cB$.

Section \ref{sect:corrfnsfromECFT}
describes how the conjecture
can give a means to construct the correlation functions of fields on 
$\ExC$ as correlation functions in the 2d CFT on the two-dimensional space
$\Sigma$.
This will require extending the observables of the 2d CFT
from products of local fields over finite sets of points in $\Sigma$ 
--- e.g., products of a finite set of vertex operators --- 
to products over integral 0-currents.
Such an {\it extended} 2d conformal field theory (ECFT)
transcribes directly from the quasi Riemann surface associated to the two-dimensional space $\Sigma$ to the 
quasi Riemann surfaces associated to the spaces $\ExC$.

Section \ref{sect:perturbationtheory}
is a formal discussion of the equivalence between perturbation theory
in the space-time theory and perturbation theory in the 2d theory
---
first for the free $n$-form theory
on $M$ and
the free $1$-form theory on $\Sigma$,
then for perturbations in general.
In the course of the discussion, a modicum of evidence for the 
conjecture is found.
The discussion is formal in the sense 
that there are no distance scales,
neither on $M$ nor on $\Sigma$.
I hope that, eventually, a connection can be made between the cutoff scale 
in space-time and the cutoff 2d scale,
so that non-conformal quantum field theories of extended objects can 
be constructed from non-conformal 2d quantum field theories,
both governed by the 2d renormalization group.

Sections \ref{sect:gaugetheory} and \ref{sect:connectingtheExC}
return to the classical theory of the free $n$-form as a free 
$1$-form on $\Ex$.
Section \ref{sect:gaugetheory} presents the local gauge symmetry
in space-time as a specialization of
local gauge symmetry in the bundle $\Ex\rightarrow \cB$.
The local gauge transformations in the fibers $\Ex$ are the global 
symmetries of the 2d theory on $\Ex$.
The intent is to provide a prototype pattern 
for gauge symmetry in the general class of quantum field theories of 
extended objects.
In particular, I hope that 2d quantum field theories with nonabelian 
global symmetry will give quantum field theories 
of extended objects in space-time 
with nonabelian local gauge symmetry in space-time.
Section \ref{sect:connectingtheExC} takes up the 
multitude of the fibers $\Ex$.
The theory of extended objects should be a single entity
woven from all the analog 2d theories on the $\Ex$.
Again, the intent is to make a prototype argument that might be 
adapted to the general case.

Section \ref{sect:Explorations} explores, incautiously, some consequences of the 
conjecture on quasi Riemann surfaces.
First, to meet the technical requirements of a quasi Riemann surface,
an ordinary Riemann surface $\Sigma$ needs to be modified slightly --- 
``augmented'' --- to a space $\Sigma_{+}$.
The group $\Aut(\QSigmaplus)$ of automorphisms of the quasi 
Riemann surface $\QSigmaplus$ associated to $\Sigma_{+}$
is contained in a larger group $G(\Sigma_{+})$
which is essentially the group of automorphisms of
the integral 1-currents in $\Sigma$.
There is a universal bundle
$\cQ(0)\rightarrow\PB(0)$
of quasi Riemann surfaces over the 
homogeneous space $\PB(0) = G(\Sigma_{+})/\Aut(\cQ(\Sigma))$.
For every conformal manifold $M$ with the homology data
of $\Sigma$,
the bundle $\QM \rightarrow \PBM$ of quasi Riemann surfaces 
associated to $M$ is 
embedded in a natural way in the universal bundle.
The universal bundle of quasi Riemann surfaces becomes the natural
setting for the quantum field theories of extended objects.

Section \ref{sect:mathematicalquestions} lists the main mathematical
questions that need to be resolved.  Section \ref{sect:furthersteps}
lists some of the further steps that might be taken, most of which
require assuming that the mathematical questions are resolved
favorably.  Section \ref{sect:hist-refs-questions} asks advice about
the history of the ideas used and about references.

\section{More on currents in $M$}
\label{sect:moreoncurrentsinM}

\subsection{Intersection of currents}

The intersection form is the bilinear form
on smooth currents
\eq
I_{M}(\xi_{1},\,\xi_{2}) = 
\int_{M} d^{d}x\;
\epsilon_{\mu_{1}\cdots\mu_{k_{1}}\nu_{1}\cdots\nu_{k_{2}}}\; 
\frac1{k_{1}!}
\xi_{1}(x)^{\mu_{1}\cdots \mu_{k_{1}}}
\frac1{k_{2}!}
\xi_{2}(x)^{\nu_{1}\cdots \nu_{k_{2}}}
\en
for $k_{1}+k_{2}=d$, and zero for $k_{1}+k_{2}\ne d$.
The intersection form extends to generic pairs of integral currents (``in 
general position'').
On singular currents, the intersection form
agrees with the intersection number of the corresponding singular 
chains.
The intersection form on currents depends only on the orientation of 
$M$.  It is independent of the conformal 
structure.

For $d=2n$, the intersection form satisfies
\ateq{2}{
\label{eq:intersection1}
I_{M}(\xi_{1},\xi_{2}) &= (-1)^{k_{1}} 
I_{M}(\xi_{2},\xi_{1})
\\[1ex]
\label{eq:intersection2}
I_{M}(\partial\xi_{1},\xi_{2}) &= (-1)^{k_{1}}
I_{M}(\xi_{1},\partial\xi_{2})
\\[1.5ex]
\label{eq:intersection3}
I_{M}(*\xi_{1},\xi_{2}) &= (-1)^{n} I_{M}(\xi_{1},*\xi_{2})
\,,\qquad&
k_{1}&=k_{2}=n
\\[1ex]
I_{M}(\xi,*\xi) &> 0
\,,\quad\xi \ne 0
\,,\qquad&
k_{1}&=k_{2}=n
\,.
}
The last equation fixes the relation between the sign of $*$ and the orientation of 
$M$.
The positive definite quadratic form $I_{M}(\xi,*\xi)$  is independent 
of the orientation .

\subsection{The operator $J=\epsilon_{n}{*}$ on $n$-currents}

The properties of the Hodge $*$-operator and of the intersection 
form $I_{M}(\xi_{1},\xi_{2})$ depend on the parity of $n$. 
In particular, the $*$-operator on $n$-currents satisfies $*^{2}= 
(-1)^{n}$.
This is a problem if the spaces $\Ex$ are to look like two 
dimensional spaces, for every $n$.
Some small modifications are needed to make the
properties  uniform for all values of $n$, even and odd.

Define the {\it  $J$-operator} acting on $n$-forms and on
$n$-currents to be
\eq
\label{eq:J}
J = \epsilon_{n} *
\en
where $\epsilon_{n}$ is a number satisfying
\eq
\epsilon_{n}^{2}=(-1)^{n-1}
\en
so that
\eq
\label{eq:Jsq=1}
J^{2}=-1
\en
for very value of $n$.
One possible choice of the numbers $\epsilon_{n}$ is
\eq
\epsilon_{n} = 
\left \{
\begin{array}{ll}
1\,,
\quad
& n \text{ odd,}
\\[2ex]
i\,, & n \text{ even.}
\end{array}
\right .
\en

For $n$ even, $J$ is imaginary,
so this modification requires allowing currents 
to be complex.
For the sake of uniformity in $n$, 
we will take the currents to be complex for all $n$.
For $n$ odd, there will be a complex conjugation symmetry.
For $n$ even, there will be a  symmetry combining complex conjugation
with reversal of orientation.
Discussion 
of complex conjugation and reality conditions
will be left until section \ref{app:complexconjugation}.

\subsection{The chiral projection operators $P_{\pm}$}

Define the chiral projection operators acting on $n$-currents and on 
$n$-forms,
\eq
P_{\pm} = \frac12 \left (1 \pm i^{-1}J \right )
\,.
\en
$P_{+}$ projects on the self-dual $n$-currents,  $P_{-}$ on the anti-self-dual 
$n$-currents.

\subsection{The skew-hermitian intersection form $\IM{\bar 
\xi_{1}}{\xi_{2}}$ on currents}
\label{sect:hermitianintersectionform}

Define the {\it skew-hermitian intersection form} on currents by
\eq
\label{eq:hermitianintersectionform}
\IM{\bar\xi_{1}}{\xi_{2}} = 
\epsilon_{n,k_{2}-n} I_{M}(\bar\xi_{1}, \xi_{2})
\,,
\en
where the numbers $\epsilon_{n,k}$ are
\eq
\epsilon_{n,k} = (-1)^{nk+k(k+1)/2} \,\epsilon_{n}^{-1}
\,.
\en
The constants $\epsilon_{n,k}$ are chosen so that
the skew-hermitian intersection form satisfies
\ateq{2}{
\label{eq:hermitianintersection1}
\IM{\bar\xi_{1}}{\xi_{2}}  &= 
- \overbar{\IM{\bar\xi_{2}}{\xi_{1}}}
\\[1ex]
\label{eq:hermitianintersection2}
\IM{\overbar{\partial\xi_{1}}}{\xi_{2}} &= - 
\IM{\bar\xi_{1}}{\partial\xi_{2}}
\\[1ex]
\label{eq:hermitianintersection3}
\IM{\bar\xi_{1}}{\xi_{2}}  &=  0
\,,\qquad&
k_{1}&+k_{2}\ne 2n
\\[1ex]
\label{eq:hermitianintersection4}
\IM{\overbar{J \xi_{1}}}{ \xi_{2}}  &= 
-\IM{\bar \xi_{1}}{ J\xi_{2}} 
\,,\qquad&
k_{1}&=k_{2}=n
\\[1ex]
\label{eq:hermitianintersection5}
\IM{\bar \xi}{ J\xi}  &> 0
\,,\quad\xi \ne 0\,,
&
k_{1}&=k_{2}=n
\,.
}
The skew-hermitian intersection form  on 
$n$-currents is block diagonal in the chiral decomposition,
\eq
\IM{\bar \xi_{1}}{\xi_{2}} = 
\IM{\overbar{P_{+}\xi_{1}}}{P_{+}\xi_{2}}+\IM{\overbar{P_{-}\xi_{1}}}{P_{-}\xi_{2}}
\,,\qquad
k_{1}=k_{2}=n
\,.
\en

\section{Quantum field theory of the free $n$-form on $M$}
\label{sect:qftofthefreenform}

The quantum field theory of the free $n$-form is described by its
partition function and its two-point correlation functions.  The
two-point functions are determined by their Schwinger-Dyson equations,
which will be written here in terms of the $J$-operator and the
skew-intersection form $\IM{\bar\xi_{1}}{\xi_{2}}$ on currents in $M$, with no mention of $n$.
This means that $J$ and $\IM{\bar\xi_{1}}{\xi_{2}}$
encode all the geometric data needed to construct the quantum field 
theory.

The partition function has interesting dependence on the space-time
manifold $M$ and its conformal structure.  The derivatives of the 
partition function wrt the parameters describing $M$ can be derived from the 
two-point functions, so the S-D equations on the two-point functions
completely determine the quantum theory.
For now we are only
interested in the quantum field theory on a fixed space-time.  The
partition function is left for later.

The Schwinger-Dyson equations for the $n$-form fields and the gauge 
potentials are obtained in Appendix 
\ref{app:euclean-n-form} by deriving them
in the free $n$-form theory on $\Reals^{d}$.

\subsection{The chiral fields}
Take the $n$-form field $F(x)$ to be a complex field
$F=F_{1}+i F_{2}$
with the global $U(1)$ symmetry $F\rightarrow e^{i\alpha}F$.
The reality condition $F= \bar F$ will be applied later.

The chiral $n$-form fields are
\eq
F_{\pm}= P_{\pm}F
\,,\qquad
F_{\pm}(\xi) = F(P_{\pm}\xi)
\,.
\en
The chiral gauge potentials are the $(n{-}1)$-forms solving
\eq
dA_{\pm} =F_{\pm}
\,.
\en
The euclidean adjoint fields are
\eq
F^{\dagger}_{\pm}(\bar \xi) = \overbar{F_{\mp}(\xi)}
\,,\qquad
A^{\dagger}_{\pm}(\bar \xi) = \overbar{A_{\mp}( \xi)}
\,.
\en
The euclidean adjoint fields
are defined so that,  on $\Reals^{d}$, the euclidean adjoint field is the Wick rotate of the 
Minkowski space adjoint field.
This is worked out in Appendix \ref{app:euclean-n-form}.

Use the index notation $F_{\alpha}$, $A_{\alpha}$, $\alpha=\pm$.

\subsection{The Schwinger-Dyson equations in terms of 
$\IM{\bar \xi_{1}}{\xi_{2}}$ }

The quantum fields $F_{\alpha}(x)$ are distributions that are smeared against 
smooth (complex) $n$-currents $\xi$ to give observables $F_{\alpha}(\xi)$.
The $A_{\alpha}(x)$ are smeared against smooth $(n{-}1)$-currents.
The two-point functions are determined by the Schwinger-Dyson 
equations (derived in Appendix \ref{app:euclean-n-form}),
\aeq{
\label{eq:SDFF}
\expval{F^{\dagger}_{\bar \alpha}(\bar\xi_{1})F_{\beta}(\partial\xi_{2})} &= 
- 2\pi i\, \gamma_{\bar\alpha\beta} \IM{\overbar{\partial\xi_{1}}}{ \xi_{2}}
\\[1ex]
\label{eq:SDAF}
\expval{A^{\dagger}_{\bar \alpha}(\bar \xi_{0})F_{\beta}(\partial\xi_{2})}
&=  -2\pi i \gamma_{\bar\alpha\beta} \IM{\bar\xi_{0}}{ \xi_{2}}
}
\eq
\gamma_{\bar ++}=1
\,,\quad
\gamma_{\bar +-}=0
\,,\quad
\gamma_{\bar -+}=0
\,,\quad
\gamma_{\bar --}= -1
\,.
\en
Note that these S-D equations are consistent with
$d A_{\pm} = F_{\pm}$,
but they are not consistent with
$F_{\pm} = P_{\pm} F$.
The identities $F_{\pm} = P_{\pm} F$ hold only up to contact terms ---  they hold as field 
equations, but not as equations on the distributional correlation functions.
The only ambiguity in the two-point functions is in the 
contact term in 
$\expval{F^{\dagger}_{\bar\alpha}(\bar\xi_{1})F_{\beta}(\xi_{2})}$,
\eq
\expval{F^{\dagger}_{\bar\alpha}(\bar\xi_{1})F_{\beta}(\xi_{2})} 
\rightarrow
\expval{F^{\dagger}_{\bar\alpha}(\bar\xi_{1})F_{\beta}(\xi_{2})} 
+ 2\pi i (\Delta\gamma_{\bar\alpha\beta}) \IM{\bar \xi_{1}}{ \xi_{2}}
\,.
\en
Either  $d A_{\pm} = F_{\pm}$ or $F_{\pm} = P_{\pm} F$ can be satisfied 
in the distributional correlation functions, but not both.

The only geometric data used in the quantization of the free $n$-form 
on $M$
is the $J$-operator acting on $\cD_{n}(M)$ and
the skew-hermitian intersection form 
$\IM{\bar \xi_{1}}{\xi_{2}}$ restricted to the 
subspace
\eq
%\label{eq:currentsubspace}
\cD_{n-1}(M)\oplus\cD_{n}(M)\oplus\cD_{n+1}(M)
\,.
\en
The skew-hermitian intersection form depends only on the manifold
structure of $M$.
The $J$-operator depends on the conformal structure of $M$.

\section{$\ExC$ as an almost-complex space}
\label{sect:ExCalmostcomplex}

For $n$ even, the operator $J=\epsilon_{n}{*}$ is imaginary, so we have to complexify
$\Ex$ in order to get a space where $J$ acts on the tangent vectors.
There is no need to complexify when $n$ is odd, but we do so 
anyway for the sake of uniformity in $n$.

A straightforward, natural complexification is
\eq
\ExC = \Ex \oplus i \partial\cD^{\integral}_{n}(M)
\en
The tangent space at each point is the complex vector space $\cV_{n}\oplus i \cV_{n}$.
$J$ acts on each tangent space and satisfies $J^{2}=-1$,
so $\ExC$ is an almost-complex space with almost complex structure 
$J$.

The pushdown maps $\Pi_{*}$ extend to $\ExC$, producing complex 
currents in $M$.
The pushdown map $\Pi_{*}^{1,n-1}$ taking $1$-currents in $\ExC$ to $n$-currents in 
$M$ is compatible with 
the $J$ operators on 1-currents in $\ExC$ and
on complex $n$-currents in $M$,
\eq
\Pi_{*}^{1,n-1} J =  J \Pi_{*}^{1,n-1}
\en
The dual pull-up map $\Pi_{*}$ takes complex $n$-forms on $M$ to 
complex $1$-forms on $\ExC$, and is compatible with the $J$ operators 
on the forms,
\eq
J\, \Pi^{*} = \Pi^{*} J
\,.
\en

\section{The Schwinger-Dyson equations on $\ExC$}
\label{sect:theqftonEx}

We pull the Schwinger-Dyson equations up from $M$ to $\ExC$.
The fields on $\ExC$ are the space-time fields pulled up from $M$,
\ateq{2}{
j_{\alpha} &= \Pi^{*} F_{\alpha}
\qquad &
\phi_{\alpha} &= \Pi^{*} A_{\alpha}
\\
j_{\alpha}(\eta) &=  F_{\alpha}(\Pi_{*}^{1,n-1} \eta)
\qquad &
\phi_{\alpha}(\eta) &=  A_{\alpha}(\Pi_{*}^{0,n-1} \eta)
\\
j_{\alpha} &= P_{\alpha} j
\qquad &
d \phi_{\alpha} &= j_{\alpha}
\\
j^{\dagger}_{\pm}(\bar\eta) &= \overbar{j_{\mp}(\eta)}
\qquad &
\phi^{\dagger}_{\pm}(\bar\eta) &= \overbar{\phi_{\mp}(\eta)}
\,.
}
The S-D equations 
(\ref{eq:SDFF}--\ref{eq:SDAF}) become
\aeq{
\label{eq:SDjj}
\expval{j^{\dagger}_{\bar\alpha}(\bar\eta_{1})j_{\beta}(\partial\eta_{2})} &= 
- 2\pi i\, \gamma_{\bar\alpha\beta} \Pi^{*}\IM{\overbar{\partial\eta_{1}}}{ \eta_{2}}
\\[1ex]
\label{eq:SDphij}
\expval{\phi^{\dagger}_{\bar\alpha}(\bar\eta_{0})j_{\beta}(\partial\eta_{2})}
&= -2\pi i \gamma_{\bar\alpha\beta} \Pi^{*}\IM{\bar\eta_{0}}{ \eta_{2}}
\,,
}
where $\Pi^{*}\IM{\bar\eta_{1}}{\eta_{2}}$ is the 
the skew-hermitian intersection form pulled up from $M$,
\eq
\Pi^{*}\IM{\bar\eta_{1}}{\eta_{2}}
= \IM{\overbar{\Pi_{*}\eta_{1}}}{\Pi_{*}\eta_{2}}
\,.
\en
The S-D equations (\ref{eq:SDjj}--\ref{eq:SDphij}) on $\ExC$ are formally
identical to the S-D equations (\ref{eq:SDFF}--\ref{eq:SDAF}) for the
free 1-form
on a Riemann surface $\Sigma$,
with $J$ taking the place of the almost complex structure of $\Sigma$
and $\Pi^{*}\IM{\bar\eta_{1}}{ \eta_{2}}$ taking the place of 
the skew-hermitian intersection form
$\ISigma{\bar\eta_{1}}{ \eta_{2}}$ of $\Sigma$.

The analogy between the field theory on $\Ex$ and the 2d field 
theory now holds on the quantum level.
We have a {\it quantum} field theory on each  $\ExC$
that has the same form as the 2d {\it quantum} field theory.

\section{Quasi two-dimensionality of $\ExC$}
\label{sect:quasi2d}

The only geometric data used in the quantization
on $\ExC$
is the $J$-operator acting on $\cD_{1}(\ExC)$ and
the skew-hermitian  form 
$\Pi^{*}\IM{\bar \eta_{1}}{\eta_{2}}$ on
\eq
%\label{eq:currentsubspace}
\cD_{0}(\ExC)\oplus\cD_{1}(\ExC)\oplus\cD_{2}(\ExC)
\,.
\en
$\Pi^{*}\IM{\bar\eta_{1}}{\eta_{2}}$
vanishes unless the pushed-down currents
intersect in $M$, i.e., unless
\eq
(k_{1}+n-1) + (k_{2}+n-1) = d
\en
which is
\eq
k_{1}+k_{2} =2
\,,
\en
just as for the skew-hermitian intersection form of a two-dimensional manifold.
The $J$-operator on 1-currents and the skew-hermitian form $\Pi^{*}\IM{\bar\eta_{1}}{\eta_{2}}$
have exactly the properties
(\ref{eq:Jsq=1}, \ref{eq:hermitianintersection1}--\ref{eq:hermitianintersection5})
of the $J$-operator and the skew-hermitian intersection form of a 
Riemann surface.
The strict positivity property 
(\ref{eq:hermitianintersection5}) will hold if we mod out by the 
null space of the skew-hermitian form on the 1-currents.

This structure --- spaces of $0$-currents, $1$-currents, and $2$-currents
with a $J$-operator and a skew-hermitian form
---
is to be the geometric setting for 
quantum field theory of extended objects.

\section{Quasi Riemann surfaces}
\label{sect:quasiRiemannsurfaces}

I will try to formulate an abstract definition of {\it quasi Riemann surface}
that encompasses the geometry of currents on ordinary Riemann surfaces
and also of currents on the spaces $\ExC$.
The idea is to include in the definition all the properties that apply
both to the currents in a Riemann surface
and to the currents in the spaces $\ExC$.
The definition will be not be entirely precise.
I will mention some of the properties, but probably not all of them.
And I will remark one key technical gap.

Then I will conjecture: (1) that quasi Riemann surfaces are 
classified up to isomorphism by the homology data --- the 
homology along with the skew-hermitian 
form and the $J$-operator in the middle dimension, 
and (2) that every isomorphism class contains a two-dimensional 
conformal space, which is a Riemann surface when the
homology data is appropriate.
The homology of a connected Riemann surface $\Sigma$ is $H_{1}(\Sigma)$.
The middle homology of $\ExC$ is $H_{n}(M)$.
For example, if $M=S^{2n}$ then each $\ExC$ is isomorphic to
$S^{2}$ as a quasi Riemann surface, if the conjecture is true.

Further, I will suppose that every 2d conformal field theory (2d CFT)
can be installed in a natural way
on any two-dimensional quasi Riemann surface $\Sigma$,
in particular on the quasi Riemann surface corresponding to an 
ordinary Riemann surface, e.g., $S^{2}$.
This means extending the observables from products of ordinary local 
quantum fields 
over a finite collection of points in the two-dimensional space
to products of local fields over an integral 0-current in the 
two-dimensional space.
The integral 0-currents are the extended objects 
in the two-dimensional space.
The extension of a 2d CFT from Riemann surfaces to the two-dimensional quasi Riemann surfaces
might be called a 2d {\it extended} conformal field theory (2d ECFT).

The mathematical conjecture will then allow the 2d ECFT to be 
installed on each of the $\ExC$, via an isomorphism of quasi Riemann 
surfaces between $\ExC$ and the corresponding two-dimensional quasi 
Riemann surface.
This will give a conformal field theory of extended objects in $M$ 
for every 2d CFT.

\subsection{Definition}

A quasi Riemann surface is to consist
of abelian groups $\cQ^{\integral}_{0}$, $\cQ^{\integral}_{1}$, 
$\cQ^{\integral}_{2}$.
Each $\cQ^{\integral}_{k}$ is  contained in the
corresponding complex vector space $\cQ_{k} = \Complexes 
\otimes_{\Integers} \cQ^{\integral}_{k} $.
The $\cQ^{\integral}_{k}$ are complete metric spaces.
There is a boundary operator $\partial$ and an augmented de 
Rham complex
\eq
\begin{diagram}
0 	&\lTo	& \cQ^{\integral}_{-1} &\lTo^{\partial}	& \cQ^{\integral}_{0} 
&\lTo^{\partial}	& \cQ^{\integral}_{1} 
&\lTo^{\partial}	& \cQ^{\integral}_{2} 
&\lTo^{\partial}	& \cQ^{\integral}_{3} 
&\lTo	& 0
\\
&& \dInto && \dInto&& \dInto&& \dInto&& \dInto
\\
0 	&\lTo	& \cQ_{-1} &\lTo^{\partial}	& \cQ_{0} 
&\lTo^{\partial}	& \cQ_{1} 
&\lTo^{\partial}	& \cQ_{2} 
&\lTo^{\partial}	& \cQ_{3} 
&\lTo	& 0
\end{diagram}
\en
with
\eq
\cQ^{\integral}_{-1}  = \Integers
\,,\quad
\cQ_{-1}  = \Complexes
\,,\qquad
\cQ^{\integral}_{3} = \Integers \oplus i \Integers
\,,\quad
\cQ_{3} = \Complexes\,.
\en
Call $\cQ_{k}$ the $k$-space of the quasi Riemann surface,
and call its elements the $k$-currents.

The imaginary parts are 
to cope with manifolds $M$ of dimension $d=2n$ with $n$ even.
If only $n$ odd were considered,
then all the imaginary parts could be dropped.

There are pushdown maps
\eq
\Pi_{*}^{j,k}\colon \cD^{\integral}_{j}(\cQ^{\integral}_{k})
\rightarrow \cQ^{\integral}_{j+k}
\en
satisfying
\eq
\partial \Pi_{*}^{j,k} = 
\Pi_{*}^{j-1,k} \partial
+ \Pi_{*}^{j,k-1} \partial_{*} 
\,.
\en
The induced maps
on the homology groups of the augmented de Rham 
complex of integral currents $\cD^{\integral}_{j}((\cQ^{\integral}_{k})_{0})$ 
on the space of integral $k$-cycles,
\eq
\Pi_{*}^{j,k}\colon
H_{j}((\cQ^{\integral}_{k})_{0})
\rightarrow
H_{j+k}(\cQ^{\integral})
\,,
\en
should be isomorphisms.

There is a linear operator $J$ on $\cQ_{1}$
and there is a nondegenerate skew-hermitian form $\IQ{\bar\eta_{1}}{\eta_{2}}$
on $\oplus_{k} \cQ_{k}$.
They satisfy
\eq
J^{2} = -1
\en
and
\ateq{2}{
\label{eq:Sigmahermitianintersection1}
\IQ{\bar\eta_{1}}{\eta_{2}}  &= 
- \overbar{\IQ{\bar\eta_{2}}{\eta_{1}}}
\\[1ex]
\label{eq:Sigmahermitianintersection2}
\IQ{\overbar{\partial\eta_{1}}}{\eta_{2}} &= - 
\IQ{\bar\eta_{1}}{\partial\eta_{2}}
\\[1ex]
\label{eq:Sigmahermitianintersection3}
\IQ{\bar\eta_{1}}{\eta_{2}}  &=  0
\,,\qquad&
k_{1}&+k_{2}\ne 2
\\[1ex]
\label{eq:Sigmahermitianintersection4}
\IQ{\overbar{J \eta_{1}}}{ \eta_{2}}  &= 
-\IQ{\bar \eta_{1}}{ J\eta_{2}} 
\,,\qquad&
k_{1}&=k_{2}=1
\\[1ex]
\label{eq:Sigmahermitianintersection5}
\IQ{\bar \eta}{ J\eta}  &> 0
\,,\quad\eta \ne 0\,,
&
k_{1}&=k_{2}=1
\\[1ex]
\label{eq:Sigmahermitianintersection6}
\IQ{\bar\eta_{1}}{ \eta_{2}}  &= 
-\bar\eta_{1} \eta_{2}
\,,\qquad&
k_{1}&=-1, \; k_{2}=3
\,.
}
More properties that might be included in the definition of quasi 
Riemann surface
are discussed below, in section \ref{sect:definitionmore}.

We will say that a quasi Riemann surface is {\it connected} when
\eq
H_{0}(\cQ^{\integral}) = H_{2}(\cQ^{\integral}) = 0
\,.
\en
So all the homology of the augmented de Rham complex
is in the middle dimension, which is dimension 1,
$H_{2}(\cQ^{\integral})$.
The quasi Riemann surface associated to 
a Riemann surface $\Sigma$
will be connected iff 
$\Sigma$ is connected.  
The quasi Riemann surfaces associated to
a manifold $M$
will be connected
iff $H_{n-1}(M)=0$.
To avoid (minor) complications, I will assume connectedness.  So Riemann surfaces 
$\Sigma$ will be assumed connected.  And $H_{n-1}(M)=0$ will be 
assumed for space-time manifolds $M$.

Note that $\cQ_{1}$ decomposes as a Hilbert space into three orthonormal subspaces
\eq
\cQ_{1} = (\partial \cQ_{2}) \oplus \cQ_{1,H} \oplus (J \partial 
\cQ_{2})
\en
where $\cQ_{1,H}$ is the space of harmonic 1-currents,
\eq
\cQ_{1,H} =(\Ker \partial) \cap (J \Ker \partial)
\,,\qquad
J \cQ_{1,H} = \cQ_{1,H}
\,,\qquad
\Ker \partial = (\partial \cQ_{2}) \oplus \cQ_{1,H} 
\,.
\en

\subsection{$\ExC$ as a quasi Riemann surface}
\label{sect:ExCasaquasiRS}

To interpret $\ExC$ as a quasi Riemann surface,
let $\cQ_{k}$ be the $k$-currents in $\ExC$ modulo the null spaces 
of the skew-hermitian form 
$\Pi^{*}\IM{\bar\eta_{1}}{\eta_{2}}$,
\eq
\cQ^{\integral}_{k} = \cD^{\integral}_{k}(\ExC)/\cN^{\integral}_{k}
\,,\qquad
\cQ_{k} = \cD_{k}(\ExC)/\cN_{k}
\,,\qquad
k = 0,1,2,3
\,.
\en
The skew-hermitian form is
\eq
\IQ{\bar\eta_{1}}{\eta_{2}} = \Pi^{*}\IM{\bar\eta_{1}}{\eta_{2}}
= \IM{\overbar{\Pi_{*}\eta_{1}}}{\Pi_{*}\eta_{2}}
\,,
\en
which is manifestly nondegenerate.
Take $J=\epsilon_{n}*$ to be the $J$-operator already defined
on $\cD^{\integral}_{1}(\ExC)$.
The skew-hermitian form 
factors through the pushdown maps $\Pi_{*}(M)$ of $M$ (so written to 
distinguish them from the pushdown maps of $\cQ$),
so the spaces $\cQ^{\integral}_{k}$ are spaces of currents in $M$ modulo the 
appropriate null spaces,
\aeq{
\cQ^{\integral}_{0} 
&= \mathop\oplus_{k\in\Integers}
\cD^{\integral}_{n-1}(M)_{k\partial\xi} \oplus
i \partial \cD^{\integral}_{n}(M)
\\[0.5ex]
\cQ^{\integral}_{1} 
&= 
\cD^{\integral}_{n}(M) \oplus i  \cD^{\integral}_{n}(M)
\\[0.5ex]
\cQ^{\integral}_{2} &= 
\left(\cD^{\integral}_{n+1}(M) \oplus i \cD^{\integral}_{n+1}(M)
\right )
/ \cN^{\integral}_{n+1}(M,\partial\xi)
\\[0.5ex]
\cQ^{\integral}_{3} &= 
\left(\cD^{\integral}_{n+2}(M) \oplus i \cD^{\integral}_{n+2}(M)
\right )
/\cN^{\integral}_{n+2}(M,\partial\xi)
\\[1ex]
\cQ_{0}&= \cD_{n-1}(M)_{\Complexes\partial\xi}
\\[0.5ex]
\cQ_{1} &= \cD_{n}(M)
\\[0.5ex]
\cQ_{2} &= 
\cD_{n+1}(M)/\cN_{n+1}(M,\partial\xi)
\\[0.5ex]
\cQ_{3} &= 
\cD_{n+2}(M)/\cN_{n+2}(M,\partial\xi)
}
where
\aeq{
\cN^{\integral}_{n+1}(M,\partial\xi)
&= \left \{
\xi_{2} \in \cD^{\integral}_{n+1}(M)\oplus i  \cD^{\integral}_{n+1}(M)
:
\:\IM{\bar\xi_{1}}{\xi_{2}} = 0\,,
\:\:
\forall \xi_{1} \in \cQ^{\integral}_{0}
\right \}
\\[1ex]
\cN^{\integral}_{n+2}(M,\partial\xi)
&= \left \{
\xi_{2} \in \cD^{\integral}_{n+2}(M)\oplus i  \cD^{\integral}_{n+2}(M)
:
\:\:\IM{\overbar{\partial\xi}}{\xi_{2}} = 0
\right \}
\\[1ex]
\cN_{n+1}(M,\partial\xi)
&= \left \{
\xi_{2} \in \cD_{n+1}(M):
\:\:\IM{\bar\xi_{1}}{\xi_{2}} = 0\,,
\:\:
\forall \xi_{1} \in \cD_{n-1}(M)_{\Complexes\partial\xi}
\right \}
\\[1ex]
\cN_{n+2}(M,\partial\xi)
&= \left \{
\xi_{2} \in \cD_{n+1}(M):
\:\:\IM{\overbar{\partial\xi}}{\xi_{2}} = 0
\right \}.
}
The elements of $\cD^{\integral}_{n-1}(M)_{\partial\xi}$ are of the 
form $\xi_{1}= \xi + \xi_{0}$ for arbitrary integral $(n{-}1)$-cycle 
$\xi_{0}$.  Therefore
\eq
\cN^{\integral}_{n+1}(M,\partial\xi)
= \partial \cN^{\integral}_{n+2}(M,\partial\xi)
\,,\qquad
\cN_{n+1}(M,\partial\xi)
= \partial \cN_{n+2}(M,\partial\xi)
.
\en
$\cN^{\integral}_{n+2}(M,\partial\xi)$ is
the subgroup of (imaginary) integral $(n{+}2)$-currents that do not intersect
the $(n{-}2)$-boundary $\partial\xi$.
$\cN^{\integral}_{n+1}(M,\partial\xi)$ is
the subgroup of (imaginary) integral $(n{+}1)$-boundaries that do not link
the $(n{-}2)$-boundary $\partial\xi$.

The boundary operator acts on $\eta\in\cQ^{\integral}_{0}$ by
\eq
\partial_{M} \eta  = (\partial\eta) \partial\xi
\en
where the lhs is the boundary operator on currents in $M$
acting on $\eta$ considered as an integral $(n{-}1)$-current in $M$.

The problem is to define $1\in \cQ^{\integral}_{3}$.
It must satisfy
\eq
\IQ{1}{1} = -1
\,,
\en
where, in the skew-hermitian form, on the left is $1\in \cQ^{\integral}_{-1}$
and on the right is $1\in \cQ^{\integral}_{3}$.
So $1\in \cQ^{\integral}_{3}$ must be represented 
by an $(n{+}2)$-current in $M$
\eq
\xi_{3} \in \cD^{\integral}_{n+2}(M) \oplus i \cD^{\integral}_{n+2}(M)
\en
which satisfies
\eq
\IM{\overbar{\partial\xi}}{ \xi_{3}} = -1
\,.
\en
The definition of the skew-hermitian intersection form in $M$
in terms of the ordinary intersection form
given in section \ref{sect:hermitianintersectionform} says
\eq
\IM{\overbar{\partial\xi}}{ \xi_{3}} =  \epsilon_{n,2} 
I_{M}(\overbar{\partial\xi},\xi_{3})
= I_{M}(\overbar{\partial\xi},-\epsilon_{n}^{-1}\xi_{3})
\,,
\en
so $\xi_{3}$ must satisfy
\eq
I_{M}(\overbar{\partial\xi},\epsilon_{n}^{-1}\xi_{3}) = 1\,.
\en
The existence of such a current $\xi_{3}$ is a constraint on $\partial\xi$.  There must exist an 
integral $(n{+}2)$-cycle $\epsilon_{n}^{-1}\xi_{3}$ which has intersection 
number 1 with $\partial\xi$.
I believe that this is equivalent to the condition that $\partial 
\xi$ is irreducible, i.e., that
\eq
\partial \xi \ne k \partial\xi'
\,,\qquad \forall k\in \Integers,
\: \partial\xi' \in \partial \cD^{\integral}_{n-1}(M)\,,
\quad k \ne \pm 1
\,,
\en
but I do not have a proof.
Assuming this to be true,
then $\ExC$ gives a quasi Riemann surface iff
$\partial\xi$ is irreducible.

\subsection{The bundle $\QM\rightarrow\PBM$ of quasi Riemann surfaces}

$\cE^{\Complexes}_{\pm\partial\xi}$ give the same quasi Riemann 
surface,
so we can
associate the quasi-Riemann surface to the {\it integral line} $\Integers 
\partial \xi \subset \cB$.
The integral lines $\Integers \partial \xi$ are
the maximal abelian group homorphisms $\Integers \rightarrow \cB$.
Write $\PBM$ for the space of integral lines in $\cB$,
\eq
\PBM = \left \{ 
\Integers \partial \xi:
\partial \xi\in \cB\,, \partial \xi\text{ irreducible}
\right \}
\,.
\en
We might call $\PBM$ the {\it integral projective space} of $\cB$.

There is a quasi Riemann surface $\QM_{\Integers\partial\xi}$ for 
every integral line $\Integers\partial\xi\in \PBM$.
Let
\eq
\QM = \mathop\cup_{\Integers\partial\xi\in \PBM} \QM_{\Integers\partial\xi}
\en
so
\eq
\QM \rightarrow \PBM
\en
is a bundle of quasi Riemann surfaces.
% (This is abuse of notation ---
% using $\cQ$ for the general quasi Riemann surface
% and also for this bundle.)

The 0-space of the quasi Riemann surface $\cQ_{\Integers\partial\xi}$ is
\eq
\QM^{\integral}_{\Integers\partial\xi,0}
 = \mathop\oplus_{k\in\Integers}
\cD^{\integral}_{n-1}(M)_{k\partial\xi} \oplus
i \partial \cD^{\integral}_{n}(M)
\en
within which the space of 0-cycles, $\Ker \partial$, is
\eq
(\QM^{\integral}_{\Integers\partial\xi,0})_{0}
 = \cD^{\integral}_{n-1}(M)_{0} \oplus i \partial \cD^{\integral}_{n}(M)
\,.
\en
Note that the same $(\cQ_{\Integers\partial\xi,0})_{0}$
occurs in every fiber $\cQ_{\Integers\partial\xi}$.

The sum of the 0-spaces of all the fibers 
$\QM_{\Integers\partial\xi}$,
taken as subgroups within the space of $(n{-}1)$-currents in $M$,
is the full space of extended objects,
\eq
\sum_{\Integers\partial\xi\in \PBM}
\QM^{\integral}_{\Integers\partial\xi,0}
=
\cD^{\integral}_{n-1}(M) \oplus
i \partial \cD^{\integral}_{n}(M)
=
\cE^{\Complexes}
.
\en

\subsection{A Riemann surface $\Sigma$ as a quasi Riemann 
surface}
\label{sect:RiemannsurfaceasquasiRS}

Now we want to associate a quasi Riemann surface $\QSigma$
to every ordinary Riemann surface $\Sigma$.
The integral spaces
$\QSigma^{\integral}_{k}$
are complexifications of the integral $k$-currents in 
$\Sigma$,
and the $\QSigma_{k}$
are the complex $k$-currents in $\Sigma$,
\ateq{2}{
\QSigma^{\integral}_{0} &= \cD^{\integral}_{0}(\Sigma) \oplus i\partial 
\cD^{\integral}_{1}(\Sigma)
\\
\QSigma^{\integral}_{k} &= \cD^{\integral}_{k}(\Sigma) \oplus i  
\cD^{\integral}_{k}(\Sigma)
\,,\qquad
&k&=1,2
\\
\QSigma_{k} &= \cD_{k}(\Sigma)
\,,\qquad
&k&=1,2,3
\,.
}
Again, the imaginary parts are 
needed to cope with manifolds $M$ of dimension $d=2n$ with $n$ even.

The boundary operator acts on $\QSigma^{\integral}_{0}$ by
\eq
\partial\colon\eta\mapsto \int_{\eta} 1
\,,
\en
and on $\QSigma_{3}$ by
\eq
\partial\colon 1 \mapsto \Sigma
\,.
\en
The $J$-operator is $\epsilon_{1}{*}$ with $\epsilon_{1}^{2}=1$.
The skew-hermitian form
is $\IQ{\bar\eta_{1}}{\eta_{2}} =\ISigma{\bar\eta_{1}}{\eta_{2}}$.
Condition (\ref{eq:Sigmahermitianintersection6}) expresses the extension of
$\ISigma{\bar\eta_{1}}{\eta_{2}}$  to 
$\eta_{1}\in\cD_{-1}(\Sigma)$,
$\eta_{2}\in\cD_{3}(\Sigma)$,
by
\eq
\ISigma{\bar 1}{1}
= \ISigma{\overbar{\partial\delta_{z}}}{1}
= - \ISigma{\bar\delta_{z}}{\partial 1}
= - \ISigma{\bar\delta_{z}}{\Sigma}
= -1
\,.
\en

One significant technical issue will be left unresolved.
The pushdown maps
\eq
\Pi_{*}^{j,k}\colon \cD^{\integral}_{j}(\QSigma^{\integral}_{k})
\rightarrow \QSigma^{\integral}_{3}
\,,\qquad
j+k = 3
\en
must be identically zero,
as the definition of $\QSigma$ presently stands.
Somehow,
the spaces of currents $\cD^{\integral}_{j}(\QSigma^{\integral}_{k})$
will have to be augmented
to reflect the augmentation of the de Rham complex of currents in 
$\Sigma$.
This is discussed further in
section~\ref{sect:augmentSigma} below.

\subsection{Definition (more)}
\label{sect:definitionmore}

The abstract definition of quasi Riemann surface
should require all properties that are
satisfied by the concrete quasi Riemann surfaces 
$\QSigma$ and
$\cQ_{\Integers\partial\xi}$.
The list of properties might include
\begin{enumerate}
\item
Each of the complete metric spaces $\cQ^{\integral}_{k}$ should be generated as an abelian 
group by an arbitrarily small neighborhood of the identity.
\item
The tangent spaces $T_{0}\cQ^{\integral}_{k}$ should be dense 
subspaces of the vector spaces $\cQ_{k}$.
\item
The operator $J$ on $\cQ_{1}$ should preserve the tangent space $T_{0}\cQ^{\integral}_{0}$.
\item
The skew-hermitian form $\IQ{\bar\eta_{1}}{\eta_{2}}$ should be 
defined on a dense subspace of $\bar \cQ_{k} \otimes \cQ_{2-k}$.
\item
$\IQ{\bar\eta_{1}}{\eta_{2}}$ should be 
defined on a dense abelian subgroup of $\bar \cQ^{\integral}_{k} \times 
\cQ^{\integral}_{2-k}$,
where it should take values in $\Integers \oplus i\Integers$.
\item The maps $\Pi_{*}^{j,k}$ should be surjective (onto).
For $\QSigma$ and $j+k=3$, this is an unresolved issue.
\item
$\Pi_{*}^{1,0}$ should be injective on the tangent space
$T_{0}\cQ^{\integral}_{0}$
and
$\Pi_{*}^{2,0}$ should be injective on the space of 2-vectors
at $0\in \cQ^{\integral}_{0}$.
\item
The maps $\Pi_{*}^{j,k}$ should satisfy compatibility conditions,
perhaps such as that
\ateq{3}{
&\cD^{\integral}_{1}(\cD^{\integral}_{1}(\cQ^{\integral}_{0}))
& &\rightarrow  \cD^{\integral}_{2}(\cQ^{\integral}_{0})
&&\rightarrow \cQ^{\integral}_{2}
\\
\text{and}\quad
& \cD^{\integral}_{1}(\cD^{\integral}_{1}(\cQ^{\integral}_{0}))
&&\rightarrow  \cD^{\integral}_{1}(\cQ^{\integral}_{1})
&&\rightarrow \cQ^{\integral}_{2}
}
should give the same result.
\end{enumerate}

\subsection{Morphisms}

A morphism $f\colon\cQ \rightarrow \cQ'$ of quasi Riemann surfaces
is a set of maps
\eq
f_{k}\colon\cQ^{\integral}_{k}\rightarrow 
\cQ^{\prime\,\integral }_{k}
\en
preserving all the structures and properties of quasi Riemann 
surfaces.
The morphism is determined by $f_{0}\colon\cQ^{\integral}_{0}\rightarrow 
\cQ^{\prime\,\integral}_{0}$.
The quasi Riemann surface conditions are constraints on $f_{0}$.
The map $f_{1}\colon\cQ^{\integral}_{1}\rightarrow 
\cQ^{\prime\,\integral}_{1}$ is the derivative of $f_{0}$,
\eq
\partial f_{1} = f_{0} \partial
\,.
\en
Extended to a linear map $f_{1}\colon\cQ_{1}\rightarrow 
\cQ^{\prime}_{1}$, it should preserve the $J$-operators
\eq
f_{1} J = J f_{1}
\en
and it should be a partial unitary transformation with respect to the 
positive definite hermitian forms,
\eq
\IQ{\bar\eta_{1}}{J\eta_{2}}
= I_{\cQ'}\expval{\overbar{f_{1}\eta_{1}},J f_{1}\eta_{2}}
\,.
\en
In addition, $f_{1}$ should preserve the kernels of the pushdown 
operator,
\eq
f_{1} \left (\Ker \Pi_{*}^{1,1} \right) \subset \Ker \Pi_{*}^{1,1}
\,.
\en
Then$f_{2}\colon\cQ^{\integral}_{2}\rightarrow 
\cQ^{\prime\,\integral}_{2}$ will be given by
\eq
\partial f_{2} = f_{1} \partial
\,.
\en
Continuity of $f_{1}$ should guarantee that the skew-hermitian 
forms are preserved in toto.
In particular, continuity should ensure that
\eq
f_{2}\partial 1 = \partial 1\,.
\en
Alternatively, a morphism
is determined by $f_{1}\colon\cQ^{\integral}_{1}\rightarrow 
\cQ^{\prime\,\integral}_{1}$, subject
to the constraints
\begin{itemize}[leftmargin=7em,labelsep=1em]
\item[{\bf M1}]
$f_{1}$ is a homorphism of abelian groups.
\item[{\bf M2}]
$f_{1}$ is continuous.
\item[{\bf M3}]
$f_{1} \left (\Ker \partial \right)\subset \Ker \partial$
\item[{\bf M4}]
$f_{1} \left (\Ima \partial\right) \subset \Ima \partial$
\item[{\bf M5}]
$f_{1} J = J f_{1}$
\item[{\bf M6}]
$\IQ{\bar\eta_{1}}{J\eta_{2}}
= I_{\cQ'}\expval{\overbar{f_{1}\eta_{1}},J f_{1}\eta_{2}}$
\item[{\bf M7}]
$f_{1*}\left (\Ker \Pi_{*}^{1,1} \right) \subset \Ker \Pi_{*}^{1,1}$
\,.
\end{itemize}

The last condition is essential.  All the quasi Riemann surfaces 
$\QxZ$ have the same $(\QxZ^{\integral})_{1}=\cD^{\integral}_{n}(M)+i 
\cD^{\integral}_{n}(M) $.
The subgroups $\Ker \partial$ and $\Ima \partial$
in $(\QxZ^{\integral})_{1}$ are the same.
They have the same $J$ and 
$\IQ{\bar\eta_{1}}{\eta_{2}}$
on $(\QxZ)_{1}$.
Only $\Ker \Pi_{*}^{1,1} $ depends nontrivially on $\Integers\partial\xi$.
Let
\eq
\cW^{\integral}_{1,1} =  \left \{ 
\eta \in \cD^{\integral}_{1}((\QxZ^{\integral})_{1})
\colon
\Pi_{*}^{1,1} \eta \in \partial (\QxZ^{\integral})_{3}
\right \}
\,.
\en
Then $\Ker \Pi_{*}^{1,1} \subset \cW^{\integral}_{1,1}$ and
\eq
\cW^{\integral}_{1,1}/ \Ker \Pi_{*}^{1,1} = \Integers\,.
\en
$\cW^{\integral}_{1,1}$ is the same for all $\Integers\partial\xi$.
What distinguishes --- and characterizes --- the different $(\QxZ^{\integral})_{1}$ is the way 
that $ \Ker \Pi_{*}^{1,1}$ sits inside $\cW^{\integral}_{1,1}$.

The linear operator $f_{1}$ on $\cQ_{1}= (\partial \cQ_{2}) \oplus \cQ_{1,H} \oplus (J \partial 
\cQ_{2})$ is determined by its action on 
$\partial\cQ_{2}$ and its action on $\cQ_{1,H}$, which is the complex homology.
The action on $J\partial\cQ_{2}$ then follows because $f_{1}$ commutes 
with $J$.

\subsection{Isomorphisms and automorphisms}
\label{sect:Isoandauto}

Write $\Iso(\cQ, \cQ')$ for the isomorphisms between two quasi 
Riemann surface.
Write $\Aut(\cQ)$ for the group of automomorphisms of $\cQ$.

If $f\colon\cQ\rightarrow \cQ'$ is an isomorphism, then $f_{1}\colon 
\cQ_{1}\rightarrow \cQ^{\prime}_{1}$ is unitary.
Conversely, unitarity of $f_{1}$ should imply that a morphism $f$ is an 
isomorphism.

For a Riemann surface $\Sigma$, the group $\Conf(\Sigma)$ of conformal 
symmetries of $\Sigma$ is a subgroup of $\Aut(\QSigma)$.
For example, $\Aut(\cQ(S^{2}))$ contains $\mathbf{PSL}(2,\Complexes)$ 
as a subgroup.

For a conformal manifold $M$, the group $\Conf(M,\Integers\partial\xi)$ of 
conformal symmetries of $M$ that preserve $\pm\partial\xi$ is a subgroup 
of $\Aut(\QxZ)$.
For example, if $M=S^{d}$ and $\partial\xi$ is an $(n{-}2)$-sphere in 
$M$, then $\Aut(\QxZ)$ contains
$S(O(n{+}2,\Reals)\times O(n{-}1,\Reals))$
as a subgroup.

\subsection{Morphisms and homology}
A morphism $f\colon \cQ\rightarrow \cQ'$ maps the homology group
$H_{1}(\cQ^{\integral})$ into the homology group 
$H_{1}(\cQ^{\prime\,\integral})$.
The homology map is injective because the 
skew-hermitian form is preserved.
The free parts of the homology groups are lattices in the real homology 
groups $H_{1}(\cQ)$ and $H_{1}(\cQ')$,
which are complex Hilbert spaces, the complex structures given by 
the respective $J$-operators
and the positive definite hermitian inner product given 
by the skew-hermitian forms combined with the $J$-operators.
The map of homology groups preserves this structure.
I do not know how to deal with the torsion subgroup of the 
homology, so I will just disregard the possibility of torsion.

Call the homology of a quasi Riemann surface, along with the action 
of $J$ and the skew-hermitian form, the {\it homology data}.
An isomorphism of quasi Riemann surfaces gives an isomorphism of 
homology data,
so two quasi Riemann surfaces are isomorphic only if they have 
isomorphic middle homology data.
I will conjecture shortly that this should be `if and only if',
that quasi Riemann surfaces are classified by the homology data.

For connected quasi Riemann surfaces, all the homology is in the 
middle dimension.
For the quasi Riemann surface $\QSigma$ associated to an ordinary 
Riemann surface $\Sigma$, the middle homology of the quasi Riemann 
surface is the middle homology of $\Sigma$, doubled,
\eq
H_{1}(\QSigma^{\integral}) = H_{1}(\Sigma,\Integers)
\oplus i H_{1}(\Sigma,\Integers)
\,.
\en
For a quasi Riemann surface $\QxZ$ associated to a manifold $M$ of 
dimension $d=2n$, the  middle homology of the quasi Riemann 
surface is the middle homology of $M$, doubled,
\eq
H_{1}(\QxZ^{\integral}) = H_{1}(M,\Integers)
\oplus i H_{1}(M,\Integers)
\,.
\en
Again, the imaginary parts can be dropped if we limit ourselves to 
$n$ odd.

\subsection{Quasi-holomorphic curves}

Given a Riemann surface $\Sigma$,
define a {\it quasi-holomorphic curve},
or a {\it quasi-holomorphic $\Sigma$-curve},
to be a morphism of quasi Riemann surfaces from
$\QSigma$ to a $\QxZ$.
Define a {\it local quasi-holomorphic curve} to be a q-h
$\Sigma$ -curve with $\Sigma$ conformally equivalent to the open unit disk.

Recall that
\eq
(\QxZ)_{0} \supset \partial^{-1}(1) 
= \ExC
= 
\cD^{\integral}_{n-1}(M)_{\partial\xi}\oplus i \partial \cD^{\integral}_{n}(M)
\,.
\en
The map $f_{0}\colon \QSigma_{0}\rightarrow(\QxZ)_{0}$
is equivalent to the map
\eq
\hc\colon  \Sigma \rightarrow \ExC
\,,\qquad
C(z) = f_{0}(\delta_{z})
\,.
\en
The map $f_{0}$ can be recovered from the map $C$ by
\eq
f_{0}\colon  \sum_{i} n_{i}\delta_{z_{i}}
\mapsto\sum_{i} n_{i}C(z_{i})
\,.
\en
The map $f_{1}\colon \QSigma_{1}\rightarrow(\QxZ)_{1}$
is the derivative of $C$,
\eq
f_{1} = C_{*} \colon \cD^{\integral}_{1}(\Sigma)
\oplus i  \cD^{\integral}_{1}(\Sigma)
\rightarrow
\cD^{\integral}_{n}(M)\oplus i  \cD^{\integral}_{n}(M)
\en
It preserves the 
almost-complex structures $J$ and the skew-hermitian forms on currents,
\eq
\label{eq:qhc1}
\hc_{*} J = J \hc_{*} 
\quad
\text{on } \cD_{1}(\Sigma)
\,,
\en
\eq
\label{eq:qhc2}
\Pi^{*}\IM{\overbar{\hc_{*}\eta_{1}}}{\hc_{*}\eta_{2}} 
= \ISigma{\bar\eta_{1}}{\eta_{2}}
\,.
\en
A {\it pseudo-holomorphic curve} \cite{MR809718}
is a map from a Riemann surface $\Sigma$ to an
almost complex space that preserves the almost complex structures.
So a quasi holomorphic curve is a pseudo-holomorphic curve
in $\ExC$
that preserves the skew-hermitian forms
on currents in the middle dimension.

For $z$ a local complex coordinate on $\Sigma$,
condition (\ref{eq:qhc1}) takes the form
\eq
J \partial_{z} \hc = i \partial_{z}\hc
\,,\qquad
J \partial_{\bar z} \hc = - i \partial_{\bar z}\hc
\,.
\en
For $n$ odd, $J=\epsilon_{n}{*}$ is real,  so 
it is consistent to impose the reality condition
$\bar C = C$.
Then $C:\Sigma\rightarrow \Ex$ and we
can forgo complexifying $\Ex$.
For $n$ even, $C$ must be complex.

A q-h curve $C$ sweeps out an integral $(n{+}1)$-current
$\Pi_{*}C_{*}\Sigma$
in the space-time $M$.
An $(n{+}1)$-current intersects
an $(n{-}1)$-current
at a $0$-current.
A small local q-h curve will sweep out a small $(n{+}1)$-current in 
$M$,
so a small local q-h curve sees the local structure
of an extended object in the form of a 0-current.

\section{2d CFT on a quasi-holomorphic curve}
\label{sect:2dCFTonaqhc}

Suppose $C$ 
is a quasi-holomorphic curve in $\ExC$.
The 0-form and 1-form fields on $\ExC$
pull back to give 0-form and 1-form fields
on the Riemann surface $\Sigma$,
\ateq{2}{
C^{*}\phi_{\pm} (\eta_{0}) &= \phi_{\pm}(C_{*}\eta_{0})
\,,\quad&
\eta_{0}&\in\cD_{0}(\Sigma)
\\
C^{*}j_{\pm} (\eta_{1}) &= j_{\pm}(C_{*}\eta_{1})
\,,\quad&
\eta_{1}&\in\cD_{1}(\Sigma)
\,.
}
The fields on $\Sigma$  satisfy the field equations of the 2d CFT,
\eq
d (C^{*}\phi_{\pm}) = C^{*}j_{\pm}
\,.
\en
The 1-forms on $\Sigma$ are chiral
by the quasi-holomorphic condition (\ref{eq:qhc1}),
\eq
P_{+} (C^{*}j_{+}) = C^{*}j_{+}
\,,\qquad
P_{-} (C^{*}j_{-}) = C^{*}j_{-}
\,.
\en
Translated into the usual language of 2d CFT,
\ateq{2}{
\phi_{+} (z) &= C^{*}\phi_{+}(\delta_{z})
\,,\qquad&
\phi_{-} (\bar z) &= C^{*}\phi_{-}(\delta_{z})
\,,
\\
j_{+}(z) &= (C^{*}j_{+})^{z}(z)
\,,\qquad&
j_{-}(\bar z) &= (C^{*}j_{-})^{\bar z}(\bar z)
\,,
\\
 \partial \phi_{+} &=j_{+}(z)
\,,\qquad&
\partial \phi_{-} &= 0
\,,\quad
\\
\bar \partial \phi_{+} &= 0
\,,\qquad&
\bar \partial \phi_{-} &=j_{-}(z) 
\,.
}
We now have the classical fields of the 2d CFT on $\Sigma$.

To have the 2d {\it quantum} field theory on $\Sigma$,
we need the correlation functions to satisfy
the Schwinger-Dyson equations of the 2d CFT,
\aeq{
\label{eq:2dSD1}
\expval{(C^{*}j_{\bar \alpha})^{\dagger}(\bar\eta_{1})\, C^{*}j_{\beta}(\partial\eta_{2})} &= 
- 2\pi i\, \gamma_{\bar\alpha\beta} \ISigma{\overbar{\partial\eta_{1}}}{ \eta_{2}}
\\[1ex]
\label{eq:2dSD2}
\expval{(C^{*}\phi_{\bar \alpha})^{\dagger}(\bar \eta_{0})\,C^{*}j_{\beta}(\partial\eta_{2})}
&=  -2\pi i \gamma_{\bar\alpha\beta} \ISigma{\bar\eta_{0}}{ \eta_{2}}
\,,
}
which translates to
\eq
\bar \partial\expval{\phi^{\dagger}_{+}(z)\,j_{+}(w)} = 2\pi 
\delta^{2}(z-w)
\,,\quad\text{etc.}
\en
The second quasi-holomorphic condition (\ref{eq:qhc2}) ---
that $C$ preserves the skew-hermitian forms ---
implies that the 2d S-D equations (\ref{eq:2dSD1}--\ref{eq:2dSD2}) follow from
the S-D equations (\ref{eq:SDjj}--\ref{eq:SDphij}) on $\ExC$.
So we have the 2d CFT on the quasi-holomorphic 
curve.

The vertex operators $V(\xi)$ on $\ExC$ are exponentials of the
$\phi_{\pm}(\xi)$.  They pull back to $\Sigma$
to the corresponding exponentials of $C^{*}\phi_{\pm}$,
\eq
C^{*}V(\eta) = V(C_{*}\eta) 
\,,\qquad
\eta\in \cD^{\integral}_{0}(\Sigma)
\,.
\en
In the usual language 2d CFT, the pulled back vertex operators are the local 
fields
\eq
V(z) = C^{*}V(\delta_{z}) = V(C(z))
\,.
\en
The correlation functions of the 2-d fields,
\eq
\expval{C^{*}V_{1}(\eta_{1})\cdots
C^{*}j_{\pm}(\eta'_{1}) \cdots}
= \expval{V_{1}(C_{*}\eta_{1}) \cdots j_{\pm}(C_{*}\eta'_{1}) \cdots}
\,,
\en
are the correlation functions of the 2d CFT on $\Sigma$,
since the local properties of the correlation functions
are completely determined by the S-D equations.

For $n$ odd, the q-h curve $C$ can be taken real, a map
$\Sigma\rightarrow \Ex$.  The real $n$-form field $F(x)$ on $M$
becomes the real 1-form $j$ on $\Ex$.  The 1-form field $C^{*}j$ on
$\Sigma$ is then real.  The 2d CFT is the free theory of a real
1-form.  The 2d CFT on $\Sigma$ is exactly what was called the analog
theory on $\Ex$.
For $n$ even, the situation is more complicated.  Even if $F(x)$ is 
a real $n$-form, $C^{*}j$ must be a complex 1-form, because the q-h 
curve $C$ is necessarily complex.
The fields on $\ExC$ pulled back to $\Sigma$ comprise a subalgebra of the 2d CFT of 
the free {\it complex} 1-form.
Consider the fields of the real $n$-form theory as a 
subalgebra in the complex $n$-form theory --- 
the subalgebra generated by the vertex operators invariant under the complex 
conjugation symmetry $F\leftrightarrow \bar F$.
On $\Sigma$, this becomes a {\it different} subalgebra
of the 2d CFT of the free complex 1-form $C^{*}j$,
the subalgebra
generated by the vertex operators
invariant under the combination of 
complex conjugating and 
reversing orientation $J\rightarrow -J$,
$P_{+}\leftrightarrow P_{-}$.
More detail is given in Appendix~\ref{app:complexconjugation}.

\section{A wishful conjecture }
\label{sect:awishfulconjecture}

The project is to construct the CFT of extended objects in space-time
from the 2d CFT.
For every 2d CFT,
there is to be a CFT of extended objects
which, when pulled back to a quasi-holomorphic curve,
gives the 2d CFT on the Riemann surface.
Some theories of extended objects --- for example, orbifolds of 
free $n$-form theories --- might be constructed directly
in terms of the skew-hermitian form on currents and the $J$-operator,
so that their pull-backs to q-h curves are manifestly the corresponding 2d theories.
For the general case, however, a method is needed to construct the theory 
of extended objects from the 2d CFT.
The data of the 2d CFT on the quasi-holomorphic curves must be 
enough to construct the correlation functions of the theory of extended 
objects.

%\subsection{Classification of quasi Riemann surfaces}

The only way I can imagine realizing this project is
by means of isomorphisms
between the quasi Riemann surfaces $\QxZ$
and the quasi Riemann surface $\QSigma$.
The fields and correlation functions on $\QxZ$ 
will be constructed simply by pulling back the fields and correlation 
functions of the 2d theory on the Riemann surface $\Sigma$.

The only reason I can imagine for such isomorphisms to exist
is if {\it any} two quasi Riemann surfaces with the same
homology data are isomorphic.
The conjecture is
\begin{quote}
{\it
The isomorphism classes of 
quasi Riemann surfaces 
are classified by their
homology data.

Each isomorphism class contains a unique two-dimensional model.
For each possible set of homology data
there is a unique two-dimensional 
conformal space $\Sigma$
with that homology data,
so that $\QSigma$ belongs to the isomorphism class.
The two-dimensional conformal space $\Sigma$ is an ordinary Riemann surface when the homology 
data allows.
}
\end{quote}
I have no idea what the two-dimensional conformal space $\Sigma$ might be
when the homology data is not that of a Riemann surface,
much less how to
construction the correlation functions of a 2d CFT on such a 
space.
Prudence suggests limiting to space-times $M$ 
with homology data
such that $\Sigma$ can be a Riemann surface.
Actually, I will be more than delighted if
the conjecture can be shown to hold for trivial homology 
data,
so that it will apply to the basic case, $M=S^{d}$, $\Sigma=S^{2}$.

I do not know how to prove the conjecture.  There might be a 
route via the local quasi-holomorphic curves.
It might be supposed that
\begin{quote}
{\it
Every local quasi-holomorphic curve in $\QSigma$ 
is given by a local neigborhood in $\Sigma$,
up to automorphisms of $\QSigma$.

A local quasi-holomorphic curve in a quasi Riemann surface $\cQ$ is a 
rigid object
in the sense that it has a unique ``analytic'' continuation
to an isomorphism $\QSigma \rightarrow \cQ$
for some two-dimensional space $\Sigma$.
}
\end{quote}

If the conjecture is true, then the group $\Aut(\QSigma)$ will be
a very interesting object.
It will naturally contain the group $\Conf(\Sigma)$ of conformal symmetries of 
$\Sigma$.
It will be isomorphic to the groups $\Aut(\QxZ)$ for every conformal manifold $M$ with the 
same middle homology data as $\Sigma$, and 
every integral $(n{-}1)$-boundary $\partial\xi$ in $M$,
so it will also contain the groups $\Conf(M,\Integers\partial\xi)$ of 
conformal symmetries of $M$ that preserve $\pm\partial\xi$,
for every such $M$ and $\partial\xi$.

\section{Correlation functions from extended 2d CFT}
\label{sect:corrfnsfromECFT}

Assuming the conjecture is true,
the geometric isomorphism between each of the $\QxZ$ and $\QSigma$
can be used to construct a CFT
on each of the $\QxZ$ from {\it any} 2d CFT on $\Sigma$.
The observables on $\QxZ$ will be the pull-backs
under an isomorphism  $f\colon\QxZ\rightarrow \QSigma$,
\eq
f^{*}\Phi(\xi) = \Phi(f\xi)\,,
\en
of the 2d observables $\Phi(\eta)$ on $\QSigma$.
The correlation functions on $\QxZ$ will be given by the correlation 
functions on $\Sigma$,
\eq
\label{eq:generalconstruction}
\expval{f^{*}\Phi(\xi) \cdots}
=
\expval{\Phi(f\xi) \cdots}_{\Sigma}
\,.
\en
For this to work, the 2d CFT on $\Sigma$ needs to be extended so that
\begin{quote}
{\it
The extended 2d observables $\Phi(\eta)$ are defined on the integral currents 
$\eta\in\QSigma^{\integral}_{k}$.
The automorphism group $\Aut(\QSigma)$ acts on the vector space of
extended observables.
The correlation functions of extended observables are invariant under 
$\Aut(\QSigma)$.
}
\end{quote}

In ordinary 2d CFT, the observables are products of local fields over 
finite sets of distinct points,
\eq
\Phi(z_{1},\ldots,z_{N}) = \varphi_{1}(z_{1}) \cdots \varphi_{n}(z_{N})
\,,
\en
and linear combinations of such products.
We can regard such an observable as living on the 0-current
$\eta = \sum_{i} \delta_{z_{i}}$.
The problem is to extend observables
to integral currents.

The vertex operators 
of the 2d gaussian model
extend formally --- classically --- to the integral $0$-currents
since the fields $\phi$ and $\phi^{*}$ are 0-forms,
\eq
V_{p,p^{*}}(\eta) = e^{ip\phi(\eta)+ip^{*}\phi^{*}(\eta)}
\,.
\en
The regularization and renormalization of such extended vertex 
operators is still to be dealt with.
From the start,
I have been making an unspoken assumption, as a guiding hypothesis,
that the vertex operators of the free $n$-form CFT
can be constructed
on the integral $(n{-}1)$-currents in the space-time $M$,
as quantum fields.
I have not actually carried out this construction.
Now the proposal is to construct the extended vertex operators
as observables in the extended 2d CFT,
then transport them to the space-time $M$
via the conjectured isomorphism of quasi Riemann surfaces.

The construction of an extended 2d CFT from an ordinary 2d CFT
is one of a number of further steps towards
realizing the project 
that are listed in section~\ref{sect:furthersteps} below.

\section{Perturbation theory}
\label{sect:perturbationtheory}

The plan is to move, eventually, from conformal field theories of 
extended objects to non-conformal quantum field theories of extended 
objects, constructed from extended 2d non-conformal quantum field 
theories.
As a step in that direction, I consider perturbing a 2d CFT
and constructing the corresponding perturbation of
the CFT of extended objects in space-time.
The discussion is formal.  Nothing is said about
a relation between the 2d cutoff scale 
and the space-time cutoff scale.

A perturbation of a 2d CFT on a Riemann surface is given by
integrating a quantum field that is
a (1,1)-form on the Riemann surface.  Such perturbations arise when the
2d CFT depends on parameters, such as the parameter $R$ of the 2d 
gaussian model,
and  also when the
the 2d CFT can be perturbed to give a non-conformal 2d QFT.
In the latter case, the integral will depend on the 2d metric.
It will have to be cut off and renormalized.
The correspondence between  variations of the 2d QFT
and integrals of (1,1)-form fields is a manifestation of the action
principle.

\subsection{Varying the parameter $R$ of the gaussian model}

The parameter $R$ in the theory of a free $n$-form $F$ on $M$ is varied by inserting
in correlation functions the integral
\eq
\label{eq:Mperturbation}
\int_{M} {*}F\wedge F 
\,.
\en
The $n$-form $F$ pulls up to the 1-form $j=\Pi^{*}F$ on $\ExC$.
Suppose $\Sigma$ is a compact Riemann surface 
without boundary,
and $C$ is a quasi-holomorphic $\Sigma$-curve.
Then $C^{*}j$ is a free $1$-form on $\Sigma$.
On the Riemann surface $\Sigma$, we have the 2d gaussian model with parameter $R$.
The parameter $R$ in the 2d gaussian model is varied by inserting
in the correlation functions on $\Sigma$ the integral
\eq
\label{eq:Sigmaperturbation}
\int_{\Sigma} {*}(C^{*}j)\wedge (C^{*} j)  = \int_{\Sigma} C^{*}({*}j\wedge j)
\,,
\en
which is the integral of the (1,1)-form ${*}j\wedge j$ pulled back from 
$\ExC$ to $\Sigma$.

The insertions (\ref{eq:Mperturbation}) and 
(\ref{eq:Sigmaperturbation}) must give the same result.
They can be written
\eq
\int_{M} {*}F\wedge F  = I_{M}^{-1}\expval{\overbar{J F}, \, F}
\,,\qquad
\int_{\Sigma} {*}(C^{*}j)\wedge (C^{*} j)  = 
I_{\Sigma}^{-1}\expval{\overbar{J C^{*}j}, \, C^{*}j}
\,,
\en
where $I_{M}^{-1}\expval{\bar\omega_{1},\omega_{2}}$ is the skew-hermitian bilinear form on 
$n$-forms that is the inverse of the skew-hermitian bilinear form 
$\IM{\bar\xi_{1}}{\xi_{2}}$ on $n$-currents in $M$,
and $I_{\Sigma}^{-1}\expval{\bar\omega_{1},\omega_{2}}$ is the inverse of
$I_{\Sigma}\expval{\bar\eta_{1},\eta_{2}}$
on $1$-currents in $\Sigma$.

The identity between the two integrals (\ref{eq:Mperturbation}) and 
(\ref{eq:Sigmaperturbation}) for the variation of $R$ is 
now
\eq
I_{M}^{-1}\expval{\overbar{J F}, \, F}
=
I_{\Sigma}^{-1}\expval{\overbar{J C^{*}\Pi^{*}F}, \, C^{*}\Pi^{*}F}
=
\Pi_{*}C_{*} (I_{\Sigma}^{-1})\expval{\overbar{J F}, \, F}
\en
which will hold if and only if
\eq
\label{eq:inverseqh}
I_{M}^{-1} = \Pi_{*}C_{*} (I_{\Sigma}^{-1})
\,.
\en
Combined with the quasi-holomorphic condition on $C$,
\eq
I_{\Sigma} = C^{*} \Pi^{*} I_{M}
\,,
\en
equation (\ref{eq:inverseqh}) is equivalent to the unitarity of
\eq
\Pi_{*}C_{*} \colon \cD_{1}(\Sigma) \rightarrow \cD_{n}(M)
\,.
\en
$\Pi_{*}C_{*}$ is unitary
iff $C$ is an isomorphism of quasi Riemann surfaces.
Therefore, if the conjecture holds,
then the variation of $R$ in the 2d CFT on $\Sigma$
is equivalent to the variation of $R$ in the $n$-form theory on $M$.

This argument for the unitarity of $\Pi_{*}C_{*}$ is based on 
the structure of the $n$-form
CFT on $M$ and the $1$-form 2d CFT on $\Sigma$.
These are both 
free quantum field theories,
so it should be possible to translate the argument
into purely  mathematical terms,
amounting to a proof that, for $\Sigma$ a compact Riemann surface 
without boundary, {\it any} morphism of quasi Riemann surfaces 
$\QSigma \rightarrow \QxZ$ must be an isomorphism.

\subsection{Perturbing by a general $(1,1)$-form}

Suppose that the conjecture holds.
Then perturbing the 2d QFT on $\Sigma$ is equivalent to perturbing 
the QFT of extended objects in $M$.
The 2d perturbation is given by integrating a $(1,1)$-form over 
$\Sigma$.
The perturbation of the QFT in $M$ should be given by integrating 
an $(n,n)$-form over $M$.

Suppose $f\colon\QSigma\rightarrow \QxZ$ is an isomorphism.
Suppose $\Phi_{\Sigma}$ is a (1,1)-form on $\Sigma$.
Represent $\Phi_{\Sigma}$ as
\eq
\Phi_{\Sigma} = \sum_{a,b} c_{a,b} \overbar{J w_{a}} \wedge  w_{b}
\en
for some collection of 1-forms $w_{a}$ on $\Sigma$
and some constants $c_{a,b}$.
Then
\eq
\int_{\Sigma} \Phi_{\Sigma} =  \sum_{a,b} c_{a,b}
I_{\Sigma}^{-1}\expval{\bar w_{a},w_{b}}
\,.
\en
The linear operator
\eq
f_{1}\colon \QSigma_{1}\rightarrow (\QxZ)_{1}
\,,\qquad
\QSigma_{1} = \cD_{1}(\Sigma)
\,,\qquad
(\QxZ)_{1} = \cD_{n}(M)
\en
is invertible,
so we can construct
the $n$-forms $W_{a}$ on $M$,
\eq
W_{a} = (f_{1}^{*} )^{-1} w_{a}
\,.
\en
From the $W_{a}$ make a $(n,n)$-form on $M$,
\eq
\Phi_{M} = \sum_{a,b} c_{a,b} \overbar{J W_{a}} \wedge  W_{b}
\,.
\en
Then
\eq
\int_{M}\Phi_{M} = \sum_{a,b} c_{a,b} I_{M}^{-1}\expval{\bar W_{a},W_{b}}
= \sum_{a,b} c_{a,b}
I_{\Sigma}^{-1}\expval{\bar w_{a},w_{b}}
=
\int_{\Sigma} \Phi_{\Sigma}
\,.
\en
So the perturbation by the $(1,1)$-form $\Phi_{\Sigma}$ on $\Sigma$ and the perturbation by 
the $(n,n)$-form $\Phi_{M}$ on $M$ will have the 
same effect.

The construction of  $\Phi_{M}$ from $\Phi_{\Sigma}$ is ambiguous.
$\Phi_{M}$ is
only determined up to an $(n,n)$-form that integrates to zero on $M$.
Such $(n,n)$-forms are the redundant fields
--- the perturbations that 
only reparametrize the fields of the QFT, taking the QFT to an 
equivalent QFT.
Thus the 2d perturbation $\Phi_{\Sigma}$
maps to a class of equivalent
perturbations $\Phi_{M}$ of the space-time QFT.

\section{Gauge symmetry of the classical free $n$-form}
\label{sect:gaugetheory}

This section and the next 
go back to the classical field theory of a free $n$-form in the space-time $M$
regarded as a collection of
free 1-form theories on the fibers of the bundle $\cE\rightarrow\cB$
of integral $(n{-}1)$-currents in $M$
over the integral $(n{-}2)$-boundaries.
This section expresses the space-time gauge symmetry of the theory
in terms of a gauge symmetry over $\cB$.
The gauge symmetry group is the global symmetry group of the analog 2d 
theory.
One message is that generalization from the 2d gaussian model to 
2d quantum field theories with nonabelian global symmetry groups
might yield space-time theories with nonabelian gauge symmetry.
In section~\ref{sect:Explorations} below,
I will try to translate the structure described 
here for the classical theory
to the quantum theory on the bundle of quasi Riemann surfaces.

The $n$-form theory has local $U(1){\times}U(1)$ gauge symmetries
\eq
A\mapsto A + df
\,,\qquad
A^{*}\mapsto A^{*} + df^{*}
\en
given by pairs $f,f^{*}$ of $(n-2)$-forms on the space-time $M$.
The $1$-form theory on each $\Ex$
has a global $U(1){\times}U(1)$ symmetry.
We have seen that local gauge transformations on $M$
give global symmetry transformations in each $\Ex$.
If we are to build a space-time QFT from 2d QFT on the $\Ex$,
we will need to go in the opposite direction,
constructing the local gauge symmetries in space-time from 
the global symmetries in the $\Ex$.

\subsection{The local gauge transformations over $\cB$}

In the 2d gaussian model on a Riemann surface $\Sigma$, 
write  $T_{\Sigma}$ for the space of solutions
$\phi$, $\phi^{*}$
of
\eq
j = d\phi
\,,\qquad
j^{*} = d\phi^{*}
\,.
\en
Then $T_{\Sigma}=S^{1}\times S^{1}$.
The global symmetry group $G=U(1){\times}U(1)$ of the 2d gaussian 
model acts by
\eq
\phi \mapsto \phi + g
\,,\qquad
\phi^{*} \mapsto \phi^{*} + g^{*}
\,,
\en
so $G$ acts on $T_{\Sigma}$
as a principal homogeneous space.

Think of $T_{\Sigma}$
as the space of equivalent 2d theories
built from the 1-forms $j,j^{*}$.
The local fields, such as the vertex operators 
$V_{p,p^{*}}=e^{ip\phi+ip^{*}\phi^{*}}$,
transform in representations of $G$ which can be 
represented as functions on $T_{\Sigma}$.
We can think of the action of $G$ on the local fields
as factoring through its action on $T_{\Sigma}$.

On each non-special fiber $\Ex$ of $\cE\rightarrow \cB$ there is space of 
theories all isomorphic to the (formal analog of the) 2d gaussian model.
On a given $\Ex$,
the space of isomorphic theories
is the space of solutions  $\phi(\xi)$ and $\phi^{*}(\xi)$ on $\Ex$
-- the space of integration constants
for $\phi$ and $\phi^{*}$ on $\Ex$.
Write $\cT_{\partial\xi}$ for the space of isomorphic theories on 
$\Ex$.
The space $\cT_{0}$ of theories on the special fiber $\cE_{0}$
is a single point.
Collectively, the isomorphism classes of theories
form a fiber bundle
$\cT\rightarrow \cB$.

The isomorphism class of theories $\cT_{\partial\xi}$ has symmetry group
$\cG_{\partial\xi}$
which is isomorphic, for $\partial\xi\ne 0$, to
the global symmetry group $G= U(1){\times}U(1)$.
The group $\cG_{\partial\xi}$
acts on the isomorphism class $\cT_{\partial\xi}$ by
\eq
\phi(\xi)\mapsto \phi(\xi)+g(\partial\xi)
\,,\qquad
\phi^{*}(\xi)\mapsto \phi^{*}(\xi)+g^{*}(\partial\xi)
\,.
\en
The symmetry group $\cG_{0}$ of the special fiber is 
the trivial group.
Note that $g(\partial\xi)$ and $g^{*}(\partial\xi)$ are 0-forms 
(functions) on $\cB$.
They are not necessarily additive functions
on the abelian group $\cB$,
which is to say that
they do not necessarily come from $(n{-}2)$-forms on $M$.

The isomorphism class  $\cT_{\partial\xi}$
is a principal homogeneous space for the group $\cG_{\partial\xi}$.
Collectively, the symmetry groups $\cG_{\partial\xi}$ form a fiber bundle of groups
$\cG\rightarrow\cB$.
The sections of $\cG\rightarrow\cB$ are the local gauge 
transformations.
This construction of local gauge symmetry over $\cB$ 
would make sense for any 2d CFT (or QFT),
with any global 2d symmetry group $G$, 
abelian or nonabelian.

%\subsection{The bundle of frames}

Let $\cF_{\partial\xi} = \Iso(T_{\Sigma},\cT_{\partial\xi})$ be the 
space of equivalences, taking the scalars $\phi,\phi^{*}$ on $\Sigma$ to the 
scalars  $\phi,\phi^{*}$ on $\Ex$.
The space $\cF_{\partial\xi}$ is a principal homogeneous space both for
$G$ acting on the right, and for $\cG_{\partial\xi}$
acting on the left.
Collectively, the $\cF_{\partial\xi}$ form the fiber bundle of frames
$\cF\rightarrow\cB$,
a principal fiber bundle with structure group $G$.
If $V$ is a representation of $G$, 
the fields of charge $V$ on the $\Ex$ live in the vector bundle
$\cF\times_{G} V\rightarrow \cB$.

\subsection{The reconstruction of the space-time gauge potentials}

The group of gauge transformations of the 1-form theory over $\cB$ is
much bigger than the group of gauge transformations of the $n$-form
theory in space-time.
I will attempt to argue now that there exists a partial gauge fixing 
in the 1-form theory over $\cB$
that reduces the large gauge 
group over $\cB$ to the smaller space-time gauge group.
The argument seems to depend crucially on the 
global symmetry group $G=U(1){\times} U(1)$ being abelian,
in concert with the abelian group structure of $\cB$ and of $\cE$.
Also, the argument is only classical.
So the argument very well might not have any general significance.
For 2d theories with nonabelian global symmetry groups $G$,
there will be the nonabelian local gauge symmetry over $\cB$,
but perhaps without any
reduction to ordinary local nonabelian gauge symmetry
in space-time.

Suppose $\phi(\xi)$, $\phi^{*}(\xi)$ are solutions of $d\phi=j$,
$d\phi^{*}=j^{*}$ in each fiber.  The goal is to make a gauge 
transformation
\eq
\label{eq:gaugetransftoadditive}
\tilde \phi(\xi) = \phi(\xi)+\tilde g(\partial\xi)
\,,\qquad
\tilde \phi^{*}(\xi) = \phi^{*}(\xi)+\tilde g^{*}(\partial\xi)
\en
so that $\tilde\phi$ and $\tilde\phi^{*}$ are additive,
\eq
\label{eq:additivecond}
\tilde \phi(\xi_{1}) + \tilde \phi(\xi_{2}) - \tilde \phi(\xi_{1}+\xi_{2})
= 0
\,,\qquad
\tilde \phi^{*}(\xi_{1}) + \tilde \phi^{*}(\xi_{2}) - \tilde \phi^{*}(\xi_{1}+\xi_{2})
=0
\,.
\en
Then $\tilde\phi(\xi)$ and $\tilde\phi^{*}(\xi)$ will be determined 
by their values on infinitesimal $\xi$, so will correspond to 
$(n{-}1)$-forms $A$, $A^{*}$ on space-time.  The remaining gauge 
invariance will be given by additive functions $g(\partial\xi)$,
$g^{*}(\partial\xi)$, corresponding to $(n{-}2)$-forms $A$, $A^{*}$ 
on space-time,
which is the space-time gauge invariance of the $n$-form theory.

Define
\aeq{
C(\xi_{1},\xi_{2}) &= 
\phi(\xi_{1}) + \phi(\xi_{2}) - \phi(\xi_{1}+\xi_{2})
\\
C^{*}(\xi_{1},\xi_{2}) &= 
\phi^{*}(\xi_{1}) + \phi^{*}(\xi_{2}) - \phi^{*}(\xi_{1}+\xi_{2})
\,.
}
$C(\xi_{1},\xi_{2})$ and $C^{*}(\xi_{1},\xi_{2})$  actually depend only on $\partial\xi_{1}$ and 
$\partial\xi_{2}$,
because, if $\partial\xi_{1}'=\partial\xi_{1}$, then
\aeq{
\tilde C(\xi_{1}',\xi_{2}) - \tilde C(\xi_{1},\xi_{2})
&= [ \phi(\xi_{1}')  - \phi(\xi_{1}) ] 
- [ \phi(\xi_{1}'+\xi_{2}) - \phi(\xi_{1}+\xi_{2})]
\\
&= j(\partial(\xi_{1}' -\xi_{1})) - j(\partial(\xi_{1}' 
+\xi_{2}-\xi_{1}-\xi_{2})) 
\\
&=0
\,,
}
and similarly for $\partial\xi_{2}'=\partial\xi_{2}$ and for $C^{*}$ 
in place of $C$.
So we can write
\eq
C(\xi_{1},\xi_{2}) = c(\partial\xi_{1},\partial\xi_{2})
\,,\qquad
C^{*}(\xi_{1},\xi_{2}) = c^{*}(\partial\xi_{1},\partial\xi_{2})
\,.
\en
To get the additive condition (\ref{eq:additivecond}),
we need the gauge transformation (\ref{eq:gaugetransftoadditive})
to solve
\aeq{
\label{eq:2cochain1}
c(\partial\xi_{1},\partial\xi_{2}) 
&= \tilde g(\partial\xi_{1}+\partial\xi_{2}) -  
\tilde g(\partial\xi_{2})- \tilde g(\partial\xi_{2})
\\
c^{*}(\partial\xi_{1},\partial\xi_{2})
&= \tilde g^{*}(\partial\xi_{1}+\partial\xi_{2}) -  
\tilde g^{*}(\partial\xi_{2})- \tilde g^{*}(\partial\xi_{2})
\label{eq:2cochain2}
}
for $\tilde g$ and $\tilde g^{*}$.
This is a problem in group cohomology.
The pair $c$, $c^{*}$ is a 2-cochain on the abelian group $\cB$
with coefficients in the group $G=U(1){\times}U(1)$.
Equations (\ref{eq:2cochain1}--\ref{eq:2cochain2})
are solvable iff
$c$, $c^{*}$ is trivial in the group cohomology.
But 
$c(\partial\xi_{1},\partial\xi_{2})$ and
$c^{*}(\partial\xi_{1},\partial\xi_{2})$
are symmetric in their arguments,
so  (\ref{eq:2cochain1}--\ref{eq:2cochain2})
can be solved
if the symmetric 2-cocycles are 
trivial in the group cohomology of $\cB$.
I believe that this is the case,
because $\cB$ and $G$ are abelian.

\section{Connecting the $\Ex$}
\label{sect:connectingtheExC}

The  extended objects in the free $n$-form theory
are to be described by  fields on $\cE$.
But the analog 2d theories on the fibers of $\cE\rightarrow \cB$
would seem to give only correlation functions of fields on the same 
fiber $\Ex$.
What about correlation functions between fields on different
fibers $\Ex$?
Such correlation functions will be non-zero only on gauge invariant 
observables on the individual fibers
--- only on observables invariant under the global symmetry of the 
fiber.
So we need to understand the correlation functions of products 
of invariant observables on different fibers.

Here I argue that the invariant observables are the same
on every fiber,
so correlation functions of products of invariant observables should 
be calculable in any one fiber.
The argument is specific to the free $n$-form theory and is only 
classical.
The argument uses a natural connection in the bundle $\cE \rightarrow 
\cB$, described in the next section.
My hope is that there might be a parallel argument in the general 
quantum case, using some analogous natural geometric 
structure in the bundle $\QM \rightarrow \PBM$ of quasi Riemann surfaces.

\subsection{The natural connection in 
%the principal fiber bundle
$\cD^{\integral}_{k}(M)\xrightarrow{\partial} \partial \cD^{\integral}_{k}(M)$}
%of integral $k$-currents}

Recall that $\cD^{\integral}_{k}(M) \xrightarrow{\partial} 
\partial\cD^{\integral}_{k}(M)$ is a principal fiber bundle for the 
additive abelian group $\cD^{\integral}_{k}(M)_{0}$ of integral $k$-cycles.
A connection in $\cD^{\integral}_{k}(M) \xrightarrow{\partial} 
\partial\cD^{\integral}_{k}(M)$ is a collection of linear maps
lifting tangent vectors $v_{0}$ in the base at $\partial\xi$
to tangent vectors $v$ in the total space at $\xi$,
one such linear map for every point $\xi$ in the total space.
These linear maps must be compatible with the bundle structure,
i.e., must satisfy $\partial_{*} v=v_{0}$ and 
must be invariant under translations of $\xi$ in the fiber over $\partial\xi$.

A tangent vector in the base $\partial\cD^{\integral}_{k}(M)$
is a perturbation by an infinitesimal element $v_{0}$ in 
$\partial\cD^{\integral}_{k}(M)$.
We have seen that there is a uniquely determined minimal infinitesimal element
$v\in \cD^{\integral}_{k}(M)$
satisfying $\partial v = v_{0}$.
This $v$ is the lift of $v_{0}$.

Equivalently, $v_{0}$ corresponds to an infinitesimal integral 1-current $\eta_{0}$ 
in the base at $\partial\xi$ by
\eq
\Pi_{*}^{0,k-1}  \partial \eta_{0} =  v_{0}
\,.
\en
The pushdown of $\eta_{0}$, 
\eq
v = \Pi_{*}^{1,k-1}\eta_{0}\in \cD^{\integral}_{k}(M)
\en
corresponds to an infinitesimal integral 1-current 
$\eta$ in the total space $\cD^{\integral}_{k}(M)$ at $\xi$,
by
\eq
\Pi_{*}^{0,k}  \partial \eta =  v
\,,
\en
with $\eta$ a lift of $\eta_{0}$,
\eq
\partial_{*} \eta = \eta_{0}
\,.
\en

The lift $v_{0}\mapsto v$, or $\eta_{0}\mapsto\eta$, is translation invariant in the 
fiber, since the construction of the lift makes no mention of $\xi$.
In fact, the construction is translation invariant in the total space
$\cD^{\integral}_{k}(M)$ of the bundle.
The lift $v_{0}\mapsto v$ thus is a natural, translation invariant connection 
in the principal fiber bundle $\cD^{\integral}_{k}(M) \xrightarrow{\partial} 
\partial\cD^{\integral}_{k}(M)$,
a natural splitting of the tangent space
\eq
T_{\xi}\cD^{\integral}_{k}(M) = \cV_{k+1}(M) \oplus \cV_{k}(M)
\en
where the first summand is the vertical subspace and the second 
summand is the horizontal subspace, the tangent space in the base
$\partial\cD_{k}(M)$.

Write $D^{\nat}$ for the covariant 
derivative of the natural connection.
The curvature tensor $(D^{\nat})^{2}$ is a 2-form on the base with 
values in the translation invariant vertical vector fields.
To calculate the curvature, consider an infinitesimal 2-current 
$\eta_{2}$ in the base at $\partial \xi$.
Use the natural connection to lift the 1-current 
$\eta_{0}=\partial\eta_{2}$ to a 1-current $\eta$ in the total space 
at $\xi$.  Then $\partial\eta$ is an infinitesimal 0-current in the 
fiber.  This is is the monodromy around $\partial\eta_{2}$,
which is the curvature tensor acting on  $\eta_{2}$
to give the vertical tangent vector
\eq
v_{c} = \Pi_{*}^{0,k} \partial\eta
= \Pi_{*}^{1,k-1}(\eta_{0})
= \Pi_{*}^{1,k-1}(\partial\eta_{2})
= \partial \Pi_{*}^{2,k-1} \eta_{2}
\,.
\en
We can write the curvature as
\eq
(D^{\nat})^{2} \eta_{2}
=  \Pi_{*}^{2,k-1}\eta_{2}
\en
which is an infinitesimal element in $\cD^{\integral}_{k+1}(M)$,
whose boundary
\eq
v_{c} = \partial \Pi_{*}^{2,k-1}\eta_{2}
\en
is an infinitesimal vertical perturbation in the fiber.

\subsection{The connection in the bundle of theories over $\cB$}

Specializing to $k=n{-}1$, we have a natural connection in 
$\cE\rightarrow \cB$, with covariant derivative $D^{\nat}$.
The natural connection in $\cE\rightarrow \cB$ combines with the
$0$-forms $j$, $j^{*}$ to give a connection 
in the bundle of theories $\cT\rightarrow \cB$, as follows.
An infinitesimal motion in $\cB$, from $\partial\xi$ to 
$\partial\xi'$, 
lifts to a translation from the fiber $\cE_{\partial\xi}$ to the fiber 
$\cE_{\partial\xi'}$, taking 0-currents on the first fiber to 
0-currents on the second, and pulling 0-forms back from the second to 
the first.
The 1-forms $j$ and $j^{*}$ are translation invariant,
so the natural connection pulls back solutions of $d\phi=j$, 
$d\phi^{*}=j^{*}$ on $\cE_{\partial\xi'}$ to solutions
on $\cE_{\partial\xi}$.
This is the connection in $\cT\rightarrow \cB$.

Suppose $\phi$, $\phi^{*}$ is a local section of $\cT \rightarrow 
\cB$.
That is, $\phi(\xi)$ and $\phi^{*}(\xi)$ are solutions 
in the fibers over some neighborhood in $\cB$,
a choice of integration constants in each fiber.
The covariant derivative $D^{\nat}\phi(\xi)$, or 
$D^{\nat}\phi^{*}(\xi)$, 
is, on each fiber, the difference of two 
solutions, thus is a constant on each fiber,
\eq
D^{\nat}\phi(\xi) = D\phi(\partial\xi)
\,,\qquad
D^{\nat}\phi^{*}(\xi) = D\phi^{*}(\partial\xi)
\en
giving the covariant derivative $D$ of the
connection in $\cT\rightarrow\cB$.

The curvature tensor of the connection in $\cT\rightarrow\cB$ is a 2-form on the base 
$\cB$ with values in the invariant vertical vector fields --- the pair 
of 2-forms on $\cB$
\ateq{2}{
(D^{\nat})^{2} \phi &= (d\phi) (D^{\nat})^{2}& &= j (D^{\nat})^{2} 
\\[1ex]
(D^{\nat})^{2} \phi^{*} &= (d\phi^{*}) (D^{\nat})^{2} & &= j^{*} (D^{\nat})^{2} 
\,.
}
Here, $d\phi$ and $d\phi^{*}$ are the vertical derivatives in 
$\cE$.
When the 1-forms $j$ and $j^{*}$ on $\Ex$ come from $n$-forms $F$ and 
$F^{*}$ on $M$,
\eq
j = F\, \Pi_{*}^{1,n-1}
\,,\qquad
j^{*} = F^{*}\,  \Pi_{*}^{1,n-1}
\,,
\en
then
\ateq{2}{
(D^{\nat})^{2} \phi &= F\,  \Pi_{*}^{1,n-1} (D^{\nat})^{2} 
& &= F\, \Pi_{*}^{2,n-2}
\\[1ex]
(D^{\nat})^{2} \phi^{*} &=  F^{*}\,  \Pi_{*}^{1,n-1} (D^{\nat})^{2} 
& &= F^{*}\, \Pi_{*}^{2,n-2}
\,.
}
We might think of the bundle $\cT\rightarrow \cB$ with connection $D$ 
as an alternate representation of the
$1$-form fields  $j$ and $j^{*}$,
or of the $n$-form fields $F$ and $F^{*}$.

\subsection{Transport of observables between fibers $\Ex$}

%\subsection{Transport of observables between fibers of 
%$\cE{{\times}_{\cB}}\cT\rightarrow \cB$}

A vertex operator $e^{ip\phi(\xi)+ip^{*}\phi^{*}(\xi)}$
transforms under the symmetry group $\cG_{\partial\xi}$
in the one-dimensional representation
labelled by the charges $p,p^{*}$.
Equivalently, the vertex operator
belongs to the one-dimensional vector space
associated to the space $\cT_{\partial\xi}$ of theories on the fiber
and the one-dimensional representation of the global symmetry group $G=U(1){\times}U(1)$
labelled by $p,p^{*}$.
In general,
the observables (\ref{eq:observables}) on each fiber belong to vector 
bundles associated to the $G$-bundle which is the fiber product
\eq
\cE \times_{\cB} \cT \rightarrow \cB
\,,\qquad
(\cE \times_{\cB} \cT)_{\partial\xi} = \cE_{\partial\xi} \times \cT_{\partial\xi}
\,.
\en
The natural connection in $\cE$ combines with the connection in $\cT$ 
to give a connection in the fiber product,
which determines a parallel transport of observables from fiber to 
fiber.
The observables are invariant under this parallel transport.

One way to see this is by choosing a local section of $\cE\rightarrow 
\cB$.  This is a choice $\xi_{1}$ of a relative $(n{-}1)$-cycle in 
each fiber over a neighborhood $\cN$ in $\cB$.  The choice of the 
$\xi_{1}$ trivializes 
$\cE$ as $\cN\times \Ezero $ over $\cN$,
\eq
\partial\xi, \xi \longleftrightarrow \partial\xi, 
\xi-\xi_{1}(\partial\xi)
\,.
\en
This also trivializes $\cT$ over $\cN$ by 
singling out the local section
given by the normalization condition
\eq
\phi_{1}(\xi_{1}) = \phi^{*}_{1}(\xi_{1}) = 0
\,.
\en
Now the covariant derivative for the natural connection in $\cE$ is
\eq
D^{\nat} = d + A^{\nat}
\en
where
\eq
A^{\nat} = D^{\nat} \xi_{1}
\en
is a 1-form on $\cB$ with values in the vertical tangent space $\cV_{n}$.
The covariant derivative in $\cT$ is
\eq
D\phi = (d + A)\phi 
\,,\qquad
D\phi^{*} = (d + A^{*})\phi^{*}
\,,
\en
\eq
A = D^{nat}\phi_{1} = j A^{\nat}
\,,\qquad
A^{*} = D^{nat}\phi^{*}_{1} = (*j) A^{\nat}
\,.
\en
The total covariant derivative in the fiber product, acting on an 
observable of charges $p$, $p^{*}$ is
\eq
D^{\tot} = d + A^{\nat}d_{V} - ip A - i p^{*} A^{*}
\,,
\en
where $d_{V}$ is the vertical derivative in the trivialization of $\cE$.

The covariant constancy of the observables implies that the 
observables are annihilated by the curvature tensor $(D^{\tot})^{2}$.
This is just $j = d\phi$, ${*}j = d\phi^{*}$.
Thus the gauge structure -- the bundle $\cT\rightarrow \cB$
and its connection $D$ -- encodes the classical theory of extended objects
of the free $n$-form.

The charged observables are associated to nontrivial representations 
of the symmetry group $G$ and live in vector bundles associated to 
the fiber product bundle $\cE\times_{\cB}\cT$.  The connection that 
transports the charged observables between fibers is dynamical.
But neutral observables --- observables that are $G$-invariant, that 
are associated to the trivial $G$ representation --- do not see the 
gauge bundle $\cT$.  The neutral observables are transported from 
fiber to fiber by the natural connection in $\cE$, which is 
independent of the dynamics, i.e., kinematical.  The neutral 
observables are, essentially, the products of vertex operators with zero total 
charges,
\eq
e^{ip_{1}\phi(\xi_{1})+ip_{1}^{*}\phi^{*}(\xi_{1})}
\cdots
e^{ip_{N}\phi(\xi_{N})+ip_{N}^{*}\phi^{*}(\xi_{N})}
\,,\qquad
\sum p_{i} = \sum p_{i}^{*} = 0
\,.
\en

Assume that the gauge symmetry is unbroken, so that only the gauge invariant 
observables have non-zero expectation values.  Gauge invariant 
observables are products of neutral observables on fibers.
Neutral observables can be transported from fiber to fiber, 
independent of the dynamics.  So expectation values of gauge 
invariant observables can be calculated entirely on any single fiber.

This argument might seem to imply that all calculations can be done 
in a single fiber, for example in the distinguished fiber over  $0\in \cB$.
But this is too extreme.  Certainly, in a 2d quantum field theory 
there is interesting information to obtain about the charged observables.
In the space $\cE$ of extended objects,
the algebra of charged observables will depend on the fiber.
On the other hand, products of charged observables from different 
fibers will always have zero expectation values.
So the theory of extended objects 
can be considered as the collection of theories
on the fibers.

\section{Explorations}
\label{sect:Explorations}

In this section, for simplicity, the imaginary parts of 
the quasi Riemann surfaces are omitted, and all quasi Riemann 
surfaces are assumed connected.
And the conjecture is presumed to hold.

\subsection{The $\QM_{\Integers\partial\xi}$}

As described in section~\ref{sect:ExCasaquasiRS},
each of the quasi Riemann surfaces $\cQ = \QM_{\Integers\partial\xi}$ 
shares a common {\it core,}
\eq
\cQ^{\integral}_{1} = \cD^{\integral}_{n}(M)
\,,\qquad
(\cQ^{\integral}_{0})_{0} = \cD^{\integral}_{n-1}(M)_{0}
= \partial \cD^{\integral}_{n}(M)
\,.
\en
The last identity expresses the connectedness
of the quasi Riemann surfaces,
\eq
H_{0}(\cQ) = H_{n-1}(M) = 0
\,.
\en
Consider the restriction of $\Pi_{*}^{1,1}$ to
the space of integral 1-cycles in the space of integral 1-cycles
of $\cQ$,
\eq
\Pi_{*}^{1,1}\colon \cD^{\integral}_{1}((\cQ^{\integral}_{1})_{0})_{0} \rightarrow \cQ^{\integral}_{2}
\,,
\en
which satisfies
\eq
\partial \Pi_{*}^{1,1} = 0
\,,
\en
so
\eq
\Pi_{*}^{1,1} \cD^{\integral}_{1}((\cQ^{\integral}_{1})_{0})_{0}  = 
(\cQ_{2})_{0}
=\partial \cQ_{3}
\simeq \Integers
\,.
\en
Define
\eq
\Lambda(\Integers\partial\xi)^{\integral}
= \Ker \Pi_{*}^{1,1} \subset \cD^{\integral}_{1}((\cQ^{\integral}_{1})_{0})_{0} 
\,.
\en
Then
\eq
\cD^{\integral}_{1}((\cQ^{\integral}_{1})_{0})_{0}  /\Lambda(\Integers\partial\xi)^{\integral}
= (\cQ_{2})_{0}
=\partial \cQ_{3}
\simeq \Integers
\,.
\en
The subgroup $\Lambda(\Integers\partial\xi)^{\integral}$
is a {\it co-line} in 
$\cD^{\integral}_{1}((\cQ^{\integral}_{1})_{0})_{0}$,
where a co-line is defined to be
a minimal subgroup among those whose quotient is $\Integers$.

The map $\Pi_{*}^{1,1}$ acts in two stages
\eq
\cD^{\integral}_{1}((\cQ^{\integral}_{1})_{0})_{0} 
=\cD^{\integral}_{1}(\cD^{\integral}_{n}(M)_{0})_{0}
\rightarrow
\cD^{\integral}_{n+1}(M)_{0}
\rightarrow
(\cQ_{2})_{0}
\,.
\en
Write the first stage
\eq
\Pi_{*}^{1,1}(M) \colon
\cD^{\integral}_{1}(\cD^{\integral}_{n}(M)_{0})_{0}
\rightarrow
\cD^{\integral}_{n+1}(M)_{0}
\,,
\en
and define the subgroup
\eq
\Lambda^{\integral}_{M} = \Ker \Pi_{*}^{1,1}(M) 
\subset \cD^{\integral}_{1}((\cQ^{\integral}_{1})_{0})_{0} 
\,.
\en
Then
\eq
\Lambda^{\integral}_{M} 
\subset
\Lambda(\Integers\partial\xi)^{\integral}
\,.
\en
In fact, {\it any} co-line in $ 
\cD^{\integral}_{1}((\cQ^{\integral}_{1})_{0})_{0}$ that contains 
$\Lambda^{\integral}_{M}$ is 
$\Lambda(\Integers\partial\xi)^{\integral}$ for some
$\Integers\partial\xi \in \PBM$,
\eq
\PBM = \left \{
\text{co-lines } \Lambda^{\integral} :
\Lambda^{\integral}_{M} \subset \Lambda^{\integral}\subset  \cD^{\integral}_{1}((\cQ^{\integral}_{1})_{0})_{0} 
\right \}
\,.
\en

\subsection{Reconstruct $\cQ$ from $\cQ_{1}$}

All of the considerations of the previous section up to the 
introduction of $\Lambda_{M}$
apply to {\it any} quasi Riemann surface $\cQ$.
Every quasi Rieman surface $\cQ$ has a distinguished co-line
\eq
\Lambda^{\integral} = \Ker \Pi_{*}^{1,1} \subset \cD^{\integral}_{1}((\cQ^{\integral}_{1})_{0})_{0}
\,.
\en
Conversely, every co-line in 
$\cD^{\integral}_{1}((\cQ^{\integral}_{1})_{0})_{0}$
corresponds to a quasi Riemann surface.
That is,
any quasi Riemann surface $\cQ$
can be reconstructed from the data on its 1-subspace
$\cQ^{\integral}_{1}$.
The {\it core} of the quasi Riemann surface is
$\cQ^{\integral}_{1}$,
with the skew-hermitian form and the $J$-operator,
and with the two subgroups of boundaries and cycles,
\eq
\partial \cQ^{\integral}_{2} \subset (\cQ^{\integral}_{1})_{0}
\subset \cQ^{\integral}_{1}
\,.
\en
From the core data we reconstruct
\eq
(\cQ^{\integral}_{0})_{0} = 
\cQ^{\integral}_{1}/(\cQ^{\integral}_{1})_{0}
\en
and
\eq
\cQ^{\integral}_{2}/(\cQ^{\integral}_{2})_{0}
= 
\partial \cQ^{\integral}_{2}
\,.
\en
Only one line's worth of data is left to specify
in order to reconstruct all of $\cQ$.

Define
\eq
\ZQ = \cD^{\integral}_{1}((\cQ^{\integral}_{1})_{0})_{0}
\en
The last piece of data is a co-line
\eq
\Lambda^{\integral} \subset \ZQ
\,.
\en
Now we reconstruct
\eq
(\cQ^{\integral}_{2})_{0} = \ZQ/\Lambda^{\integral}
\en
and we finish the construction of 
$\cQ^{\integral}_{0}$
using duality under the skew-hermitian intersection form.
The map $\Pi_{*}^{1,1}$ is given by
\eq
\Pi_{*}^{1,1}\colon \ZQ 
\rightarrow \ZQ/\Lambda^{\integral}
= \cQ^{\integral}_{2}
\,.
\en
So a quasi Riemann surface $\cQ$ is specified by
the 1-space $\cQ^{\integral}_{1}$ along with the core data
\begin{itemize}
\item the skew-hermitian form on $\cQ^{\integral}_{1}$,
\item the $J$-operator on $\cQ_{1}$,
\item the subgroups $\partial \cQ^{\integral}_{2} \subset 
(\cQ^{\integral}_{1})_{0}\subset \cQ^{\integral}_{1}$,
\end{itemize}
plus
\begin{itemize}
\item a co-line $\Lambda^{\integral} \subset \ZQ$.
\end{itemize}

\subsection{Augment $\Sigma$}
\label{sect:augmentSigma}

Now we want to use the conjecture to 
equate the $\QM_{\Integers\partial\xi}$
with the quasi Riemann surface $\QSigma$
associated to a two-dimensional conformal space $\Sigma$.
We want to do this in a way that preserves the
commonality of the cores of all the $\QM_{\Integers\partial\xi}$.

First, however, we have to construct $\QSigma$.
As pointed out in section~\ref{sect:RiemannsurfaceasquasiRS},
there is a flaw in
the straightforward construction of a quasi-Riemann surface 
\eq
0 	\leftarrow \cQ^{\integral}_{-1}
\xleftarrow{\partial} \cQ^{\integral}_{0} 
\xleftarrow{\partial} \cQ^{\integral}_{1} 
\xleftarrow{\partial} \cQ^{\integral}_{2} 
\xleftarrow{\partial} \cQ^{\integral}_{3} 
\leftarrow  0
\en
from an 
ordinary Riemann surface $\Sigma$,
as the augmented de Rham complex
\eq
0 	\leftarrow \Integers
\xleftarrow{\partial} \cD^{\integral}_{0}(\Sigma)
\xleftarrow{\partial} \cD^{\integral}_{1}(\Sigma)
\xleftarrow{\partial} \cD^{\integral}_{2}(\Sigma)
\xleftarrow{\partial} \Integers
\leftarrow  0
\,.
\en
The problem is that we need
the maps $\Pi_{*}^{j,k}$ on the currents in the 
$(\cQ^{\integral}_{k})_{0}$ 
to induce isomorphisms of homology,
\eq
H_{2}((\cQ^{\integral}_{0})_{0}) = H_{1}((\cQ^{\integral}_{1})_{0}) = 
H_{2}(\cQ^{\integral}) = 0\,.
\en
However, the homology groups
$H_{j}(\cD^{\integral}_{k}(\Sigma)_{0})$
are isomorphic to the  homology groups 
$H_{j+k}(\Sigma)$ of the {\it un-augmented} de Rham complex,
\eq
H_{2}(\cD^{\integral}_{0}(\Sigma)_{0}) = H_{1}(\cD^{\integral}_{1}(\Sigma)_{0}) = 
H_{2}(\Sigma) = \Integers\,.
\en
That is, there exist
\eq
\mu_{1,1} \in \cD^{\integral}_{1}(\cD^{\integral}_{1}(\Sigma)_{0})_{0}
\,,\qquad
\mu_{2,0} \in \cD^{\integral}_{2}(\cD^{\integral}_{0}(\Sigma)_{0})_{0}
\en
with
\eq
\Pi_{*}^{1,1}\mu_{1,1} = \Sigma
\,,\qquad
\Pi_{*}^{2,0}\mu_{2,0} = \Sigma
\en
so neither $\mu_{1,1}$ nor $\mu_{2,0}$ can be a boundary.
They generate the non-trivial homology groups.

We need to construct an ``augmentation'', $\Sigma_{+}$, of $\Sigma$ 
such that the ordinary de Rham complex of $\Sigma_{+}$ is
\eq
0 	\leftarrow \cD^{\integral}_{-1}(\Sigma_{+})
\xleftarrow{\partial} \cD^{\integral}_{0}(\Sigma_{+})
\xleftarrow{\partial} \cD^{\integral}_{1}(\Sigma_{+})
\xleftarrow{\partial} \cD^{\integral}_{2}(\Sigma_{+})
\xleftarrow{\partial} \cD^{\integral}_{3}(\Sigma_{+})
\leftarrow  0
\,,
\en
with
\eq
\cD^{\integral}_{-1}(\Sigma_{+}) \simeq \Integers
\,,\qquad
\cD^{\integral}_{3}(\Sigma_{+}) \simeq \Integers
\en
and homology
\eq
H_{1}(\Sigma_{+}) = H_{1}(\Sigma)
\,,\qquad
H_{k}(\Sigma_{+}) = 0
\,,\quad k\ne 1
\,.
\en
In some sense, we want to make $\Sigma_{+}$ from $\Sigma$ by adding 
three new integral currents, 
\aeq{
\cD^{\integral}_{3}(\Sigma_{+}) &= \Integers\eta_{3}
\\
\cD^{\integral}_{2}(\cD^{\integral}_{1}(\Sigma_{+})_{0}) &= 
\cD^{\integral}_{2}(\cD^{\integral}_{1}(\Sigma)_{0})
\oplus  \Integers \mu_{2,1} 
\\
\cD^{\integral}_{3}(\cD^{\integral}_{1}(\Sigma_{+})_{0}) &= 
\cD^{\integral}_{3}(\cD^{\integral}_{0}(\Sigma)_{0})
\oplus   \Integers \mu_{3,0} 
}
satisfying
\eq
\partial \eta_{3} = \Sigma
\,,\qquad
\partial \mu_{2,1} = \mu_{1,1}
\,,\qquad
\partial \mu_{3,0} = \mu_{2,0}
\en
\eq
\Pi_{*}^{2,1}\mu_{2,1} = \eta_{3}
\,,\qquad
\Pi_{*}^{3,0}\mu_{3,0} = \eta_{3}
\,.
\en
Now $\Sigma$, $\mu_{1,1}$, and $\mu_{2,0}$ are boundaries.  The 
bothersome homology groups are killed.
I will have to suppose that a precise construction of $\Sigma_{+}$ can be 
made.

The de Rham complex of $\Sigma_{+}$ now gives
an actual quasi-Riemann surface $\QSigmaplus$.
The defining co-line is
\eq
\Lambda(\Sigma_{+})^{\integral} = \Ker \Pi_{*}^{1,1} 
= \partial \cD^{\integral}_{2}(\cD^{\integral}_{1}(\Sigma)_{0})
\subset
\partial \cD^{\integral}_{2}(\cD^{\integral}_{1}(\Sigma_{+})_{0})
=
\cD^{\integral}_{1}(\cD^{\integral}_{1}(\Sigma_{+})_{0})_{0}
\,.
\en

\subsection{Isomorphisms of the $\QM_{\Integers\partial\xi}$ to 
$\QSigmaplus$}
\label{sect:IsoQMZxQSigmaplus}

Pick a point $\Integers\partial\xi\in \PBM$.  Pick an isomorphism
\eq
f \colon \QM_{\Integers\partial\xi} \rightarrow \QSigmaplus
\en
which is determined by its isomorphism  of the core structures,
\eq
f_{1}\colon \QM^{\integral}_{\Integers\partial\xi,1} \rightarrow \QSigmaplus^{\integral}_{1}
\,,
\en
of the defining co-lines,
\eq
f_{1*}\; \Lambda(\Integers\partial\xi)^{\integral}
 = \Lambda(\Sigma_{+})^{\integral}
\,.
\en
Now consider another point $\Integers\partial\xi'\in \PBM$.
The core structures of $\QM_{\Integers\partial\xi'}$
are the same as those of $\QM_{\Integers\partial\xi}$,
so $f_{1}$ is also an isomorphism
\eq
f_{1} \colon \QM_{\Integers\partial\xi',1} \rightarrow \QSigmaplus_{1}
\,,
\en
preserving the core structures.
However, the defining co-lines of
$\QM_{\Integers\partial\xi'}$
and $\QM_{\Integers\partial\xi}$
are different, so
\eq
f_{1*}\; \Lambda(\Integers\partial\xi')^{\integral}
\ne \Lambda(\Sigma_{+})^{\integral}
\,,
\en
so $f_{1}$ does {\it not} give an isomorphism of quasi Riemann surfaces
between $\QM_{\Integers\partial\xi'}$
and $\QSigmaplus$.
Its image is a quasi Riemann surface that has the same core as 
$\QSigmaplus$, but a different defining co-line.

\subsection{The universal bundle $\cQ(0) \rightarrow \PB(0)$ of quasi Riemann surfaces}

Define
\eq
\cZ = \cD^{\integral}_{1}((\QSigmaplus_{1} )_{0})_{0}
\en
and
\eq
\PB(0) = \left \{ 
\text{co-lines } \Lambda \subset 
\cZ
\right \}
\en
so
\eq
\Lambda(\Sigma_{+})^{\integral} \in \PB(0)
\en
is a distinguished point.
Let $\cQ(0)_{\Lambda}$ be the quasi Riemann surface with the same 
core structure as $\QSigmaplus$ but with defining co-line $\Lambda$.
Then
\eq
\cQ(0) = \mathop\cup\limits_{\Lambda\in\PB(0)}\; \cQ(0)_{\Lambda}
\en
forms a bundle of quasi Riemann surfaces
\eq
\cQ(0) \rightarrow \PB(0)
\,.
\en

\subsection{Embed $\QM\rightarrow \PBM$ in the universal bundle $\cQ(0)\rightarrow \PB(0)$}

Consider again the isomorphism $f \colon \QM_{\Integers\partial\xi} 
\rightarrow \QSigmaplus$
of section~\ref{sect:IsoQMZxQSigmaplus}.
Now
\eq
f_{1}\colon  \QM^{\integral}_{\Integers\partial\xi',1} \rightarrow 
\cQ(0)^{\integral}_{\Lambda',1}
\,,\qquad
\Lambda' = f_{1*}\Lambda(\Integers\partial\xi')^{\integral}
\en
preserves the defining co-lines, so gives an isomorphism
\eq
f \colon  \QM_{\Integers\partial\xi'} \rightarrow 
\cQ(0)_{\Lambda'}
\,.
\en
Thus we have the embedding
\eq
\begin{diagram}
\QM &\rTo^{f}	& \cQ(0)
\\
\dTo&  & \dTo
\\
\PBM & \rTo^{f_{1*}}	& \PB(0)
\end{diagram}
\en
which is natural, up to the choice of the point 
$\Integers\partial\xi\in\PBM$ to be mapped to 
$\Lambda(\Sigma_{+})^{\integral}\in \PB(0)$,
and up to the choice of the isomorphism $f$.

In terms of the subgroup
\eq
f_{1*} \Lambda^{\integral}_{M} \subset \cZ
\,,
\en
the embedding of $\PBM$ is
\eq
f_{1*}\PBM = 
\left \{
\text{co-lines }
\Lambda \subset  \cZ
:
f_{1*} \Lambda^{\integral}_{M} \subset \Lambda
\right \}
\en
and the embedding of $\QM$ is the restriction of $\cQ(0)$.
So $M$ is characterized by the subgroup 
$f_{1*} \Lambda^{\integral}_{M} \subset \cZ$,
again up to the choice of $\Lambda(\Sigma_{+})^{\integral}$
and $f$.

\subsection{Homogeneity of $\cQ(0)\rightarrow \PB(0)$}

Let
\eq
G(\Sigma_{+}) = \Aut(\QSigmaplus^{\integral}_{1})
\en
be the group of automorphisms of the core structure of $\QSigmaplus$,
the continuous abelian group automorphisms that preserve the 
skew-hermitian form, the $J$-operator, and the subspaces of cycles 
and boundaries.

Assuming that the conjecture holds,
all of the quasi Riemann surfaces
$\cQ(0)_{\Lambda}$
in the universal bundle
are isomorphic as quasi Riemann surfaces.
Every isomorphism between
$\cQ(0)_{\Lambda}$
and $\cQ(0)_{\Lambda'}$
is given by an element of $G(\Sigma_{+})$.
Therefore $G(\Sigma_{+})$ acts transitively on $\PB(0)$.
The subgroup of $G(\Sigma_{+})$ that fixes $\Lambda$ is
$\Aut(\cQ(0)_{\Lambda})$.
Therefore $\PB(0)$ is the homogeneous space
\eq
\PB(0) = G(\Sigma_{+})/\Aut(\QSigmaplus)
\en
And the universal bundle $\cQ(0)\rightarrow \PB(0)$
is the bundle on $G(\Sigma_{+})/\Aut(\QSigmaplus)$ associated to the action of 
$\Aut(\QSigmaplus)$ on $\QSigmaplus$.

% uncomment to remove subsections from toc from this point
%\addtocontents{toc}{\protect\setcounter{tocdepth}{1}}

\section{Mathematical questions}
\label{sect:mathematicalquestions}

This section is simply a list of the main mathematical questions that 
the project depends on,
with references to the relevant sections of the paper
where some background can be found.
Apart from these more or less specific questions,
all of the mathematical arguments in the paper need tightening.
For the most part, I have not specified topologies on vector spaces 
of currents, or domains of definition of forms on spaces of currents.
I have supposed the context to determine what is appropriate.

Recall from section \ref{sect:intro1} that $M$ is a compact, real, oriented manifold of even 
dimension $d=2n$, endowed with
a conformal class of Riemannian metrics ---
or at least with a Hodge $*$-operator acting in the middle dimension,
on $n$-forms.

Recall from section \ref{sect:GMT} that $\cD^{\integral}_{k}(M)$
is the space of integral $k$-currents in $M$.

\subsection{Does the Hodge $*$-operator act on $T_{0}\cD^{\integral}_{n-1}(M)_{0}$ ?}

$\cD^{\integral}_{n-1}(M)_{0}$ is the space of integral 
$(n{-}1)$-cycles in $M$.  Its tangent space at $0$ can be identified 
with a subspace of the flat $n$-currents in $M$,
\eq
T_{0}\cD^{\integral}_{n-1}(M)_{0} 
= \cV_{n} \subset \cD^{\flat}_{n}(M)
\,.
\en
The Hodge $*$-operator acts on the flat $n$-currents.
Does it preserve $\cV_{n}$?

Appendix~\ref{app:GOT} contains what I believe is the germ of a proof 
that the answer is `yes'.  I think that only some detail needs to 
be filled in.

This question was asked in section \ref{sect:Hodgestaracts},
based on definitions in section \ref{sect:GeometryofDkM}.
The project depends on an affirmative answer.

The space $\Ex = \cD^{\integral}_{n-1}(M)_{\partial\xi}$ 
is the fiber over $\partial\xi$ of the bundle
\eq
\cD^{\integral}_{n-1}(M) \xrightarrow{\partial}
\partial \cD^{\integral}_{n-1}(M) \subset \cD^{\integral}_{n-2}(M)
\,.
\en
All of the vertical tangent spaces in the fibers are the same, 
\eq
T_{\xi}\Ex  = T_{0}\cD^{\integral}_{n-1}(M)_{0}
= \cV_{n} \subset \cD^{\flat}_{n}(M)
\,.
\en
The action of the Hodge $*$-operator
on the tangent spaces of $\Ex$ is
used in sections \ref{sect:moreoncurrentsinM} and \ref{sect:ExCalmostcomplex}
to define an operator $J=\epsilon_{n} *$, $J^{2}=-1$, on the tangent spaces
that makes $\Ex$ --- or its complexification $\ExC$ when $n$ is even ---
into an almost-complex space.

\subsection{Are quasi Riemann surfaces classified by their homology data?}

In section \ref{sect:moreoncurrentsinM}, the intersection form on currents in $M$
is modified slightly to become a skew-hermitian form $\IM{\bar\xi_{1}}{\xi_{2}}$.
In sections \ref{sect:quasi2d} and \ref{sect:quasiRiemannsurfaces},
the skew-intersection intersection form is pulled up to 
$\ExC$.
The almost-complex structure $J$ and the 
skew-hermitian form on currents in $\ExC$
have exactly the properties of
the almost-complex structure and the skew-hermitian intersection form 
on currents in a Riemann surface.
This structure
is codified in the definition of {\it quasi Riemann surface}
in section \ref{sect:quasiRiemannsurfaces}.

The quasi Riemann surface associated to an ordinary Riemann surface 
is written $\QSigma$.  The quasi Riemann surface associated to the $\ExC$ 
are written $\QM_{\Integers\partial\xi}$.  They form a bundle 
$\QM\rightarrow \PBM$ over the integral projective space $\PBM$ of 
the space $\cB$.

In section \ref{sect:awishfulconjecture}, a `wishful'' conjecture is
proposed: that quasi Riemann surfaces are classified up to isomorphism
by their homology data, and that each isomorphism class contains the 
quasi Riemann space $\QSigma$ of a two
dimensional conformal space $\Sigma$.
The space $\Sigma$ is a Riemann surface when
the homology data is that of a Riemann surface.

For the {\it connected} case,
which is $H_{0}(\cD^{\integral}_{n-1}(M))=\Integers$,
the homology data consists of the integral homology group
$H_{n}(M)$ in the middle dimension,
with the skew-hermitian form derived from the intersection form
and with the almost-complex structure.

If the conjecture should fail, there would still be the question, 
do quasi-holomorphic curves exist?  A quasi-holomorphic
curve would be a morphism of quasi Riemann surfaces from some 
$\QSigma$ to $\QM_{\Integers\partial\xi}$.

When the homology data is not that of a Riemann surface,
the conjecture supposes a two-dimensional conformal space $\Sigma$
whose integral currents form a quasi Riemann surface
with that homology data.
What are these two-dimensional conformal spaces?

If the conjecture holds, is there any possibility, say in the basic
case, $M=S^{d}$, $\Sigma = S^{2}$, of actually writing an isomorphism
between $\QM$ and $\QSigma$.

If the conjecture holds, then the automorphism group $\Aut(\cQ)$ of a 
quasi Riemann surface $\cQ$ depends only on the homology data of $\cQ$.
What can be said about $\Aut(\cQ)$,
beyond the elementary comments made in
section \ref{sect:Isoandauto}?

If the conjecture holds, what can  be said about the bundle
of quasi Riemann surfaces $\QM\rightarrow\PBM$
associated to $M$?
It will have structure group $\Aut(\QSigma)$
where $\Sigma$ is the Riemann surface, or two-dimensional conformal 
space, with the same homology data as $M$.
The homotopy groups of the bundle $\cE\rightarrow \cB$
are given by the homology groups of $M$ \cite{MR0146835}.
Presumably there are analogous results
on the homotopy groups of
the bundle $\QM\rightarrow\PBM$.

\subsection{Can a Riemann surface be augmented?}

In section~\ref{sect:RiemannsurfaceasquasiRS}, it is pointed out that 
there is an obstacle to forming a quasi Riemann surface
from the currents in a Riemann surface $\Sigma$.
The homology groups $H_{j}$ of the spaces
$\cD^{\integral}_{k}(\Sigma)_{0}$
of integral cycles in $\Sigma$
are not isomorphic to the homology groups $H_{j+k}$ of the {\it 
augmented} de Rham complex of $\Sigma$.

In section~\ref{sect:Explorations}, it is supposed that a Riemann surface $\Sigma$
--- or, more generally, one of the conjectured two-dimensional conformal 
spaces --- 
can be augmented to a space $\Sigma_{+}$ whose de Rham complex of integral currents is 
the {\it augmented} de Rham complex of the original space $\Sigma$,
which has homology only in the middle dimension, dimension 1,
\eq
0 	\leftarrow \cD^{\integral}_{-1}(\Sigma_{+})
\xleftarrow{\partial} \cD^{\integral}_{0}(\Sigma_{+})
\xleftarrow{\partial} \cD^{\integral}_{1}(\Sigma_{+})
\xleftarrow{\partial} \cD^{\integral}_{2}(\Sigma_{+})
\xleftarrow{\partial} \cD^{\integral}_{3}(\Sigma_{+})
\leftarrow  0
\,,
\en
with
\eq
\cD^{\integral}_{-1}(\Sigma_{+}) \simeq \Integers
\,,\qquad
\cD^{\integral}_{3}(\Sigma_{+}) \simeq \Integers
\en
and homology
\eq
H_{1}(\Sigma_{+}) = H_{1}(\Sigma)
\,,\qquad
H_{k}(\Sigma_{+}) = 0
\,,\quad k\ne 1
\,.
\en
The augmentation would kill the homology  of $\Sigma$
in dimension 2 by adding an integral 3-current $\eta_{3}$ whose 
boundary is $\Sigma$, $\partial\eta_{3}=\Sigma$.
This would have to be done
in such a way as to kill the homology groups
$H_1(\cD^{\integral}_{1}(\Sigma)_{0})$
and $H_2(\cD^{\integral}_{0}(\Sigma)_{0})$.

Does a Riemann surface $\Sigma$  have such an augmentation $\Sigma_{+}$?

\subsection{What can be said about the universal bundle of quasi 
Riemann surfaces?}

The ``explorations'' of section~\ref{sect:Explorations}
lead to a universal bundle of quasi Riemann surfaces
$\cQ(0) \rightarrow \PB(0)$
for each value of the homology data,
where $\PB(0)$ is a homogeneous space,
\eq
\PB(0) = G(\Sigma_{+})/\Aut(\QSigmaplus)
\,,
\en
and $\cQ(0)$ is the bundle over $\PB(0)$ associated to the action of 
$\Aut(\QSigmaplus)$ on the quasi Riemann surface $\QSigmaplus$.
The group $G(\Sigma_{+})$ is the group of automorphisms of 
$\cD^{\integral}_{1}(\Sigma_{+})$ --- the continuous abelian group automorphisms 
that preserve the skew-hermitian form and the $J$-operator,
and the subgroups of cycles and boundaries.

The homogeneous space $\PB(0)$ is also described as the space of 
integral co-lines in 
\eq
\cZ= \cD^{\integral}_{1}(\cD^{\integral}_{1}(\Sigma)_{0})_{0}
\,.
\en
The conjecture leads to the association of  a subgroup of 
$\cZ$ to the manifold $M$, and identifies $\PB(M)$ with the subset of integral co-lines in $\cZ$ 
that contain that subgroup.  So $\PB(M)$ is embedded in $\PB(0)$, and 
the bundle of quasi Riemann surfaces $\QM\rightarrow \PB(M)$ is
the restriction of $\cQ(0) \rightarrow \PB(0)$.

This is to be the universal setting for extended conformal quantum field 
theory and perhaps, eventually, extended non-conformal quantum field 
theory.
Anything about its structure could be useful.  In particular, the 
group of conformal symmetries of $M$, $\Conf(M)$, acts on 
$\QM\rightarrow \PB(M)$.  Is $\Conf(M)$ a conjugacy class of 
subgroups of $G(\Sigma_{+})$?

\subsection{How much function theory can be done on a quasi Riemann 
surface?}

This is a vague question.  Two-dimensional conformal field theory uses
a fair amount of the the theory of functions --- and conformal tensors
--- on Riemann surfaces.  How much of that can be done on a quasi
Riemann surface?  Extended conformal field theory on quasi Riemann
surfaces will presumably have need of it.  The Schwinger-Dyson
equations (\ref{eq:SDjj}--\ref{eq:SDphij}) on $\ExC$ would seem to be
a starting point.

\subsection{Are there mathematical applications?}

This is presumptuous, considering the amount of conjecture and 
supposition.
Still, if all can be made rigorous,
there should be mathematical applications.
The setting is rather general --- a space $M$ that has forms and 
currents, of even dimension $d=2n$, with a $*$-operator acting in the 
middle dimension.
The association of the bundle $\QM\rightarrow \PBM$ of quasi Riemann 
surfaces to each such $M$
should 
offer opportunities for constructing invariants of $M$
and for investigating the geometry of $M$.
The embedding in a universal bundle 
of quasi Riemann surfaces
might offer opportunities for
studying the space of such spaces $M$.

\section{Further steps}
\label{sect:furthersteps}

This section lists some  possible further steps.
The project depends on answers to
at least the first three of the mathematical questions 
listed in section~\ref{sect:mathematicalquestions}.
However, it might be possible to make some progress
--- in particular, constructing {\it extended} conformal 
field theories on quasi Riemann surfaces --- 
without resolving the conjecture on the equivalence of quasi Riemann 
surfaces.
And it might be possible to take some additional formal steps
assuming the conjecture,
without having the actual isomorphisms in hand and without knowing 
the automorphism groups of the quasi Riemann surfaces.

\subsection{Extended CFT on quasi Riemann surfaces}

A 2d extended conformal field theory (2d ECFT) is an ordinary 2d CFT
on Riemann surfaces $\Sigma$ extended to the quasi Riemann surfaces
$\QSigma$.  As discussed in section~\ref{sect:corrfnsfromECFT}, the
observables are to be extended from products of local fields over
finite sets of points in $\Sigma$ to products over integral
$0$-currents in $\Sigma$.  The extension should use only the geometric
structures of the quasi Riemann surface, so that the resulting ECFT
will be invariant under the automorphism group $\Aut(\QSigma)$ of the
quasi Riemann surface.

One might start by trying to extend the 2d gaussian model
on $\Sigma = S^{2}$,
where we have explicit expressions for
the correlation functions of the ordinary vertex operators.
The 0-form fields $\phi_{\pm}$ are linear functionals on $0$-currents.
Vertex operators $V_{i}(z)$, which are exponentials of the
$\phi_{\pm}$, give ordinary observables of the form
\eq
\Phi(\eta) = V_{1}(z_{1}) V_{2}(z_{2}) \cdots V_{N}(z_{N})
\,,\qquad
\eta = \sum_{i} \delta_{z_{i}}
\,.
\en
associated to the 0-current $\eta$.
The extension to $V(\eta)$ defined at arbitrary integral 
0-currents $\eta$ 
will be an observable of a novel sort in the 
2d CFT.  It will be a product of ordinary 2d vertex operators at a fractal 
set of points in the Riemann surface $\Sigma$.

Then one might try to carry out on the extended 2d gaussian model the 
usual constructions that are performed on the ordinary CFT, such 
as the 
$\Integers_{2}$ orbifold and the $SU(2)\times SU(2)$ symmetry in the 
$R=1$ model.

In the end, we want a general construction of a 2d ECFT from an 
ordinary 2d CFT.
The space of extended observables should be
some kind of limit
\eq
\cH = \lim_{N\rightarrow\infty}
\;
\mathop\oplus_{\{z_{1},\ldots,z_{N}\}}
\left (
\mathop\otimes_{z\in\{z_{1},\ldots,z_{N}\}}
\cH_{z}
\right )
\en
where $\cH_{z}$ is the vector space of local fields at the point 
$z\in\Sigma$, including the identity field,
such that the extended observables have the form of functions
$\Phi(\eta)$ on $\cD^{\integral}_{0}(\Sigma)$.
Among the questions that arise --- what is the role of the abelian 
group structure of $\cD^{\integral}_{0}(\Sigma)$
in the general ECFT?

It might be possible to describe the extended observables on $\QSigma$
by their manifestations on the local q-h curves.  The local q-h curves
in $\QSigma$ are given, conjecturally, by the local coordinate
neighborhoods in $\Sigma$, up to automorphisms of $\QSigma$.  For each
q-h curve $C$, an extended observable will appear as a state $\psi(C)$
in the state space of the radial quantization of the 2d CFT on the
unit complex disk.  The collection of states $\psi(C)$, subject to
coherence conditions, might serve to characterize the extended
observables.

The stress-energy tensor $T(z)$, $\bar T(\bar z)$ of the 2d CFT
on the Riemann surface $\Sigma$ has to be lifted to $\QSigma$.
In the free $n$-form/2d gaussian model, they lift to fields
on $\ExC$,
\eq
T_{++}(\xi) = -\frac12 j^{\dagger}_{+}(\bar\xi)j_{+}(\xi)
\,,\qquad
T_{--}(\xi) = -\frac12 j^{\dagger}_{-}(\bar\xi)j_{-}(\xi)
\,,
\en
where $j_{\pm}$ are the chiral 1-form fields on $\ExC$.
Pulled back along a q-h curve, $T_{++}$  and $T_{--}$ become the 
analytic and anti-analytic stress-energy tensors on the unit disk.
For each local q-h curve,
the usual pair of Virasoro algebras will act on the states.
Collectively, these should form a large Lie algebra acting on the 
space of extended observables.
An operator product algebra of the extended observables should be 
built from a product of the operator product algebras on the q-h 
curves.
And perhaps there should be an inner product on the space of extended 
observables, so that there is a representation by operators on 
Hilbert space.
Eventually,
there might be an axiomatic formulation of extended conformal field 
theory.

\subsection{Gauge invariance}

If the conjecture holds, the ECFT on $\QSigma$ is to be transported 
by isomorphisms to each of the fibers of the bundle of quasi Riemann 
surfaces $\QM \rightarrow \PBM$.
That bundle is embedded in the universal bundle $\cQ(0)\rightarrow 
\PB(0)$.
The {\it core} of each fiber is the same.
The 1-spaces in each fiber are the same,
\eq
\QM_{\Integers\partial\xi,1} = \cD^{\integral}_{n}(M)
\,,\qquad
\cQ(0)_{\lambda,1} = \cD^{\integral}_{1}(\Sigma_{+})
\,.
\en
The 0-cycles in each fiber are the same,
\eq
(\QM^{\integral}_{\Integers\partial\xi,0})_{0} = \cD^{\integral}_{n-1}(M)_{0}
\,,\qquad
(\cQ(0)^{\integral}_{\lambda,0})_{0} = \cD^{\integral}_{0}(\Sigma_{+})_{0}
\,,
\en
as are the dual spaces of 2-currents.
For convenience, I am omitting the complexifications that are needed 
when $n$ is even.

The remaining data of the quasi Riemann surfaces, that distinguishes 
the fibers, is the orientation of the integral co-line in
$\cZ= \cD^{\integral}_{1}(\cD^{\integral}_{1}(\Sigma)_{0})_{0}$
which determines the integral line of 2-cycles
\eq
(\QM^{\integral}_{\Integers\partial\xi,2})_{0}
\quad
\text{ or }
\quad
(\cQ(0)^{\integral}_{\lambda,2})_{0}
\en
and its dual line
\eq
\QM^{\integral}_{\Integers\partial\xi,0}/(\QM^{\integral}_{\Integers\partial\xi,0})_{0}
\quad
\text{ or }
\quad
\cQ(0)^{\integral}_{\lambda,0}/(\cQ(0)^{\integral}_{\lambda,0})_{0}
\,.
\en

We might call the extended observables on the core of a quasi Riemann 
surface the {\it core} of the ECFT.
The cores of the ECFTs on the fibers of $\QM\rightarrow\PBM$ must be 
all the same, because they are the same fields of extended objects in 
$M$.  It is natural to require the same in the universal bundle
$\cQ(0)\rightarrow \PB(0)$.
For the 2d gaussian model, the core ECFT is the algebra 
generated by the integrals  of the $1$-forms $j_{\pm}$ over integral $1$-cycles.

On each fiber, there will be some freedom in the extension of the 
core ECFT to an ECFT on the entire quasi Riemann surface.
The is the local gauge symmetry over $\PBM$ or, universally, over 
$\PB(0)$.  In the free $n$-form/2d gaussian model, this is the 
freedom to choose the zero-modes of the 
0-form fields $\phi_{\pm}$, as described in 
section~\ref{sect:gaugetheory} for the classical theory.
We get a bundle of theories $\cT(M)\rightarrow \PBM$ or, universally,
$\cT(0) \rightarrow \PB(0)$,
which are the ECFTs extending the core ECFT.

It seems, then, that extended CFT will have additional structure 
beyond what is given in ordinary 2d CFT.
There should always be a core ECFT and a set of extensions to the 
full quasi Riemann surface.
Consider for example the 2d gaussian model at $R=1$, where the two 
1-form fields $j_{\pm}$ of the 
$U(1)\times U(1)$ current algebra grow to the six 1-form fields of a $SU(2)\times SU(2)$ 
current algebra.
The core ECFT of the 2d gaussian model is generated by the integrals 
of $j_{\pm}$ over 1-cycles.  The space of extensions is a principal 
homogeneous space for the global symmetry group $U(1)\times U(1)$.
But, at $R=1$, the core ECFT can be expanded to the algebra 
generated by integrals of the $SU(2)\times SU(2)$ 1-form fields over 
1-cycles.  The space of extensions will be a principal 
homogeneous space for the global symmetry group $SU(2)\times SU(2)$.
Different ECFTs can be produced from the same ordinary CFT, depending on how 
the core of the ECFT is chosen.

Is the set of extensions always a homogeneous space for a group?
That group  would be the global symmetry group.
Is the set of extensions always a principal homogeneous 
space for the global symmetry group?  Some examples might clarify.

Is there some general method to gauge-fix over $\PBM$, partially, 
so that the remaining gauge symmetry is a local gauge symmetry in the 
space-time $M$,
as described in section~\ref{sect:gaugetheory} for the classical 
$n$-form theory?

\subsection{Connecting the $\QM_{\partial\xi}$}

We need a general picture of how the ECFTs on the fibers of 
$\QM\rightarrow\PBM$ fit together to make one quantum field theory.
Or perhaps we can fit together all the ECFTs on the fibers of
the universal bundle $\cQ(0)\rightarrow \PB(0)$
to form a single universal quantum field theory
that specializes to any space-time $M$ with the appropriate homology 
data.
The conformal symmetry group of space-time, $\Conf(M)$, will act on 
$\QM\rightarrow\PBM$, and perhaps on $\cQ(0)\rightarrow \PB(0)$,
so we certainly need a single theory on which $\Conf(M)$ will be 
represented.

I am not sure how to adapt the argument of section~\ref{sect:connectingtheExC}
to the general case.
In section~\ref{sect:Explorations},
$\PB(0)$ is described as the homogeneous space
$G(\Sigma_{+})/\Aut(\QSigma)$,
and $\cQ(0)\rightarrow \PB(0)$ is described
as an associated fiber bundle,
with $G(\Sigma_{+})$ acting as a symmetry group
on the fiber bundle.
The action of $G(\Sigma_{+})$ should provide a natural connection in $\cQ(0)\rightarrow \PB(0)$.
But I do not see that it fixes the common core of the fibers,
so I do not see how it will give a connection in the bundle
$\cT(0)\rightarrow \PB(0)$ of theories.
And I am not sure what the result should look like.
The result of section~\ref{sect:connectingtheExC} for the $n$-form 
theory is that gauge invariant observables in the bundle can be 
transported to any one single fiber.
Perhaps a better statement would be that gauge invariant observables 
can be transported to the common core of all the fibers
---
that the non-zero correlation functions of the space-time theory
are the correlation functions of the core ECFT.

\subsection{The local fields in space-time}

Given an extended CFT,
there should be a general way to identify the 
local fields on the  space-time $M$,
acted on by the conformal symmetry group $\Conf(M)$.
Taking $M=S^{d}$, it should be possible to
derive the spectrum of scaling dimensions
from the data of the underlying 2d CFT.
Perhaps the local fields can be identified 
from the extended observables on the infinitesimally small
integral currents.

\subsection{Partition functions and variation of conformal structure}

If the conjecture holds, then it seems reasonable to suppose that not
only the correlation functions but also the partition function of the
space-time conformal field theory will be given by the corresponding
2d CFT. If in fact there does exist a two-dimensional conformal space
$\Sigma$ for every set of homology data, the partition function of
each theory will be a function of the homology data -- or a section of
a line bundle over the space of homology data.  The identity between
the partition functions of the two theories can be demonstrated by
showing that they have the same variations with respect to the
homology data, and then showing that they have the same behavior at
transition singularities where the homology group changes.

In the two-dimensional conformal space, a variation of the homology 
data is given by an infinitesimal perturbation of the almost-complex 
structure $J$,
which is expressed by a Beltrami differential,
which is a (-1,1)-tensor $h^{z}_{\bar z}$.
Somehow, if there is to be a two-dimensional conformal space for every set of 
homology data, the usual conditions on the Beltrami differentials on a Riemann 
surface must be modified, because the homology data has more directions of 
variation than there are moduli of the Riemann surface.
Perhaps the augmentations $\Sigma_{+}$  of  Riemann surfaces $\Sigma$
discussed in section~\ref{sect:augmentSigma}
might have different moduli.

The variation of the partition function is given by the integral over 
$\Sigma$ of the (1,1)-form 
$\expval{T_{zz}h^{z}_{\bar z}}dz d\bar z$ where $T_{zz}$ is the holomorphic
$(2,0)$-component of the stress-energy tensor.
The considerations of section~\ref{sect:perturbationtheory}
might be used to equate this variation of the 2d partition function
with the variation of the space-time partition function with respect 
to the homology data of $M$.

\subsection{Non-conformal extended quantum field theory}

If extended CFTs in space-time can be built successfully from 2d CFTs,
next will be to extend the correspondence to non-conformal quantum
field theories by establishing an equivalence between renormalized
perturbation theory around an extended CFT in space-time and
renormalized perturbation theory around the corresponding 2d CFT.
Space-time will be $M = \Reals^{d}$.  The corresponding
two-dimensional space will be, presumably, $\Sigma = \Reals^{2}$.
Perturbation theory in space-time will be renormalized with respect 
to the euclidean metric on $\Reals^{d}$ in order to construct a 
quantum field theory with euclidean symmetry, to be Wick-rotated to 
Minkowski space $\Reals^{1,d-1}$.

Some way is needed to transfer the euclidean metric on $\Reals^{d}$ 
to the two-dimensional space $\Sigma$.
Perhaps the simplest route is to put a specific metric space structure on the quasi 
Riemann surface $\QM_{\Integers\partial\xi}$
using the euclidean metric on $\Reals^{d}$,
then transfer the metric space structure to $\QSigma$ via an 
isomorphism of quasi Riemann surfaces
(assuming the conjecture).
I do not know of any guarantee that the induced metric space 
structure on $\QSigma$ would come from a Riemannian metric on 
$\Sigma$.
So it would be necessary to investigate the renormalization of 
perturbations of an ECFT with respect to a metric 
space structure on $\QSigma$.
A very naive hope would be that this is the same as renormalization of 
perturbations of the underlying 2d CFT on $\Sigma$ with respect to a 
metric on $\Sigma$,
giving a correspondence between non-conformal 2d quantum field 
theories on $\Sigma$ and non-conformal quantum field theories of 
extended objects in $\Reals^{d}$.
Then one might wonder how 
the operator representation of the 2d QFT
is related to the operator representation in the Minkowski space-time 
QFT,
how the 2d S-matrix might be related to the space-time S-matrix,
how 2d integrability might be manifested in the space-time QFT.

A fanciful prospect is an integrable, asymptotically free 
2d QFT with nonabelian global symmetry
corresponding to an asymptotically free 4d quantum field theory of 
extended objects with nonabelian local gauge symmetry.

\section{Questions about history and references}
\label{sect:hist-refs-questions}

I would appreciate advice on the history of ideas germane to this work, and on 
their proper citation.  To be safe, I have cited the basic 
works on
the free 2-form in d=4
dimensions \cite{Maxwell:1865zz,Born:1926:QIG,Dirac:1931kp}.  

The proximate influences on the present work were
\begin{itemize}
\item G.~Moore's frequent comments that a QFT cannot be understood 
without understanding its extended objects,
%  --- not asking for a QFT 
%  of extended objects, but keeping extended objects in the forefront
\item N.~Seiberg's September 23, 2014 Rutgers seminar on his paper
with Gaiotto, Kapustin, and  Willett
on generalized charges in quantum field theory,
{\it Generalized global symmetries} \cite{Gaiotto:2014kfa}.
\item many conversations with S.~Thomas and C.~Keller about conformal 
field theory in dimensions $d>2$, and
\item especially a comment of S.~Thomas suggesting that nonabelian 
structure might be found in the extended objects,
after the realization \cite{FriedanKellerThomas2015}
 that nonabelian operator algebras of self-dual 
$n$-forms were impossible in $d>2$ dimensions,
\item a search of the World Wide Web for `spaces of cycles',
which yielded a number of pointers to
geometric measure theory,
especially Gromov's 2015 note
{\it Morse Spectra, Homology Measures, Spaces of Cycles and Parametric Packing Problems}
\cite{Gromov20150416}.
\end{itemize}
%
%I especially need advice on the history of

%\vskip2ex
%\noindent{\bf The 2d gaussian model}

In statistical mechanics, the 2d gaussian model as the free 1-form
with compact symmetry group $U(1){\times} U(1)$ should perhaps be
attributed to Mermin-Wagner-Hohenberg and
Berezinskii-Kosterlitz-Thouless.

In string theory, the free 1-form in 2d was the basic tool of the 
world-sheet technology.
The decomposition of the free 1-form into chiral components and
the vertex operator representation of the ``extended objects''
were done in string theory.
The compactification of the target to a circle of radius $R$ was the 
implementation of Kaluza-Klein theory.

I am not sure where the name {\it 2d gaussian model} came from.  Free
massless scalar field theory is known in statistical mechanics as the
{\it gaussian model}.  In 2d, the logarithmic infrared divergence 
forces compactification of the target space to a circle of radius $R$.
Maybe I learned to call this the {\it 2d gaussian model} from Leo 
Kadanoff in the early 1980s.

%\vskip2ex
%\noindent{\bf The free $n$-form in $d=2n$ dimensions}

Who first noted the analogy between the free 1-form in 2d and Maxwell's theory in 
4d and the generalization to the free $n$-form in $d=2n$ dimensions?
The line integrals of gauge potentials in nonabelian gauge theory are 
called Wilson loop operators.
Where was it noticed that these are the operators describing extended 
objects in the free $n$-form theory?

There is a large mathematics and physics literature on the free $n$-form
theories in $d=2n$ dimensions with $n$ odd, especially the theories of
self-dual $n$-forms ---
in particular,
Hopkins and Singer,
{\it Quadratic functions in geometry, topology, and {M}-theory}
\cite{MR2192936}
and
Freed, Moore, and Graeme Segal,
{\it Heisenberg groups and noncommutative fluxes} \cite{Freed2007236}.
These papers are too far over my head for me to tell
if they are germane to the present work.
I do notice that
the latter paper discusses the analogy between the free $n$-form theory and the 2d 
gaussian model,
and some of its calculations seem to resemble those on gauge 
invariance in section \ref{sect:gaugetheory}.
In any case, if the present project works out,
there will surely be points of contact.

\vskip2ex
\noindent{\bf Geometric measure theory}

I have cited what I understand to be the basic references for
geometric measure theory \cite{MR0123260,MR1794185,MR0146835}, but my
grasp of the subject is weak and
superficial.

\section*{Acknowledgments}
I am grateful to
Scott Thomas and Christoph Keller
for stimulating discussions.
This work was supported by the New High Energy Theory Center (NHETC) of 
Rutgers, The State University of New Jersey.
%%%%%%%%%%%%%%%%%%%%%%%%%%%%%%%%%%%%%%%%%%%%%%%%%%%%%%%%%%%%%%%%%%%%%%%%%%%%%
%                                                                           %
%                                                                           %
%   The U.S. Department of Energy Office of High Energy Physics (DOE HEP)   %
%   declined to support this work.                                          %
%                                                                           %
%%%%%%%%%%%%%%%%%%%%%%%%%%%%%%%%%%%%%%%%%%%%%%%%%%%%%%%%%%%%%%%%%%%%%%%%%%%%%

\addtocontents{toc}{\protect\setcounter{tocdepth}{1}}

\addtotoc{Acknowledgments}{}

\newpage
 
\appendix
\appendixpage
% \vskip2ex

\section{Construction of a path of integral currents 
%in the style of
a la
{\it Game of Thrones}}
%\section{A path of integral 
%currents -- a {\it Game of Thrones} construction}
%\section{``{\it Game of Thrones}'' construction}
\label{app:GOT}

This section gives a construction that should be
the germ of a proof that the Hodge $*$-operator acts on the 
tangent spaces of 
the space $\cD^{\integral}_{n-1}(M)_{0}$ of integral $(n{-}1)$-cycles in a conformal 
manifold $M$ of dimension $d=2n$.

As in section \ref{sect:Hodgestaracts} above,
we consider the path of singular $1$-cycles in $\Reals^{4}$,
\eq
\xi(\epsilon) = \partial \left [\delta(x^{1})\delta(x^{2}) \theta_{[0,1]}(x^{3}) \theta_{[0,\epsilon]}(x^{4})
\hat e_{3}\wedge \hat e_{4}\right ]
\,,
\en
where $\theta_{[a,b]}$ is the characteristic function of the 
interval $[a,b]\subset \Reals$.
The $1$-cycles $\xi(\epsilon)$ are the boundaries
of rectangles in the 3-4 plane in $\Reals^{4}$,
shrinking to the interval $[0,1]$ in the 3-axis.
The tangent vector to the path $\xi(\epsilon)$ at $\epsilon=0$ 
is the flat 2-current
\eq
\dot \xi = 
\delta(x^{1})\delta(x^{2}) \theta_{[0,1]}(x^{3}) \delta(x^{4})
\hat e_{3}\wedge \hat e_{4}
\,.
\en
The Hodge $*$-operator on $\dot \xi$ is
\eq
\label{eq:dotxistar}
* \dot \xi = 
\delta(x^{1})\delta(x^{2}) \theta_{[0,1]}(x^{3}) \delta(x^{4})
\hat e_{1}\wedge \hat e_{2}
\,.
\en
The flat $2$-current $* \dot \xi$ is not tangent to any path of 
singular $1$-cycles.

Here, a path $\xiint_{1}(\epsilon)$ of {\it integral} $1$-cycles
is constructed whose tangent vector is
\eq
\dot\xi^{\integral}_{1} = * \dot \xi\,.
\en
The rest of the proof might then go:
\begin{enumerate}
\item Use the rotations in $\Reals^{4}$ that leave the 3-axis fixed to get
paths of integral cycles with tangent vectors of the form
\eq
\delta(x^{1})\delta(x^{2}) \theta_{[0,1]}(x^{3}) \delta(x^{4}) \,t^{\mu\nu}
\en
for arbitrary 2-vector $t^{\mu\nu}$.
\item Scale and rotate in $\Reals^{4}$ to get paths with tangent
vectors $\theta_{I} t^{\mu\nu}$,
where $\theta_{I}$ is the 
characteristic 0-current of any interval $I$
in $\Reals^{4}$,
arbitrarily small, and where $t^{\mu\nu}$ is any 2-vector.
This space of flat 2-currents is manifestly closed under the action 
of the Hodge $*$-operator.
\item
Using coordinates in $M$,
take limits of
linear combinations
of the paths constructed in the previous step
to get paths of integral 1-currents
with tangent vector
any flat $2$-current in $M$
that is supported on an integral $1$-current in $M$.
\item
Thus the tangent space at $0$ of the space of integral $1$-cycles
in $M$ is exactly
the space of flat $2$-currents in $M$
that are supported on the integral $1$-currents in $M$.
\item
The Hodge $*$-operator
manifestly acts as a bounded operator on the space of flat $2$-currents 
that are supported on the singular 1-crrents,
so it acts on the tangent space.
\end{enumerate}
Generalizing the construction of
the path $\xiint_{1}(\epsilon)$ to arbitrary $d=2n$
will be straightforward,
and the rest of the proof is the same as in $d=4$ dimensions.

The construction of the path $\xiint_{1}(\epsilon)$ 
takes place within $\Reals^{3}\subset \Reals^{4}$.
First, an integral 2-current $\xi^{\integral}_{2}$ is constructed,
supported in the unit cube $([0,1])^{3}\subset \Reals^{3}$.
Then a path of integral 2-currents $\xi^{\integral}_{2}(\epsilon)$ is constructed
by scaling $\xi^{\integral}_{2}$ in the 1-2 plane, under
\eq
(x^{1},x^{2},x^{3})\mapsto
(\epsilon^{\frac12}x^{1},\epsilon^{\frac12}x^{2},x^{3})
\,.
\en
As $\epsilon\rightarrow 0$, the original 2-current 
$\xi^{\integral}_{2}$ is squashed onto the interval $[0,1]$ in the 3-axis.
The construction is designed so that
$\lim_{\epsilon\rightarrow 0}\xi^{\integral}_{2}(\epsilon) = {*}\dot\xi$.
So the path of integral 1-cycles
$\xiint_{1}(\epsilon)=\partial\xi^{\integral}_{2}(\epsilon)$
has tangent vector $\dot \xi^{\integral}_{1}= {*}\dot\xi$.

The integral 2-current $\xi^{\integral}_{2}$ is a fractal.  Its 
construction is illustrated in the attached animations 
\verb!GOT1.gif! and \verb!GOT2.gif!,
which are also available at\\
\hspace*{4em}\verb!http://www.physics.rutgers.edu/pages/friedan/GOT/!.\\
Animated gifs can be viewed in most web browsers.
The animated gifs were made with Sagemath \cite{SageMath:2015:6.7} and ImageMagick \cite{ImageMagick}.
%\cite{sage}
The visualized construction bears a slight resemblance to
a part of the title sequence animation of the television show
{\it Game of Thrones} \cite{GOT}.

Define the 2-current $S(y^{1},y^{2},y^{3};a)$ to be the square in the 
1-2 plane with 
corner at $\vec y = (y^{1},y^{2},y^{3})$ and side $a$,
\eq
S(\vec y;a) =  
\theta_{[y_{1},y_{1}+a]}(x^{1})
\theta_{[y_{2},y_{2}+a]}(x^{2})
\delta(x^{3}-y^{3}) 
\; \hat e_{1}\wedge \hat e_{2}
\,.
\en
Define an operator $R({b})$ on such squares that splits the square 
into 4 quadrants
and lifts the quadrants in the 3-direction,
not lifting the first quadrant, lifting the second 
quadrant by $b/4$, lifting the third quadrant by $2b/4$, and 
lifting the fourth quadrant by $3b/4$,
\eq
R({b})S(\vec y;a) = \sum_{i=1}^{4} S(\vec y_{i};a/2)
\en
where
\ateq{2}{
\vec y_{4} &= \vec y +\frac{a}2 \hat e_{2} +\frac{3b}4 \hat e_{3}
\,,
\qquad\qquad& 
\vec y_{3} &= \vec y +\frac{a}2 (\hat e_{1}+\hat e_{2}) +\frac{2b}4 \hat e_{3}
\\
\vec y_{1} &= \vec y
\,,
& \vec y_{2} &= \vec y +\frac{a}2 \hat e_{1} +\frac{b}4 \hat e_{3}
\,.
}
Then extend $R(b)$ to sums of squares by linearity over the integers.

Let $\xi^{\integral}_{2,0}$ be the square in the 1-2 plane of side $1$ with 
corner at the origin,
\eq
\xi^{\integral}_{2,0} = S(\vec 0;1)
\en
Define a sequence of integral $2$-currents 
$\xi^{\integral}_{2,0},\,\xi^{\integral}_{2,1},\,\xi^{\integral}_{2,2}\,,\ldots$ 
by
\eq
\xi^{\integral}_{2,k+1} = R(4^{-k}) \xi^{\integral}_{2,k}
\,,
\en
so $\xi^{\integral}_{2,k}$ is a sum of $4^{k}$ squares in the 1-2 plane,
each of area $4^{-k}$,
at heights
\eq
x^{3} = j 4^{-k}, \quad j=0,1,2,\ldots, 4^{k}-1
\,.
\en 
The total area is left unchanged by $R(b)$, so
\eq
\norm{\xi^{\integral}_{2,k}}_{\flat} = 1
\,.
\en
The operator $R(b)$ acting on a square sweeps out a 3-chain in the 
shape of a 4-step spiral staircase with square steps of heights $0$, $b/4$, 
$2b/4$, and $3b/4$.
Let $S_{3}$ be the 3-current representing the staircase
and let $S_{2}$ be the 2-current representing the vertical sides 
of the staircase, so
\eq
R(b) S(\vec y; a) - S(\vec y,a) = \partial S_{3} - S_{2}
\en
The  volume of the staircase is
\eq
M_{3}(S_{3})
= \frac{b}{4}\left( \frac{a}2 \right)^{2} + \frac{2b}{4}\left( \frac{a}2 \right)^{2}
+ \frac{3b}{4}\left( \frac{a}2 \right)^{2}
= \frac{3}{8} b a^{2}
\,.
\en
The vertical sides consist of $24-6=18$ rectangles of  vertical side $b/4$
and horizontal side $a/2$, so the area of the vertical sides is 
\eq
M_{2}(S_{2})
= 18 \frac{b}4 \frac{a}2 = \frac{9}{4}ba
\en
Therefore
\eq
\norm{R(b) S(\vec y; a) - S(\vec y,a) }_{\flat} \le  
M_{2}(S_{2})+M_{3}(S_{3})
= \frac{9}{4}ba + \frac{3}{8} b a^{2}
\en
In $\xi^{\integral}_{2,k}$, there are $4^{k}$ squares, each with $a =  2^{-k}$, so
\eq
\norm{\xi^{\integral}_{2,k+1} - \xi^{\integral}_{2,k}}_{\flat}
\le  4^{k}
\left [\frac{9}{4}4^{-k} 2^{-k}
+ \frac{3}{8} 4^{-k} (2^{-k} )^{2}
\right ]
= \frac{9}{4} 2^{-k} + \frac{3}{8} 4^{-k}
\,,
\en
so $\xi^{\integral}_{2,k}$ is a Cauchy sequence.
The space of integral currents with bounded norm and bounded support is 
compact, so Cauchy sequences converge.
Let
\eq
\xi^{\integral}_{2} = \lim_{k\rightarrow\infty} \xi^{\integral}_{2,k}
\,.
\en
%%%%%%%%%%%%%%%%%%%%%%%%%%%%%%%%%%%%%%%%%%%%%%%%%%%%%%%%%%%%%%%%%%%%%%
%                                                                    %
%   We also have                                                     %
%   \eq                                                              %
%   \norm{\eta_{k}(t_{1})-\eta_{k}(t_{2})}_{\flat} = |t_{1}-t_{2}|   %
%   \en                                                              %
%   so $\eta(t)$ should be well-behaved in $t$.                      %
%                                                                    %
%%%%%%%%%%%%%%%%%%%%%%%%%%%%%%%%%%%%%%%%%%%%%%%%%%%%%%%%%%%%%%%%%%%%%%

Scale $\xi^{\integral}_{2}$ and the $\xi^{\integral}_{2,k}$
by $\epsilon^{\frac12}$ in the 1-2 plane
to get
\eq
\xi^{\integral}_{2}(\epsilon)
=
\lim_{k\rightarrow\infty} \xi^{\integral}_{2,k}(\epsilon)
\en
with
\eq
\norm{\xi^{\integral}_{2}}_{\flat} =
\norm{\xi^{\integral}_{2,k}}_{\flat} = 1
\,.
\en
The integral 2-current $\xi^{\integral}_{2,k}(\epsilon)$ 
consists of the square 
$[0,\epsilon^{\frac12}]\times[0,\epsilon^{\frac12}]$
in the 1-2 plane divided into a checkerboard of
$4^{k}$ squares each of area $4^{-k} \epsilon$,
each small square raised to one of the 
evenly distributed heights,
\eq
x^{3} = j 4^{-k}, \quad j=0,1,2,\ldots, 4^{k}-1
\,.
\en 
As $\epsilon\rightarrow 0$, each of the $4^{k}$ small squares
is squashed onto the 3-axis.  For $\omega$ a smooth 2-form, 
\eq
\int_{\xi^{\integral}_{2}(\epsilon)} \omega
=\lim_{k\rightarrow\infty}
\int_{\xi^{\integral}_{2,k}(\epsilon)} \omega
\approx\lim_{k\rightarrow\infty} \sum_{j=0}^{4^{k}-1}  4^{-k} \epsilon\, \omega_{12}(0,0,4^{-k}j) 
\approx\epsilon \int_{0}^{1} dx^{3}\;  \omega_{12}(0,0,x^{3}) 
\,,
\en
so
\eq
\lim_{\epsilon\rightarrow 0}\frac{\xi^{\integral}_{2}(\epsilon)}{\epsilon}
= 
\delta(x^{1}) \delta(x^{2}) \theta_{[0,1]}(x^{3})
 \; \hat e_{1}\wedge\hat e_{2}
= {*} \dot \xi
\,.
\en
Therefore the tangent vector at $\epsilon=0$ to the path of integral 
1-cycles
\eq
\xiint_{1}(\epsilon) = \partial \xi^{\integral}_{2}(\epsilon)
\en
is 
\eq
\dot \xi^{\integral}_{1} = {*} \dot \xi\,.
\en

\section{The free complex $n$-form on euclidean $\Reals^{d}$}
\label{app:euclean-n-form}

In this section, the Schwinger-Dyson equations are written for the 
chiral fields $F_{\pm}(x)$ and $A_{\pm}(x)$ 
of the free complex $n$-form quantum field theory on
euclidean $\Reals^{d}$, $d=2n$.
The S-D equations on $M=\Reals^{d}$  determine the S-D equations on any manifold $M$,
by dimensional analysis.
There are some arbitrary choices: (1) the overall 
normalization of the two-point functions,
and (2) the contact terms in the two-point functions.
The overall normalization is fixed by matching to a standard convention 
in the 2d theory.
The contact terms are fixed by imposing symmetry.

The notation is as in sections \ref{sect:intro1}, \ref{sect:intro2}, and \ref{sect:moreoncurrentsinM}.

\subsection{Adjoints of $F$ and $F^{*}$}

Wick rotate to Minkowski space.
Write $x^{i}$, $i=1,\ldots,d-1$ for the spatial coordinates.
Write $x^{d}$ for euclidean time and $x^{0}$ for Minkowski space 
time, with $x^{d}=ix^{0}$.

The magnetic field (up to normalization) is $F_{i_{1}\ldots i_{n}}(x)$.  The electric field 
(up to normalization) is $F_{i_{1}\ldots i_{n-1}0}(x)= i 
F_{i_{1}\ldots i_{n-1}d}(x)$.
For real $F$, the magnetic and electric fields are self-adjoint, so,
for complex $F$,
\eq
\label{eq:EMadjoints}
F_{i_{1}\ldots i_{n}}^{\dagger}(x) = \bar F_{i_{1}\ldots i_{n}}(x)
\,,\qquad
F_{i_{1}\ldots i_{n-1} d}^{\dagger}(x) = - \bar F_{i_{1}\ldots i_{n-1} d}(x)
\,.
\en
The Hodge $*$-operator acts by
\eq
{*}F_{\mu_{1}\ldots\mu_{n}} = \frac1{n!}
\epsilon_{\mu_{1}\ldots\mu_{n}}{}^{\nu_{1}\ldots\nu_{n}} F_{\nu_{1}\ldots\nu_{n}}
\en
so
\eq
{*}F_{i_{1}\ldots i_{n}} = \frac1{(n-1)!}
\epsilon_{i_{1}\ldots i_{n}}{}^{j_{1}\ldots j_{n-1}d} F_{j_{1}\ldots 
j_{n-1}d}
\,,\qquad
{*}F_{i_{1}\ldots i_{n-1}d} = \frac1{n!}
\epsilon_{i_{1}\ldots i_{n-1}d}{}^{j_{1}\ldots j_{n}} F_{j_{1}\ldots j_{n}}
\en
so
\eq
({*}F_{i_{1}\ldots i_{n}})^{\dagger} = -{*}\bar F_{i_{1}\ldots i_{n}}
\,,\qquad
({*}F_{i_{1}\ldots i_{n-1}d})^{\dagger} = {*}\bar F_{i_{1}\ldots i_{n-1}d}
\en
so, if we define the dual $n$-form
\eq
F^{*} = \delta_{n} i^{-1} {*} F
\,,\qquad
\delta_{n} = \pm 1\,,
\en
then $F^{*}$ has the same self-adjointness properties as $F$.
That is, in Minkowski space,
\eq
F^{\dagger}(x)=\bar F(x)
\,,\qquad
F^{*\dagger}(x) = \bar F^{*}(x) = - \overbar{F^{*}}(x)
\,.
\en
The choice of $\delta_{n}=\pm 1$ will be left arbitrary in this 
section.
In the body of the paper,
$\delta_{n}=1$ is used.

% The best choice might be $\delta_{n} = -1$, since
% \eq
% F^{*}_{i_{1}\ldots i_{n}}
% = (-\delta_{n}) \frac1{(n-1)!}
% \epsilon_{i_{1}\ldots i_{n}}{}^{j_{1}\ldots j_{n-1}d}
% F_{j_{1}\ldots j_{n-1}0}
% \,.
% \en

\subsection{Notation: adjoints of euclidean fields}

Define the adjoint of a euclidean field to be the Wick-rotate of the 
adjoint of the Minkowski field:
\eq
F^{\dagger}(x) = \bar F(x)
\,,\qquad
F^{*\dagger}(x) = \bar F^{*}(x) = - \overbar{F^{*}}(x)
\,.
\en
Reflection positivity of the euclidean correlation functions is then
\eq
\expval{F^\dagger(Rx) F(x)} > 0
\,,\qquad
Rx \ne x
\,,
\en
where $R$ is the reflection $x^{d}\rightarrow -x^{d}$.

\subsection{The chiral fields $F_{\pm}$ and $A_{\pm}$ and their 
adjoints}
The chiral components of $F$ and $A$ and their adjoints are
\aeq{
F_{\pm} 
&= \frac12 \left ( 1 \pm i^{-1}J\right ) F
= \frac12 \left ( 1 \pm i^{-1}\epsilon_{n}{*}\right ) F
= \frac12 (F \pm \epsilon_{n} \delta_{n} F^{*})
\,,
\\
A_{\pm} 
&=\frac12 (A \pm \epsilon_{n} \delta_{n} A^{*})
\\[1ex]
F_{\pm}^{\dagger} 
&= \frac12 (F^{\dagger} \pm  \bar \epsilon_{n} \delta_{n} F^{*\dagger})
= \frac12 (\bar F \mp  \bar\epsilon_{n} \delta_{n} \overbar{F^{*}})
=\overbar{F_{\mp}}
\,,
\label{eq:Fadj}
\\
A_{\pm}^{\dagger}  &= \overbar{A_{\mp}}
\,.
\label{eq:Aadj}
}

\subsection{The two-point functions and the Schwinger-Dyson equations}

$F(x)$ has scaling dimension $n$, so its two-point functions are linear 
combinations of two invariants, one of which is a pure contact term.
The non-contact invariant is
\eq
G (\bar \xi_{1},\xi_{2})
= \int d^{n}x \int d^{n}y \;\frac1{n!}\bar\xi_{1}^{\mu_{1}\ldots 
\mu_{n}}(x)
G (x,y)_{\mu_{1}\ldots \mu_{n};\nu_{1}\ldots \nu_{n}}
\frac1{n!}\xi_{2}^{\nu_{1}\ldots \nu_{n}}(y)
\nonumber
\en
\eq
G (x,y)_{\mu_{1}\ldots \mu_{n};\nu_{1}\ldots \nu_{n}}
= \int \frac{d^{d}p}{(2\pi)^{d}}
\;
e^{ip(x-y)}
G(p)_{\mu_{1}\ldots \mu_{n};\nu_{1}\ldots \nu_{n}}
\label{eq:G}
\en
\eq
G(p)_{\mu_{1}\ldots \mu_{n};\nu_{1}\ldots \nu_{n}}
=
\frac{1}{p^{2}}
\frac1{(n-1)!}
\mathrm{Alt}_{\mu} \mathrm{Alt}_{\nu}
(p_{\mu_{1}}p_{\nu_{1}} \delta_{\mu_{2}\nu_{2}}\cdots  
\delta_{\mu_{n}\nu_{n}})
\,.
\nonumber
\en
The pure contact invariant is given in terms of the intersection form,
\eq
I(\bar \xi_{1},{*}\xi_{2})
= \int d^{n}x \;\frac1{n!}\bar\xi_{1}^{\mu_{1}\ldots 
\mu_{n}}(x) \mathrm{Alt}_{\mu}
(\delta_{\mu_{1}\nu_{1}}\cdots  \delta_{\mu_{n}\nu_{n}})
\frac1{n!}\xi_{2}^{\nu_{1}\ldots \nu_{n}}(x)
\,.
\en
They satisfy
\eq
\overbar{G (\bar \xi_{1},\xi_{2})}
=G (\bar \xi_{2},\xi_{1})
\,,\qquad
\overbar{I_{M}(\bar \xi_{1},{*}\xi_{2})}
=I_{M}(\bar \xi_{2},{*}\xi_{1})\,,
\en
\eq
G (\bar \xi_{1},\partial\xi_{2}) = 0
\,,\qquad
G (\overbar{\partial\xi_{1}},\xi_{2})=0
\,,
\en
\eq
G (\bar \xi_{1},\xi_{2}) + G (\overbar{{*}\xi_{1}},{*}\xi_{2})
= I_{M}(\bar\xi_{1},{*}\xi_{2})
\,.
\label{eq:GIidentity}
\en
A simple way to derive the last equation, (\ref{eq:GIidentity}), is by calculating
\eq
G (p)_{1,2,\ldots,n;1,2,\ldots,n} = 
\frac{1}{p^{2}}\sum_{i=1}^{n} p_{i}^{2}
\,,\qquad
G (p)_{n+1,n+2,\ldots,d;n+1,n+2,\ldots,d} = 
\frac{1}{p^{2}}\sum_{i=n+1}^{d} p_{i}^{2}
\,.
\label{eq:Gcalc}
\en
Equation (\ref{eq:GIidentity}) is equivalent to
\eq
G (\overbar{P_{+} \xi_{1}},P_{+}\xi_{2})
= \frac{i}2 \IM{\bar\xi_{1}}{P_{+}\xi_{2}}
\,,\qquad
G (\overbar{P_{-} \xi_{1}},P_{-}\xi_{2})
= - \frac{i}2 \IM{\bar\xi_{1}}{P_{-}\xi_{2}}
\en

The first S-D equations are imposed by fiat,
\eq
\expval{F^{\dagger}(\bar\xi_{1}) F(\partial\xi_{2})} = 0
\,,\qquad
\expval{F^{*\dagger}(\bar\xi_{1}) F^{*}(\partial\xi_{2})} = 0
\,.
\en
These determine two of the two-point functions,
\eq
\expval{F^{\dagger}(\bar\xi_{1}) F(\xi_{2})} = B_{n} G (\bar 
\xi_{1},\xi_{2})
\,,\qquad
\expval{F^{*\dagger}(\bar\xi_{1}) F^{*}(\xi_{2})} = B_{n} G (\bar 
\xi_{1},\xi_{2})
\en
where $B_{n}$ is a real constant.  
By (\ref{eq:Gcalc}), reflection positivity implies
\eq
B_{n}>0\,.
\en
The remaining two-point functions have the form
\aeq{
\expval{F^{*\dagger}(\bar\xi_{1}) F(\xi_{2})} &= B_{n} i^{-1} \delta_{n}\left[
 G (\overbar{{*}\xi_{1}},\xi_{2})
+b_{n} I_{M}(\bar\xi_{1},\xi_{2})
\right ]
\\
\expval{ F^{\dagger}(\bar\xi_{1})F^{*}(\xi_{2})}&= B_{n} i^{-1}  \delta_{n}\left[
 G (\bar\xi_{1},{*}\xi_{2})
+\bar b_{n} I_{M}(\xi_{2},\bar\xi_{1})
\right ]
}
for some complex constant $b_{n}$.  
The two-point functions of the chiral components are then
\aeq{
\expval{ F^{\dagger}_{+}(\bar\xi_{1})F_{-}(\xi_{2})}
&=
B_{n} \left[
- \frac{i}4 (1+b_{n}+\bar b_{n}) \IM{\bar\xi_{1}}{\xi_{2}}
\right ]
\\
\expval{ F^{\dagger}_{-}(\bar\xi_{1})F_{+}(\xi_{2})}
&=
B_{n} \left[
\frac{i}4 (1+b_{n}+\bar b_{n}) \IM{\bar\xi_{1}}{\xi_{2}}
\right ]
\\
\expval{ F^{\dagger}_{+}(\bar\xi_{1})F_{+}(\xi_{2})}
&=
B_{n} \left[
G (\overbar{P_{-}\xi_{1}},\xi_{2})
+\frac{i}4 (1-b_{n}+\bar b_{n}) \IM{\bar\xi_{1}}{\xi_{2}}
\right ]
\\
&=
B_{n} \left[
G (\bar\xi_{1},P_{+}\xi_{2})
-\frac{i}4 (1+b_{n}-\bar b_{n}) \IM{\bar\xi_{1}}{\xi_{2}}
\right ]
\\
\expval{ F^{\dagger}_{-}(\bar\xi_{1})F_{-}(\xi_{2})}
&=
B_{n} \left[
G (\overbar{P_{+}\xi_{1}},\xi_{2})
 - \frac{i}4 (1-b_{n}+\bar b_{n}) \IM{\bar\xi_{1}}{\xi_{2}}
\right ]
\\
&=
B_{n} \left[
G (\bar\xi_{1},P_{-}\xi_{2})
 + \frac{i}4 (1+b_{n}-\bar b_{n}) \IM{\bar\xi_{1}}{\xi_{2}}
\right ]
}
The most symmetric choice is $b_{n}=-\frac12$, giving
\aeq{
\expval{ F^{\dagger}_{+}(\bar\xi_{1})F_{-}(\xi_{2})}
&= 0
\\
\expval{ F^{\dagger}_{-}(\bar\xi_{1})F_{+}(\xi_{2})}
&= 0
\\
\expval{ F^{\dagger}_{+}(\bar\xi_{1})F_{+}(\xi_{2})}
&=
B_{n} \left[
G (\overbar{P_{-}\xi_{1}},\xi_{2})
+\frac{i}4  \IM{\bar\xi_{1}}{\xi_{2}}
\right ]
\\
&=
B_{n} \left[
G (\bar\xi_{1},P_{+}\xi_{2})
-\frac{i}4  \IM{\bar\xi_{1}}{\xi_{2}}
\right ]
\label{eq:FppG}
\\
\expval{ F^{\dagger}_{-}(\bar\xi_{1})F_{-}(\xi_{2})}
&=
B_{n} \left[
G (\overbar{P_{+}\xi_{1}},\xi_{2})
 - \frac{i}4  \IM{\bar\xi_{1}}{\xi_{2}}
\right ]
\\
&=
B_{n} \left[
G (\bar\xi_{1},P_{-}\xi_{2})
 + \frac{i}4  \IM{\bar\xi_{1}}{\xi_{2}}
\right ]
}
The S-D equations follow immediately,
\ateq{2}{
\expval{ F^{\dagger}_{+}(\bar\xi_{1})F_{+}(\partial\xi_{2})}
&= B_{n}\frac{i}4 I\expval{\bar\xi_{1},\partial\xi_{2}}
&&= - B_{n}\frac{i}4 \IM{\partial\bar\xi_{1}}{\xi_{2}}
\\
\expval{ F^{\dagger}_{+}(\overbar{\partial\xi_{1}})F_{+}(\xi_{2})}
&=- B_{n}\frac{i}4 \IM{\partial\bar\xi_{1}}{\xi_{2}}
&&= B_{n}\frac{i}4 \IM{\bar\xi_{1}}{\partial\xi_{2}}
\\
\expval{ F^{\dagger}_{-}(\bar\xi_{1})F_{-}(\partial\xi_{2})}
&= - B_{n}\frac{i}4 \IM{\bar\xi_{1}}{\partial\xi_{2}}
&&= B_{n}\frac{i}4 \IM{\partial\bar\xi_{1}}{\xi_{2}}
\\
\expval{ F^{\dagger}_{-}(\overbar{\partial\xi_{1}})F_{-}(\xi_{2})}
&=B_{n}\frac{i}4 \IM{\partial\bar\xi_{1}}{\xi_{2}}
&&= -B_{n}\frac{i}4 \IM{\bar\xi_{1}}{\partial\xi_{2}}
\,.
}
These S-D equations are compatible with $dA_{\pm}=F_{\pm}$.
Integrate them to get
\aeq{
\expval{ A^{\dagger}_{+}(\bar\xi_{0})F_{+}(\partial\xi_{2})}
&= - B_{n}\frac{i}4 \IM{\bar\xi_{0}}{\xi_{2}}
\\
\expval{ F^{\dagger}_{+}(\overbar{\partial\xi_{2}})A_{+}(\xi_{0})}
&= B_{n}\frac{i}4 \IM{\bar\xi_{2}}{\xi_{0}}
\\
\expval{ A^{\dagger}_{-}(\bar\xi_{0})F_{-}(\partial\xi_{2})}
&= B_{n}\frac{i}4 \IM{\bar\xi_{0}}{\xi_{2}}
\\
\expval{ F^{\dagger}_{-}(\overbar{\partial\xi_{2}})A_{-}(\xi_{0})}
&= -B_{n}\frac{i}4 \IM{\bar\xi_{2}}{\xi_{0}}
}
The normalization constants $B_{n}$ are fixed by matching to a 
standard convention in $d=2$ dimensions,
as described in section~\ref{sect:d=2freenform} below,
\eq
B_{n} = 8\pi
\,.
\en

\subsection{Summary}
The Schwinger-Dyson equations for the chiral fields are
\ateq{2}{
\expval{ F^{\dagger}_{\bar \alpha}(\bar\xi_{1})F_{\beta}(\partial\xi_{2})}
&= 2\pi i \gamma_{\bar \alpha\beta} \IM{\bar\xi_{1}}{\partial\xi_{2}}
&&= - 2\pi i\gamma_{\bar \alpha\beta} \IM{\partial\bar\xi_{1}}{\xi_{2}}
\\
\expval{ A^{\dagger}_{\bar \alpha}(\bar\xi_{0})F_{\beta}(\partial\xi_{2})}
&= - 2\pi i\gamma_{\bar \alpha\beta} \IM{\bar\xi_{0}}{\xi_{2}}
}
and the complex conjugate equations
\ateq{2}{
\expval{ F^{\dagger}_{\bar \beta}(\overbar{\partial\xi_{2}})F_{\alpha}(\xi_{1})}
&=- 2\pi i \gamma_{\bar \beta\alpha} 
\IM{\partial\bar\xi_{2}}{\xi_{1}}
&&=2\pi i \gamma_{\bar \beta\alpha} 
\IM{\bar\xi_{2}}{\partial\xi_{1}}
\\
\expval{ F^{\dagger}_{\bar \beta}(\overbar{\partial\xi_{2}})A_{\alpha}(\xi_{0})}
&= 2\pi i\gamma_{\bar \beta\alpha} \IM{\bar\xi_{2}}{\xi_{0}}
}
where
\eq
\gamma_{\bar ++}=1
\,,\quad
\gamma_{\bar +-}=0
\,,\quad
\gamma_{\bar -+}=0
\,,\quad
\gamma_{\bar --}= -1\,.
\en
In terms of the fields $F$, $F^{*}$, $A$, $A^{*}$,
\eq
F = F_{+}+F_{-}
\,,\quad
F^{*} = \frac1{\epsilon_{n}\delta_{n}}(F_{+}-F_{-})
\,,\quad
A = A_{+}+A_{-}
\,,\quad
A^{*} = \frac1{\epsilon_{n}\delta_{n}}(A_{+}-A_{-})
\,,
\en
the nontrivial S-D equations are
\ateq{2}{
\expval{F^{*\dagger}(\bar\xi_{1}) F(\partial\xi_{2})}
&= 4\pi i\delta_{n} I_{M}(\bar\xi_{1},\partial\xi_{2})
&&= 4\pi i \delta_{n} (-1)^{n}I_{M}(\overbar{\partial\xi_{1}},\xi_{2})
\\
\expval{F^{\dagger}(\bar\xi_{1}) F^{*}(\partial\xi_{2})}
&= 4\pi i\delta_{n}(-1)^{n} I_{M}(\bar\xi_{1},\partial\xi_{2})
&&= 4\pi i \delta_{n} I_{M}(\overbar{\partial\xi_{1}},\xi_{2})
\\[1ex]
\expval{A^{*\dagger}(\bar\xi_{0}) F(\partial\xi_{2})}
&= 4\pi i \delta_{n} (-1)^{n}I_{M}(\bar\xi_{0},\xi_{2})
\\
\expval{A^{\dagger}(\bar\xi_{0}) F^{*}(\partial\xi_{2})}
&= 4\pi i \delta_{n} I_{M}(\bar\xi_{0},\xi_{2})
}
and their complex conjugates (recalling that
$F^{\dagger}=\bar F$, $F^{*\dagger} = - \overbar{F^{*}}$),
\ateq{2}{
\expval{F^{\dagger}(\overbar{\partial\xi_{2}}) F^{*}(\xi_{1}) }
&= 4\pi i\delta_{n} I_{M}(\xi_{1},\overbar{\partial\xi_{2}})
&&= 4\pi i \delta_{n} (-1)^{n}I_{M}(\partial\xi_{1},\bar\xi_{2})
\\
\expval{F^{*\dagger}(\overbar{\partial\xi_{2}})F(\xi_{1}) }
&= 4\pi i\delta_{n}(-1)^{n} I_{M}(\xi_{1},\overbar{\partial\xi_{2}})
&&= 4\pi i \delta_{n} I_{M}(\xi_{1},\overbar{\partial\xi_{2}})
\\[1ex]
\expval{F^{\dagger}(\overbar{\partial\xi_{2}})A^{*}(\xi_{0}) }
&= 4\pi i \delta_{n} (-1)^{n}I_{M}(\xi_{0},\bar\xi_{2})
\\
\expval{F^{*\dagger}(\overbar{\partial\xi_{2}})A(\xi_{0}) }
&= 4\pi i \delta_{n} I_{M}(\xi_{0},\bar\xi_{2})
\,.
}

\subsection{$d=2$}
\label{sect:d=2freenform}

In $d=2$ dimensions, on $\Reals^{2}$,
using coordinates $z=x^{1}+i x^{2}$,
$w = y^{1}+iy^{2}$,
a standard convention is
\eq
F_{+} = j(z)dz
\,,\quad
F^{\dagger}_{+} = j^{\dagger} (z) dz
\,,\qquad
F_{-} = \bar \jmath(\bar z)d\bar z
\,,\quad
F^{\dagger}_{-} = \bar \jmath^{\dagger} (\bar z) d\bar z
\en
\eq
\expval{j^{\dagger} (z) j(w)} = \frac{-2}{(z-w)^{2}}
\,,\qquad
\expval{\bar \jmath^{\dagger}(\bar z) \jmath (\bar w) } = \frac{-2}{(\bar z-\bar w)^{2}}
\,.
\en
The identity
\eq
\partial_{\bar z} \left(\frac1{z-w}\right) = \pi \delta^{2}(z-w)
\en
is
\eq
\frac1{z-w} = \int\frac{d^{2}p}{(2\pi)^{2}}\;e^{ip(x-y)}\; \frac{(-4\pi 
i) p_{z}}{p^{2}}
\en
where
\eq
p(x-y) = p_{z} (z-w) +p_{\bar z}(\bar z-\bar w)
\,,\qquad
p^{2} = 4 p_{z}p_{\bar z}
\,.
\en
So
\eq
\frac{-2}{(z-w)^{2}} = 2 \partial_{z}
\left(\frac1{z-w}\right)
= 
 \int\frac{d^{2}p}{(2\pi)^{2}}\;e^{ip(x-y)}\; \frac{8\pi  p_{z}p_{z}}{p^{2}}
\,.
\en
Comparing to equation (\ref{eq:G}) gives
\eq
\expval{j^{\dagger} (z)dz\, j(w)dw} =
8\pi G(x,y)_{\mu;\nu} dx^{\mu}dx^{\nu}
\,.
\en
Comparing to equation (\ref{eq:FppG})
---
away from coincident points so
the contact terms can be ignored
---
gives the normalization
\eq
B_{n} = 8\pi
\,.
\en

\section{Vertex operators and the Dirac quantization condition}
\subsection{The vertex operators}

The vertex operators for the complex scalar on $\ExC$ are, in terms of $\phi$ and $\phi^{*}$,
\eq
V_{p,p^{*}}(\eta) = e^{
i p\cdot \phi(\eta) + i p^{*}\cdot \phi^{*}(\eta)
} 
\,,\qquad
\eta\in\cD^{\integral}_{0}(\ExC)
\,,
\en
\eq
p \cdot \phi(\eta) = \frac12 \left[
\bar p \phi(\eta) + p \phi^{\dagger}(\bar\eta)
\right]
\,,\qquad
p^{*}\cdot \phi^{*}(\eta) = \frac12 \left[
\overbar{p^{*}} \phi^{*}(\eta) + p^{*} \phi^{*\dagger}(\bar\eta)
\right]
\,.
\en
The reality condition is $p=\bar p$, $p^{*}=\overbar{p^{*}}$.
In terms of the chiral fields, the vertex operators are
\eq
V_{p^{+},p^{-}}(\eta) = e^{
i p^{+}\cdot \phi_{+}(\eta) + i p^{-}\cdot \phi_{-}(\eta)
} 
\,,
\en
\eq
p^{+}\cd  \phi_{+}(\eta) = \frac12\left [
\overbar{p^{+}}\phi_{+}(\eta) + p^{+} \phi_{+}^{\dagger}(\bar \eta)
\right ]
\,,\quad
p^{-}\cd  \phi_{-}(\eta) = \frac12\left [
\overbar{p^{-}}\phi_{-}(\eta) + p^{-} \phi_{-}^{\dagger}(\bar\eta)
\right ]
\,,
\en
\eq
%\label{eq:chiralcharges}
p^{+} = p + \epsilon_{n} p^{*}
\,,\qquad
p^{-} = p - \epsilon_{n}  p^{*}
\,.
\en
The vertex operators satisfy the operator product equations
\aeq{
V_{p^{+},p^{-}}(\eta_{0}) j_{\beta}(\partial \eta_{2}) &= 
- \pi i \overbar{p^{\alpha}}\gamma_{\bar\alpha\beta} 
I_{M}\expval{\bar\eta_{0}, \eta_{2}}
\,V_{p^{+},p^{-}}(\eta_{0})
\\[1ex]
j^{\dagger}_{\bar\beta}(\overbar{\partial \eta_{2}}) V_{p^{+},p^{-}}(\eta_{0})  &= 
\pi  i p^{\alpha}\gamma_{\bar\beta\alpha} I_{M}\expval{\bar\eta_{2}, 
\eta_{0}}
\,V_{p^{+},p^{-}}(\eta_{0})
}
expressing the generalized $U(1)$ charges \cite{Gaiotto:2014kfa} of the extended objects.

\subsection{The Dirac quantization condition}
\label{sect:Diracquantcond}

The Dirac quantization condition is derived from the requirement
that the correlation functions of the vertex operators be 
single-valued on $\Ex$.

Let $\xi_{0}$ be a point in $\Ex$ and
let $\eta_{0}=\delta_{\xi_{0}}$ be the 0-current in $\Ex$  representing 
$\xi_{0}$.
Let $\eta_{2}$ represent a 2-disk in $\Ex$ such that
the $(n{-}1)$-current $\xi_{0}=\Pi_{*}\eta_{0}$ and the $(n{+}1)$-current 
$\Pi_{*}\eta_{2}$ have intersection number 1 in $M$,
\eq
\Pi^{*}I_{M}(\eta_{0},\eta_{2}) = 1
\,.
\en
Then the skew-hermitian $M$-intersection form has values
\eq
\label{eq:IMvalues}
\Pi^{*}I_{M}\expval{\bar\eta_{0},\eta_{2}} = \epsilon_{n}
\,,\qquad
\Pi^{*}I_{M}\expval{\bar\eta_{2},\eta_{0}} = -\bar \epsilon_{n}
\,.
\en
Consider the product of vertex operators
\eq
V_{p^{+},p^{-}}(\delta_{\xi_{0}}) 
\;V_{q^{+},q^{-}}(\delta_{\xi_{1}}) 
\en
as $\xi_{1}$ moves around the boundary $\partial\eta_{2}$ of the disk in $\Ex$ 
represented by $\eta_{2}$.  The monodromy will be
\eq
e^{
- 
\expval{
p^{+}\cd \phi_{+}(\eta_{0}) \,q^{+}\cd  j_{+}(\partial\eta_{2})
}
- 
\expval{
p^{-}\cd \phi_{-}(\eta_{0}) \,q^{-}\cd  j_{-}(\partial\eta_{2})
}
}
\,.
\en
The S-D equations combined with (\ref{eq:IMvalues}) give
\aeq{
\expval{
p^{+}\cd \phi_{+}(\eta_{0})\, q^{+}\cd  j_{+}(\partial\eta_{2})
}
&= 
-\frac12 \pi i \left( \bar\epsilon_{n} \overbar{ p^{+}}q^{+}
+
\epsilon_{n} p^{+}\overbar{ q^{+}}\right)
\\[2ex]
\expval{
p^{-}\cd \phi_{-}(\eta_{0})\, q^{-}\cd  j_{-}(\partial\eta_{2})
}
&= 
\frac12 \pi i \left( \bar\epsilon_{n} \overbar{ p^{-}}q^{-}
+
\epsilon_{n}p^{-}\overbar{ q^{-}} \right)
}
so the Dirac quantization condition --- the condition that the correlation function be 
single-valued, that the monodromy equal 1 --- is
\eq
\frac14\left[
\overbar{ p^{+}}q^{+}\bar\epsilon_{n}
+
p^{+}\overbar{ q^{+}} \epsilon_{n}
-
 \overbar{ p^{-}}q^{-}\bar\epsilon_{n}
-
p^{-}\overbar{ q^{-}} \epsilon_{n}
\right]
\in \Integers
\en
which can be written
\eq
\label{eq:Diracquantgen}
\frac12 (\epsilon_{n}p^{+})\cd q^{+}
-
\frac12 (\epsilon_{n}p^{-})\cd q^{-}
\in \Integers
\,,
\en
where the euclidean inner product on complex charges is
\eq
p\cd  q = \frac12(\bar p q+ p \bar q)
\,.
\en
Substituting the electric and magnetic charges,
the Dirac quantization condition becomes
\eq
\label{eq:Diracquantcond1}
p\cd  q^{*} +(-1)^{n-1} p^{*}\cd q \in \Integers
\,.
\en
The Dirac quantization condition
in the real case, $p,p^{*},q,q^{*}\in \Reals$, is
\eq
p q^{*} +(-1)^{n-1} p^{*} q \in \Integers
\,.
\en
The charges lie in real lattices
\eq
p =\frac{m}{R}
\,,\quad
q =\frac{n}{R}
\,,\qquad
p ^{*}=\frac{m^{*}}{R^{*}}
\,,\quad
q^{*}=\frac{n^{*}}{R^{*}}
\en
so the Dirac quantization condition is $R R^{*} =1$.
In the complex case,
the charges $p,q$ lie in a two-dimensional lattice $L\subset\Complexes$
and the $p^{*}$, $q^{*}$ lie in a two-dimensional lattice $L^{*}\subset\Complexes$.
The Dirac quantization condition (\ref{eq:Diracquantcond1}) is the 
condition that $L$ and $L^{*}$ are dual lattices --- the euclidean 
inner-product between an element of $L$ and an element of $L^{*}$ is 
always an integer.

\subsection{$d=2$}
We can check the Dirac quantization condition in $d=2$
using the explicit formulas of section \ref{sect:d=2freenform}.
The scalar fields are given by 
\eq
\partial \phi_{+}(z) = j(z),
\,,\quad
\bar \partial \phi_{-}(\bar z) = j(\bar z)
\,.
\en
Their two-point functions are
\eq
\expval{\phi^{\dagger}_{+}(z) \phi_{+}(w)} = -2 \ln(z-w)
\,,\quad
\expval{\phi^{\dagger}_{-}(\bar z) \phi_{-}(\bar w)} = -2 \ln(\bar z-\bar w)
\,.
\en
The vertex operators
\eq
V_{p^{+},p^{-}}(z,\bar z) = e^{
i p^{+}\cdot \phi_{+}(z) + i p^{-}\cdot \phi_{-}(\bar z)
} 
\en
have two-point functions
\eq
\expval{V_{p^{+},p^{-}}(z,\bar z) \, V_{q^{+},q^{-}}(w,\bar w)}
= (z-w)^{p^{+}\cdot q^{+}} (\bar z-\bar w)^{p^{-}\cdot q^{-}} 
\en
which are single-valued iff
\eq
p^{+}\cd q^{+} - p^{-}\cd q^{-} \in \Integers
\,.
\en
The reality condition for $d=2$ is $p^{\pm}= \overbar{p^{\pm}}$,
giving
\eq
p^{+}\cd q^{+} =p^{+} q^{+}
\,,\qquad
p^{-}\cd q^{-} = p^{-} q^{-}
\,.
\en
The usual sign choices in $d=2$ are  $\epsilon_{1}=1$, $\delta_{1}=1$,
for which
\eq
p^{\pm} = \frac12 (p\pm p^{*})
\en
so the Dirac quantization condition is
\eq
p q^{*} + p^{*} q \in \Integers
\en
which, for $p=\frac{m}{R}$, $p^{*}=\frac{m^{*}}{R^{*}}$,
$q=\frac{n}{R}$, $p^{*}=\frac{n^{*}}{R^{*}}$,
gives $R R^{*} =1$.

\section{Complex conjugation and reality conditions}
\label{app:complexconjugation}

The difference between  $n$ even and $n$ odd
shows up in the class of 2d quantum field theories
that live on the quasi-holomorphic curves.
The S-D equations on the quasi-holomorphic curves,
written in terms of $J$ and the skew-hermitian 
$M$-intersection form,
are equations on complex fields $j$ and $\phi$.
For $n$ odd, both $J$ and $I_{M}\expval{\bar\eta_{1},\eta_{2}}$ are real
and $\Ex$ can remain a real space.
The quasi-holomorphic curve $C$ is a real map from $\Sigma$ as a real 
two-dimensional manifold to $\Ex$.
The reality condition $F=\bar F$ on the $n$-form field
becomes the reality condition $j = \bar j$, $\phi=\bar \phi$
on the 2d fields.
The 2d quantum field theory is the gaussian model of a single real 
1-form $j$.
For $n$ even, both $J$ and $I_{M}\expval{\bar\eta_{1},\eta_{2}}$ are 
imaginary and $\Ex$ must be complexified to $\ExC$.
The quasi-holomorphic curve $C$ will be a map from $\Sigma$
to $\ExC$.
The fields $j$ and $\phi$ on the q-h curve remain complex.
The 2d quantum field theory is a two-component gaussian model of a complex
1-form $j$.
The reality condition $F=\bar F$ on the $n$-form field becomes 
invariance under a
$\Integers_{2}$ symmetry of the 2d quantum field theory on
the quasi-holomorphic curve,
that combines complex conjugation on the fields and orientation reversal
on the Riemann surface.
Any construction on 2d field theories will preserve this 
$\Integers_{2}$ symmetry,
so, for $n$ even, the general class of 2d quantum field theories 
to be considered are the 2d theories with this symmetry.

\subsection{Complex conjugation on the fields}

Because the almost complex structure is given by $J=\epsilon_{n}{*}$ 
with $\epsilon_{n}^{2}=(-1)^{n-1}$,
and $P_{\pm} = \frac12 (1\pm i^{-1} J)$,
complex conjugation acts differently for $n$ odd and $n$ even:
\eq
{\bar J}
= 
\left \{
\begin{array}{ll}
J\,,
\quad
& n \text{ odd,}
\\[2ex]
-J\,, & n \text{ even,}
\end{array}
\right .
\qquad
{\bar P_{\pm}}
= 
\left \{
\begin{array}{ll}
P_{\mp}\,,
\quad
& n \text{ odd,}
\\[2ex]
P_{\pm}\,, & n \text{ even.}
\end{array}
\right .
\en
So, for $F$ a complex $n$-form field on $M$,
using equations (\ref{eq:Fadj}--\ref{eq:Aadj}) for the adjoint fields,
\ateq{3}{
\text{for } n \text{ odd}:&\qquad&
F^{\dagger}_{\pm}(\bar \xi)=\overbar{F_{\mp}}(\bar\xi) &= \bar F_{\pm}(\bar\xi) \,,\qquad&
A^{\dagger}_{\pm}(\bar \xi)=\overbar{A_{\mp}}(\bar\xi) &= \bar A_{\pm}(\bar\xi)
\,,
\\[1ex]
\text{for } n \text{ even}:&\qquad&
F^{\dagger}_{\pm}(\bar \xi) =\overbar{F_{\mp}}(\bar\xi) &= \bar F_{\mp}(\bar\xi) \,,\qquad&
A^{\dagger}_{\pm}(\bar \xi)=\overbar{A_{\mp}}(\bar\xi) &= \bar A_{\mp}(\bar\xi)
\,.
}

\subsection{Reality conditions on the fields}
The condition that $F$ is a real field, $F=\bar F$, is
\ateq{3}{
\text{for } n \text{ odd}:&\qquad&
F^{\dagger}_{\pm} &= F_{\pm} \,,\qquad&
A^{\dagger}_{\pm} &= A_{\pm}
\,,
\\[1ex]
\text{for } n \text{ even}:&\qquad&
F^{\dagger}_{\pm}  &= F_{\mp} \,,\qquad&
A^{\dagger}_{\pm} &= A_{\mp}
\,.
}
As a check, re-write these relations in terms of the
the usual fields $F$ and $F^{*} = i^{-1}{*}F$,
\ateq{2}{
F & = F_{+}+F_{-}\,,\qquad & A & = A_{+}+A_{-}\,,
\\
\epsilon_{n} F^{*}& = F_{+}-F_{-}\,, \qquad &  \epsilon_{n}A^{*}& = 
A_{+}- A_{-}
\,,
\nonumber
}
getting, for real $F$,
\eq
F^{\dagger}  = F
\,,\qquad
F^{*\dagger}  = F^{*}
\,,\qquad
A^{\dagger}  =  A
\,,\qquad
A^{*\dagger}  =  A^{*}
\,,
\en
which is indeed  self-adjointness of
the magnetic and electric fields and the gauge potentials.

The reality condition on
the 1-form field $j$ and the 0-form field $\phi$
on $\ExC$
is the transcription,
\ateq{3}{
\text{for } n \text{ odd}:&\qquad&
j^{\dagger}_{\pm} &= j_{\pm} \,,\qquad&
\phi^{\dagger}_{\pm} &= \phi_{\pm}
\,,
\label{eq:jrealitynodd}
\\[1ex]
\text{for } n \text{ even}:&\qquad&
j^{\dagger}_{\pm}  &= j_{\mp} \,,\qquad&
\phi^{\dagger}_{\pm} &= \phi_{\mp}
\,.
\label{eq:jrealityneven}
}
The 1-form and 0-form fields pulled back along a quasi-holomorphic 
curve $C$ to the Riemann surface $\Sigma$,
\ateq{2}{
j_{+}(z) &= j_{+}(C_{*}\delta_{z})
\,,\qquad&
\phi_{+}(z) &= \phi_{+}(C_{*}\delta_{z})
\,,
\\
j_{-}(\bar z) &= j_{-}(C_{*}\delta_{z})
\,,\qquad&
\phi_{-}(\bar z) &= \phi_{-}(C_{*}\delta_{z})
\,,
}
will satisfy the same reality condition 
(\ref{eq:jrealitynodd}--\ref{eq:jrealityneven}).

For $n$ odd, the reality condition (\ref{eq:jrealitynodd})
is the reality condition satisfied by a real 1-form 
on the Riemann surface.
So the 2d conformal field theory on a q-h curve
is the theory of a free real 1-form.

For $n$ even, the reality condition (\ref{eq:jrealityneven}) is {\it not} the reality
condition of a real 1-form field on the Riemann surface $\Sigma$.
The 2d conformal field theory on the q-h curve
is the free {\it complex} 1-form.
The reality condition on $M$
becomes a symmetry condition,
$j^{\dagger}_{\pm}= j_{\mp}$,
$\phi^{\dagger}_{\pm}= \phi_{\mp}$.
This is an anti-linear symmetry that reverses the 2d orientation.
The fields in the real theory on the space-time $M$ correspond to the invariant subalgebra of fields
in the complex 1-form theory on the Riemann surface $\Sigma$.

\subsection{Reality conditions on the vertex operators}

For the vertex operators
\eq
V_{p^{+},p^{-}}(\eta) = e^{
i p^{+}\cdot \phi_{+}(\eta) + i p^{-}\cdot \phi_{-}(\eta)
} 
\,,
\en
\eq
p^{+}\cd  \phi_{+}(\eta) = \frac12\left [
\overbar{p^{+}}\phi_{+}(\eta) + p^{+} \phi_{+}^{\dagger}(\bar \eta)
\right ]
\,,\quad
p^{-}\cd  \phi_{-}(\eta) = \frac12\left [
\overbar{p^{-}}\phi_{-}(\eta) + p^{-} \phi_{-}^{\dagger}(\bar\eta)
\right ]
\,,
\en
\eq
%\label{eq:chiralcharges}
p^{+} = p + \epsilon_{n} p^{*}
\,,\qquad
p^{-} = p - \epsilon_{n}  p^{*}
\,,
\en
The reality condition $p=\bar p$, $p^{*}=\overbar{p^{*}}$
becomes
\ateq{2}{
\text{for } n \text{ odd}:&\qquad&
p^{\pm}&= \overbar{p^{\pm}}
\,,
\\[1ex]
\text{for } n \text{ even}:&\qquad&
p^{\pm}&=\overbar{p^{\mp}}
\,.
}
For $n$ odd, the vertex operators of the real $n$-form on 
space-time are exactly the vertex operators of the real $1$-form
on the Riemann surface.
For $n$ even, the space-time vertex operators of the real $n$-form 
theory correspond to the subalgebra of the vertex operators of the 2d 
theory,
\eq
\left \{ V_{p^{+},p^{-}}:
\: p^{-} = \overbar{p^{+}}
\right \}
\,.
\en

The Dirac quantization condition on $M$ is
\eq
\label{eq:DirquantM}
\frac12 \left [ (\epsilon_{n} p^{+})\cd  q^{+} -  (\epsilon_{n} 
p^{-})\cd  q^{-} \right ]  \in \Integers
\,.
\en
The Dirac quantization condition on $\Sigma$ is
\eq
\label{eq:DirquantSigma}
\frac12 \left ( p^{+}\cd q^{+} - p^{-}\cd q^{-} \right )\in \Integers
\,.
\en
For $n$ odd, these are the same, so the
real theory on $M$ corresponds to the real theory on $\Sigma$.
For $n$ even, the Dirac quantization conditions (\ref{eq:DirquantM}) and 
(\ref{eq:DirquantSigma}) are different.
Writing $p_{\Sigma}$, $p^{*}_{\Sigma}$ for the magnetic and electric 
charges on $\Sigma$,
and writing $p_{M}$, $p^{*}_{M}$ for the charges on $M$ (formerly 
$p$, $p^{*}$),
\eq
p_{\Sigma} = \frac12 (p^{+}+p^{-}) = p_{M}
\,,\qquad
\epsilon_{1} p^{*}_{\Sigma} 
 = \frac12 (p^{+}- p^{-}) = \epsilon_{n} p^{*}_{M}
\,.
\en
For $n$ even, real magnetic charges on $M$ become imaginary magnetic 
charges in the complex 1-form theory on $\Sigma$.
The real charges $p_{M}$, $p^{*}_{M}$ on $M$ lie in dual lattices 
$L\subset \Reals$ and $L^{*}\subset \Reals$.
The corresponding charges in the complex 1-form theory on $\Sigma$ 
lie in $L\subset \Complexes$ and $i L^{*} \subset 
\Complexes$.  So the full lattice of electric charges $p_{\Sigma}$ on 
$\Sigma$ must be $L\oplus i L\subset \Complexes$, and the full 
lattice of magnetic charges $p^{*}_{\Sigma}$ must be 
$L^{*}\oplus i L^{*}\subset \Complexes$,
\eq
\label{eq:Sigmacharges}
p_{\Sigma} = \frac{m_{1}+i m_{2}}{R}
\,,\qquad
p^{*}_{\Sigma} = \frac{m^{*}_{1}+i m^{*}_{2}}{R^{*}}
\en
Complex conjugation on $M$ is
\eq
p_{\Sigma} \mapsto \bar p_{\Sigma}
\,,\qquad
p^{*}_{\Sigma} \mapsto -\overbar{p^{*}_{\Sigma}}
\,.
\en
All of the vertex operators with charges 
(\ref{eq:Sigmacharges}) are well-defined on $\Sigma$,
but only those invariant under complex conjugation on $M$ are 
well-defined on $\Ex$ and on $M$.
The q-h curve lies not in $\Ex$ but in $\ExC$, so 
vertex operators can have well-defined correlation functions on the 
q-h curve, yet not be well-defined on $\Ex$.

% \newpage
% 
% \appendix
% \appendixpage
% \vskip2ex

\bibliographystyle{ytphys}
\addtotoc{References}{\bibliography{Literature}}
%\bibliography{Literature}

\end{document}